\documentclass[preprint,12pt]{emulateapj}

\newcommand{\beq}{\begin{equation}}           
\newcommand{\eeq}{\end{equation}}             
\newcommand{\bef}{\begin{figure}}           
\newcommand{\enf}{\end{figure}}             
\newcommand{\rmd}{{\rm d}}                    
\newcommand{\delex}{\delta_{\rm ex}}       
\newcommand{\mean}[1]{\langle {#1} \rangle}     
\newcommand{\cacr}{_{\rm cc}}           
\newcommand{\degree}{\ensuremath{^\circ}}   
\newcommand{\delappr}{\ensuremath{\delta_{\rm appr}}}   
\newcommand{\delchr}{\ensuremath{\delta_{\rm chr}}}   
\newcommand{\actaa}{Acta Astron.}
\shorttitle{Extended Source in Binary Microlensing}

\begin{document}

\title{Extended-Source Effect and Chromaticity in Two-Point-Mass Microlensing}

\author{Ond\v{r}ej Pejcha and David Heyrovsk\'y}

\affil{Institute of Theoretical Physics, Charles University, Prague, Czech Republic
\email{ondrej.pejcha@gmail.com}\email{heyrovsky@utf.mff.cuni.cz}}

\begin{abstract}
We explore the sensitivity of two-point-mass gravitational microlensing to the extended nature of the source star, as well as the related sensitivity to its limb darkening. We demonstrate that the sensitive region, usually considered to be limited to a source-diameter-wide band along the caustic, is strongly expanded near cusps, most prominently along their outer axis. In the case of multi-component caustics, facing cusps may form a region with a non-negligible extended-source effect spanning the gap between them. We demonstrate that for smaller sources the size of the sensitive region extending from a cusp measured in units of source radii increases, scaling as the inverse cube root of the radius. We study the extent of different sensitivity contours and show that for a microlensed Galactic bulge giant the probability of encountering at least a $1\%$ extended-source effect is higher than the probability of caustic crossing by $40$--$60\%$ when averaged over a typical range of lens-component separations, with the actual value depending on the mass ratio of the components. We derive analytical expressions for the extended-source effect and chromaticity for a source positioned off the caustic. These formulae are more generally applicable to any gravitational lens with a sufficiently small source. Using exactly computed amplifications we test the often used linear-fold caustic approximation and show that it may lead to errors on the level of a few percent even in near-ideal caustic-crossing events. Finally, we discuss several interesting cases of observed binary and planetary microlensing events and point out the importance of our results for the measurement of stellar limb darkening from microlensing light curves.
\end{abstract}


\section{INTRODUCTION}
\label{sec:intro}

Gravitational lensing by a system of two point masses was first studied theoretically by \cite{schn_weiss86}, who mentioned that the model may be of some relevance for lensing of a distant source by a binary star in our Galaxy. At the same time they stressed that their study was not motivated by its applicability, the model was investigated rather as a specific example of a non-symmetric lens. The applicability of the model was studied by \cite{mao_paczynski91}, who estimated that $\sim 10\%$ of events observed in microlensing surveys of stars in the Galactic bulge would be strong binary-lensing events. In addition, they were the first to demonstrate the possibility of detecting microlensing by a star with a planet, as an extreme case of the two-point-mass lens.

Although the numbers of observed strong binary-lensing events are somewhat lower than the first estimate, tens of binary events have been already published by the MACHO \citep{alcocketal00} and OGLE \citep{jaroszynski02,jaroszynskietal04,jaroszynskietal06,skowronetal07} projects. Many further unpublished events are listed on the microlensing alert web pages of OGLE\footnote{\url{http://www.astrouw.edu.pl/~ogle/ogle3/ews/ews.html}} and MOA\footnote{\url{https://it019909.massey.ac.nz/moa/}}, bringing the estimated number of detected binary events to 120--160. Most of these events can be described adequately by the point-source + binary-lens model. However, events with a sufficiently dense sampling of a crossing of the binary-lens caustic require more advanced modeling. They are sensitive to the non-zero angular size of the source star as well as its surface-brightness distribution, usually described by a radial limb-darkening function. The limb-darkening sensitivity of caustic-crossing microlensing events \citep{witt95} turns them into a unique tool for studying stellar atmospheres. While single-point-mass events share the sensitivity \citep{heysalo00}, the very high alignment necessary for the point-caustic of the lens to transit the disk of the source reduces their frequency in comparison with two-point-mass events, in which caustic crossing is highly probable.

In principle, with good photometry and time sampling most caustic crossings can be used to measure the limb darkening of the source star. So far seven binary microlensing events led to a reported limb-darkening measurement: MACHO 97-BLG-28 \citep{albrowetal99a}, MACHO 97-BLG-41 \citep{albrowetal00}, MACHO 98-SMC-1 \citep{afonsoetal00}, OGLE-1999-BUL-23 \citep{albrowetal01a}, EROS BLG-2000-5 \citep{anetal02,fieldsetal03}, MOA 2002-BLG-33 \citep{abeetal03}, and OGLE-2002-BLG-069 \citep{cassanetal04,kubasetal05}. In addition, at the time of writing there are six reported planetary-mass events, most of which are strongly influenced by the limb darkening of the source. These include OGLE 2003-BLG-235 / MOA 2003-BLG-53 \citep{bondetal04}, OGLE 2005-BLG-071 \citep{udalskietal05}, OGLE 2005-BLG-390 \citep{beaulieuetal06}, OGLE 2005-BLG-169 \citep{gouldetal06}, the two-planet OGLE-2006-BLG-109 \citep{gaudietal08}, and the latest MOA-2007-BLG-192 \citep{bennettetal08}. A significant fraction of these events have been analyzed using a simplified model of the caustic crossing, because most techniques for calculating the exact light curve of an extended source with realistic limb darkening are too demanding computationally.

On the theoretical side, the original paper by \cite{schn_weiss86} studied in detail the lensing by a system of two equal masses, discussing among other aspects the structure of the caustic and the critical curve, point-source image positions and amplifications, the geometry of extended-source images, and local approximations of lensing near a caustic fold and cusp. The paper also includes light curves of a uniform extended source of various sizes crossing the caustic at a fold and a cusp. \cite{erdl_schneider93} studied lensing by two unequal masses at arbitrary distances along the line of sight (i.e., in a single lens plane or in two different planes), while \cite{witt_petters93} explored the geometry of lensing by two unequal masses with an additional shear. A range of other general works on two-point-mass lensing has been published since, of which we mention here the study by \cite{dominik99} of various limiting cases of binary lensing.

However, there have been fewer theoretical works studying in more detail the extended-source effects in two-point-mass lensing. These have been mostly limited to studies of caustic crossing using a local approximation of the caustic by a linear fold. Here we point out the works by \cite{rhie_bennett99} on fold-caustic microlensing of a linearly limb-darkened source, and the more general explorations of fold crossings by \cite{gaudi_petters02a} and \cite{dominik04}.

Only several papers deal specifically with microlensing near cusps of the caustic. \cite{schn_weiss92} studied the lensing by a generic cusp primarily for a point source, but presented also a sample amplification map for a uniform circular source. \cite{zakharov95,zakharov99} contributed simple analytical expressions for the amplifications of individual images of a point source lensed by a generic cusp. \cite{rhie02} studied point-source lensing near the cusps of a binary lens. \cite{gaudi_petters02b} explored the microlensing behavior near a generic cusp for a point source and estimated the microlensing signature of a small extended source positioned on the two main axes of the cusp.

The aim of the present work is to explore extended-source effects in two-point-mass microlensing of limb-darkened sources more generally, without resorting to local caustic approximations. Such an approach has been taken so far only in individual cases to illustrate particular effects, such as color changes around a sample binary caustic presented by \cite{han_park01}. In \S~\ref{sec:extended} we briefly introduce the basic setup of two-point-mass microlensing and the main relevant quantities. We describe the method we used for computing the microlensing amplification in \S~\ref{sec:method} together with the limb-darkening model we adopted. We explore the sensitivity to the extended nature of the source in \S~\ref{sec:sensitivity}: numerically in \S~\ref{sec:region} and analytically in \S~\ref{sec:analytical}. The extent of the sensitive area is used in \S~\ref{sec:probability} to estimate the probability of encountering an extended-source effect of a given amplitude. We study the sensitivity to differences in the limb darkening of the source star in \S~\ref{sec:chromaticity}. In \S~\ref{sec:fold} we assess the adequacy of the linear-fold approximation on the example of two observed events, followed by more general comments on other events in \S~\ref{sec:events}. We discuss the results and future prospects in \S~\ref{sec:discussion}, and summarize our main findings in \S~\ref{sec:summary}. A brief note on image amplifications by a generic cusp is added in the Appendix.

\section{AMPLIFICATION OF A POINT SOURCE AND AN EXTENDED SOURCE}
\label{sec:extended}

In a two-point-mass gravitational microlensing event the angular positions of images in the plane of the sky $\bf x$ are related to the angular source position $\bf y$ through the lens equation
\beq
{\bf y} = {\bf x} - \mu_A \frac{{\bf x} - {\bf x}_A}{|{\bf x} - {\bf x}_A|^2} - \mu_B \frac{{\bf x} - {\bf x}_B}{|{\bf x} - {\bf x}_B|^2}\,,
\label{eq:bin_lens}
\eeq
where $\mu_A$ and $\mu_B$ are the masses of the lens components relative to their total mass $M$ ($\mu_A + \mu_B = 1$), and ${\bf x}_A$ and ${\bf x}_B$ denote their locations. Positions $\bf x$, $\bf y$, ${\bf x}_A$, and ${\bf x}_B$ are measured in units of the angular Einstein radius $\theta_{\rm E}$ of the compound lens
\beq
\theta_{\rm E} = \sqrt{\frac{4GM}{c^2}\frac{D_{LS}}{D_{OS}D_{OL}}}\,,
\label{eq:eins_rad}
\eeq
where $D_{OS}$ and $D_{OL}$ are distances from the observer to the source and lens, respectively, $D_{LS} = D_{OS}-D_{OL}$, $G$ is the gravitational constant, and $c$ is the speed of light.

It is often advantageous to rewrite all vectorial positions appearing in equation~(\ref{eq:bin_lens}) in terms of complex variables $z = {\rm x}_1 + i{\rm x}_2$ and $\xi = {\rm y}_1 + i{\rm y}_2$ \citep{witt90,witt_petters93}. The corresponding form of the lens equation is
\beq
\xi = z - \frac{\mu_A}{\overline{z} - \overline{z}_A} - \frac{\mu_B}{\overline{z}-\overline{z}_B}\,,
\label{eq:bin_lens_complex}
\eeq
where the lens component positions $z_A$ and $z_B$ are defined similarly as the image position $z$.

To obtain the amplification $A_0$ of the flux from a point source passing behind the lens, we first compute the Jacobian $J({\bf x}) = \partial {\bf y}/\partial {\bf x}$ analytically using the lens equation~(\ref{eq:bin_lens}). For a given source position ${\bf y}$ we then find $n$ distinct image positions ${\bf x}_j, j=1\ldots n$, by solving equation~(\ref{eq:bin_lens}) numerically. The total amplification is the sum of the amplifications of individual images
\beq
A_0({\bf y}) = \sum_{j=1}^{n} \frac{1}{|\det J({\bf x}_j)|}\,.
\label{eq:point_mag}
\eeq

\begin{figure*}[t]
\begin{center}
\includegraphics[scale=.69]{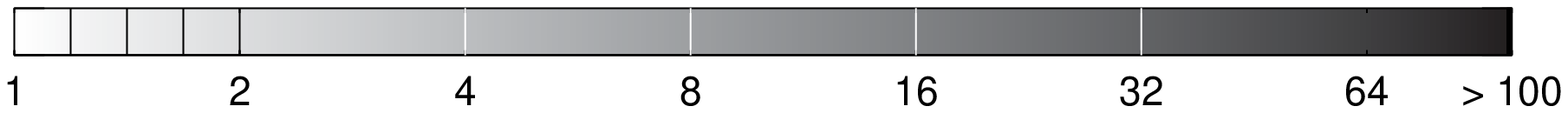}\vspace{5mm}\\
\includegraphics[scale=.31]{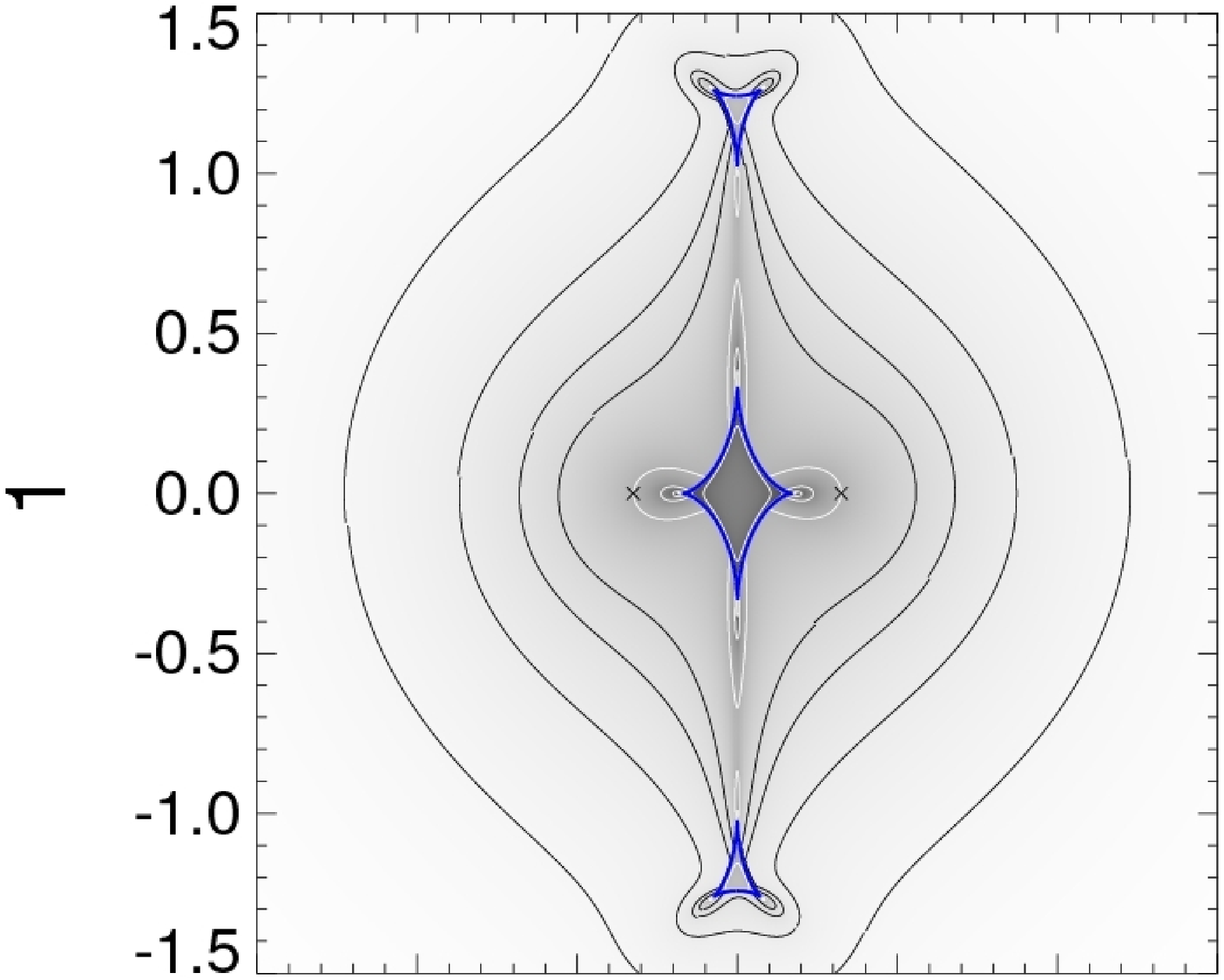}
\includegraphics[scale=.31]{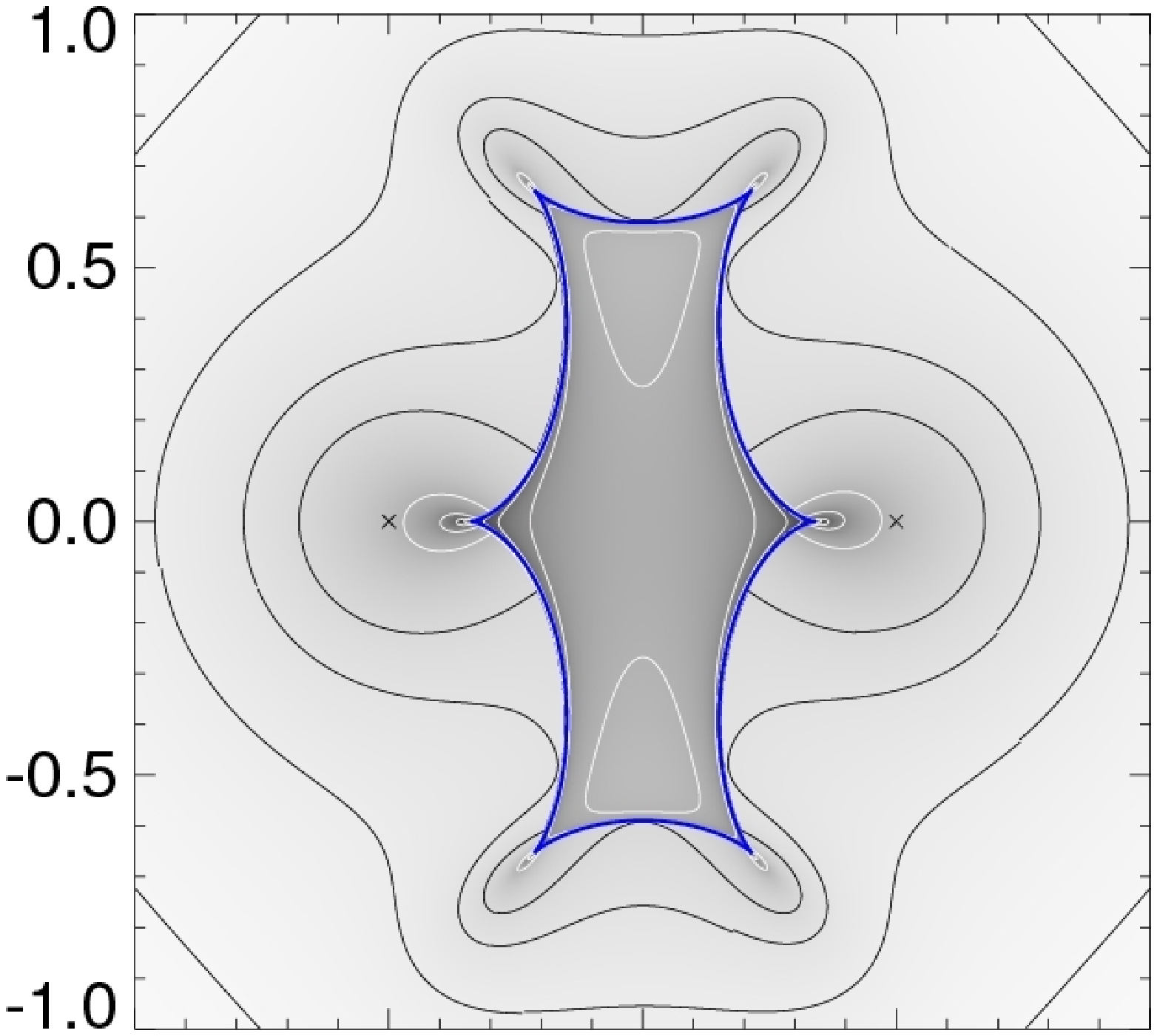}
\includegraphics[scale=.31]{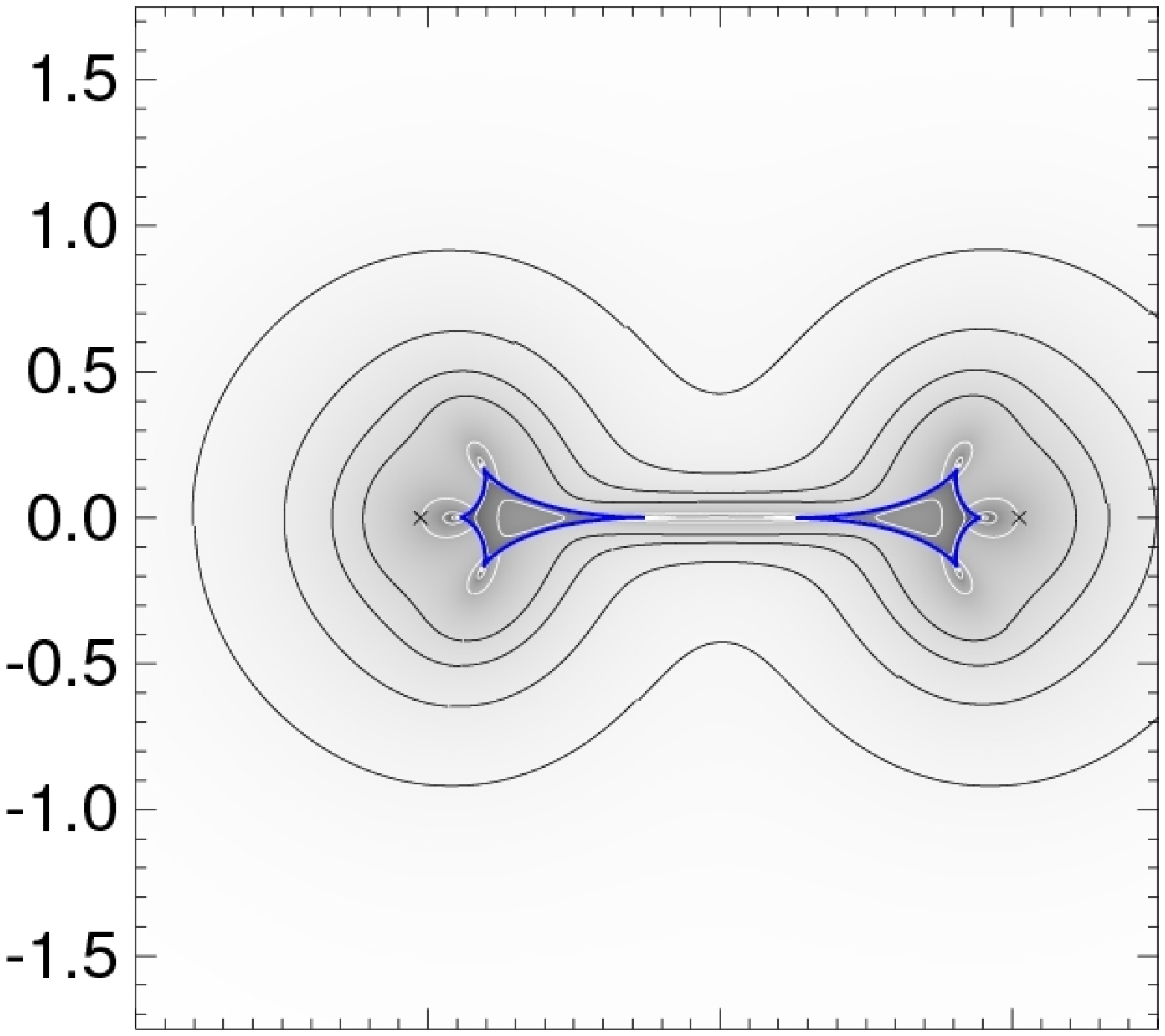}\\
\includegraphics[scale=.31]{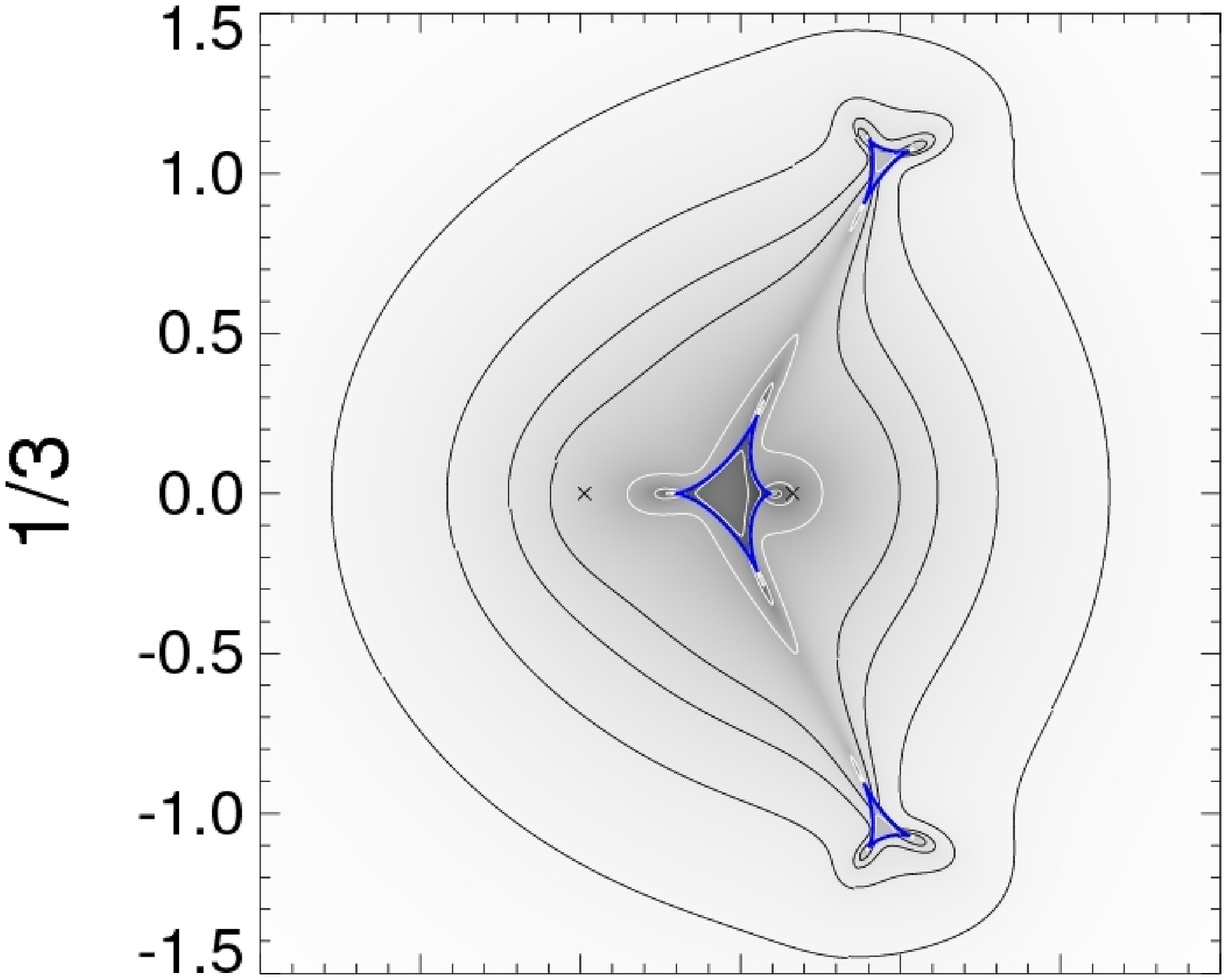}
\includegraphics[scale=.31]{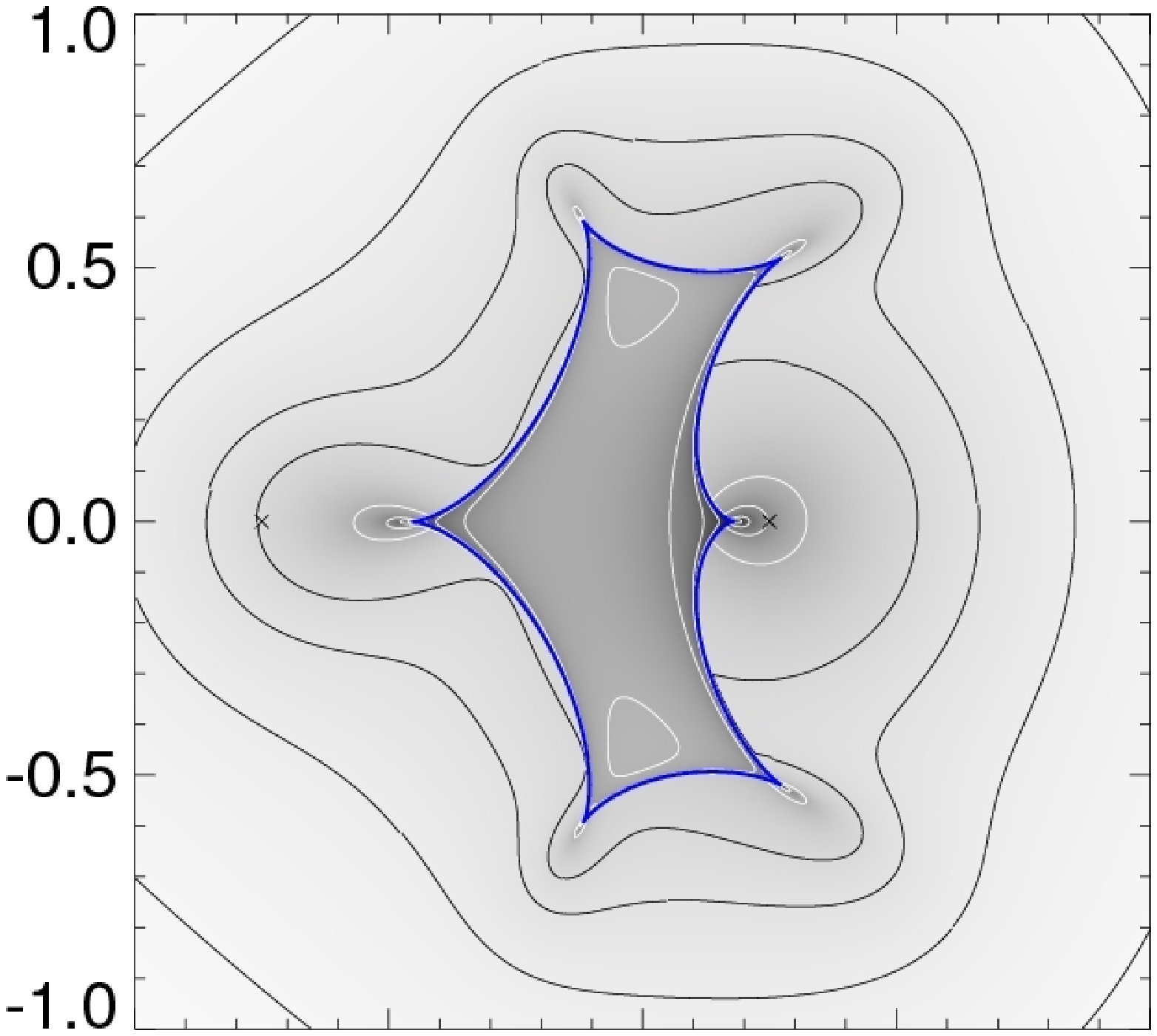}
\includegraphics[scale=.31]{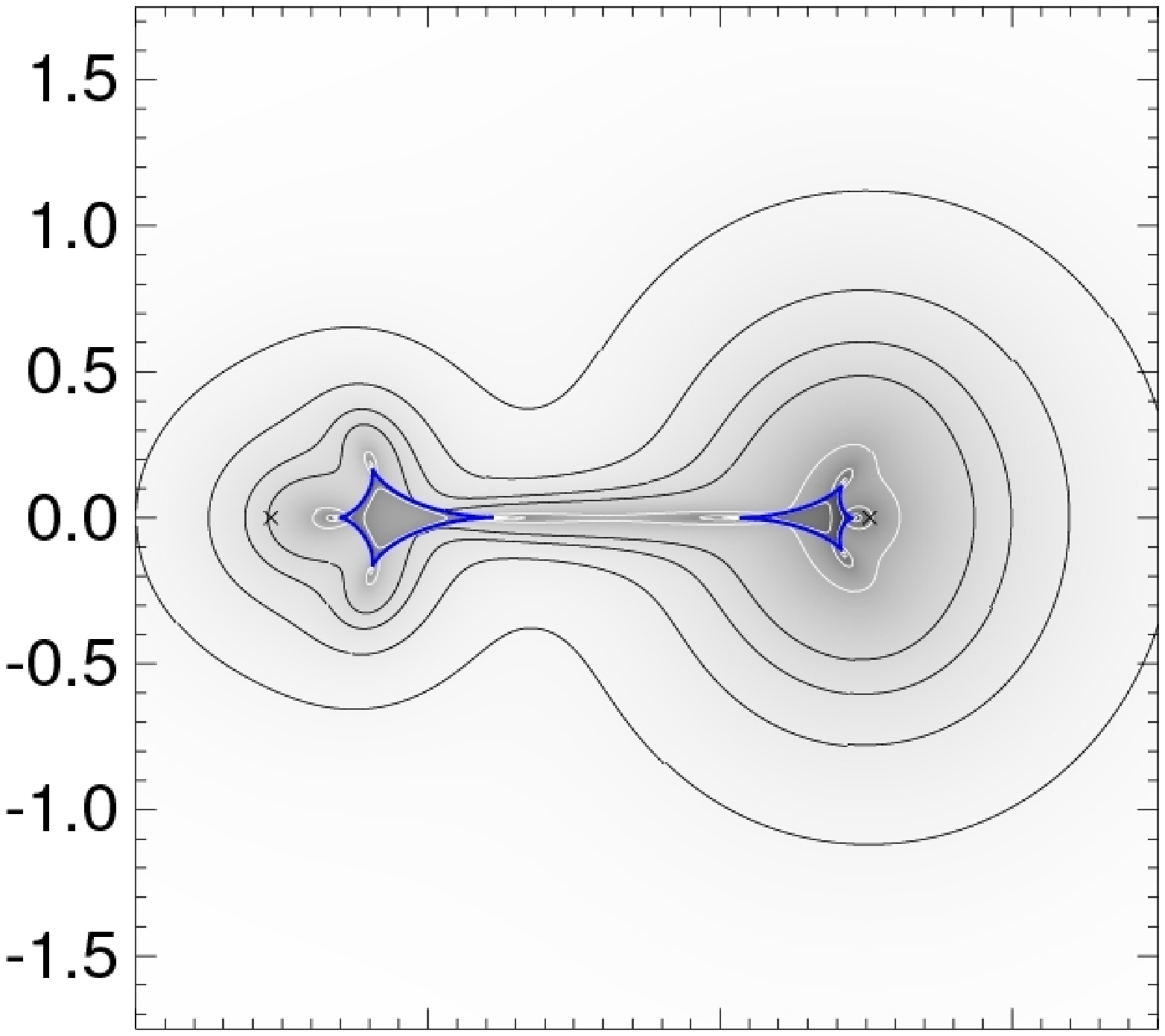}\\
\includegraphics[scale=.31]{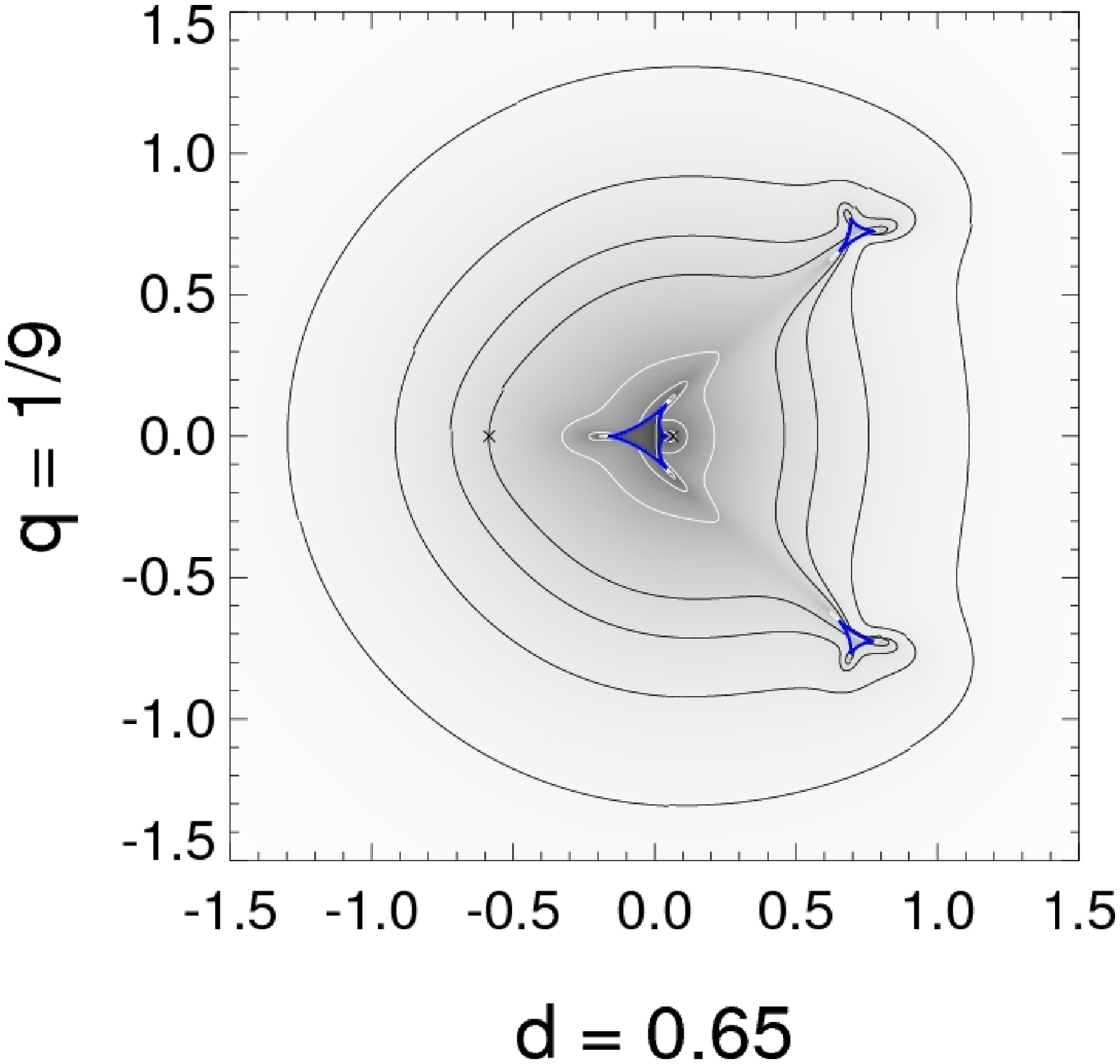}
\includegraphics[scale=.31]{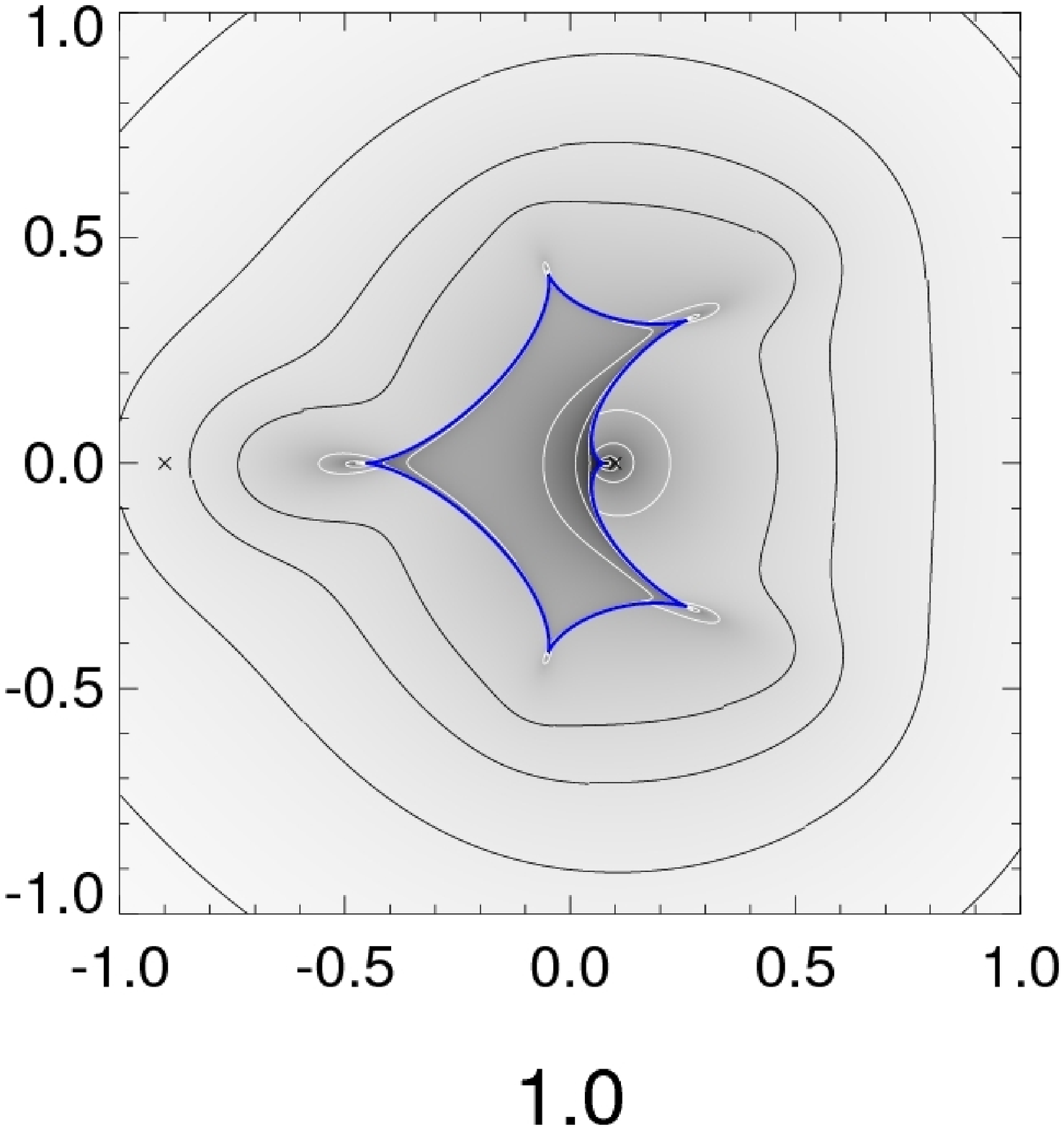}
\includegraphics[scale=.31]{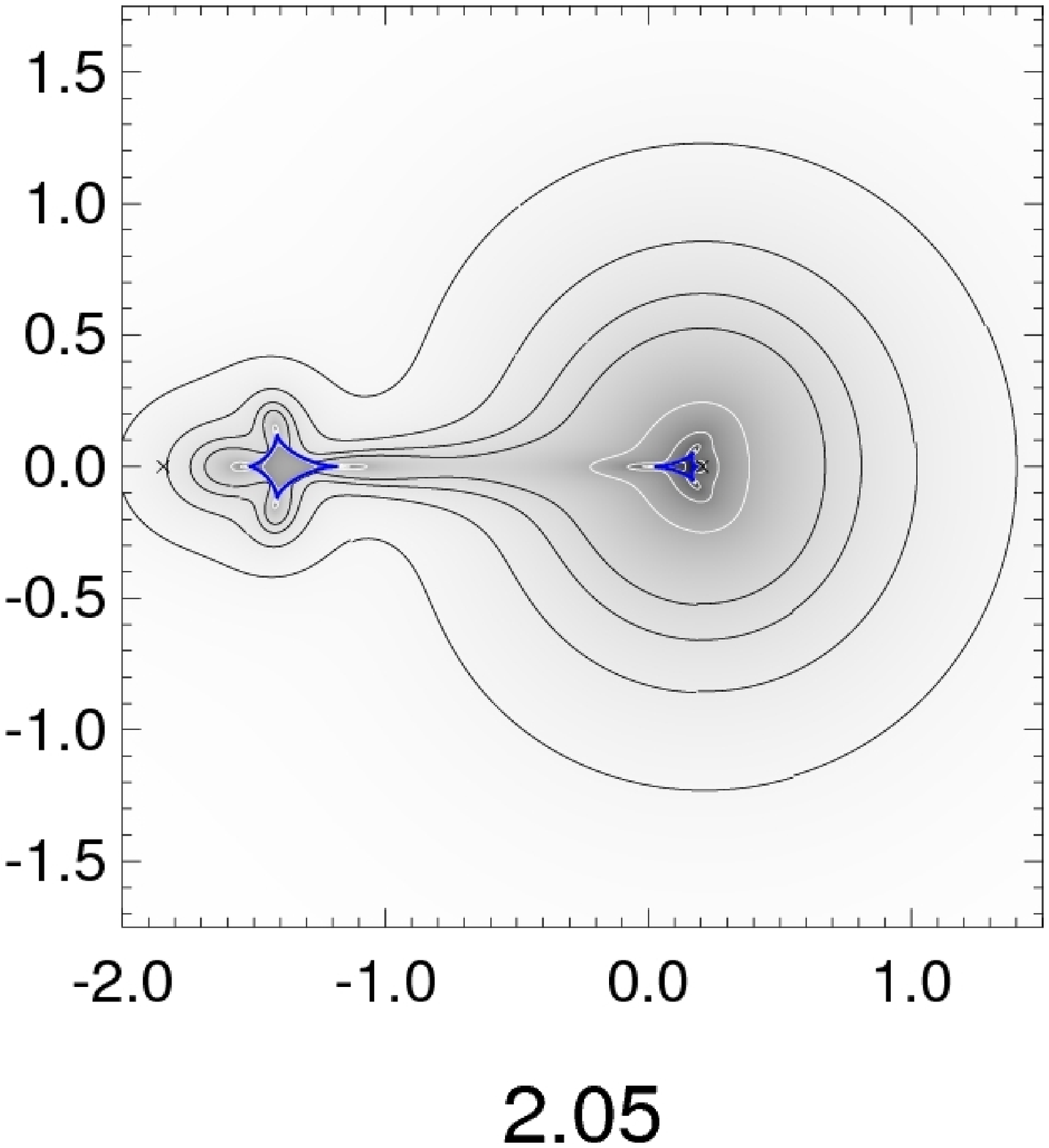}
\caption{Contour plots of point-source amplification $A_0$ by different two-point-mass lenses as a function of source position ${\bf y}$. Columns correspond to lens-component separations $d$ marked below, and each have a different scale. Rows correspond to mass ratios $q$ marked at left. The origin of each plot is placed at the center of mass of the lens; crosses denote the positions of the point masses with the heavier body placed on the right, and caustics are marked in blue. Contours are plotted for values $A_0=2^k$ increasing from the outer regions, with dark contours for $k= \{0.25, 0.5, 0.75, 1\}$ and white contours for more widely spaced values $k = \{2, 3, 4, 5\}$, as shown in the color bar.}
\label{fig:point_mag}
\end{center}
\end{figure*}

The point-source amplification $A_0$ diverges if any of the images ${\bf x}_j$ appear on the critical curve of the lens, i.e., the curve along which $\det J$ vanishes. The condition for the vanishing Jacobian can be expressed as
\beq
\det J = 1 - |\gamma|^2 = 0,
\label{eq:crit_curv}
\eeq
where $\gamma$ is the shear of the lens. In the case of a two-point-mass lens the dependence of the shear on the image position in complex notation is given by
\beq
-\overline{\gamma} = \frac{\mu_A}{(z-z_A)^2} +  \frac{\mu_B}{(z-z_B)^2} = |\gamma|e^{i\psi}\,,
\label{eq:crit_converg}
\eeq
with real phase $\psi$. Setting $|\gamma|=1$ in equation~(\ref{eq:crit_converg}) yields a fourth-order complex polynomial in $z$, the roots of which are points on the critical curve. By repeating the procedure with $\psi$ varying from 0 to $2\pi$ we obtain the full critical curve \citep{witt90}.

Two new images appear on the critical curve when the source enters the caustic of the lens. The caustic is found by a reverse mapping of the critical curve using lens equation~(\ref{eq:bin_lens}) or, equivalently, equation~(\ref{eq:bin_lens_complex}). In the case of a two-point-mass lens the caustic may consist of a single continuous curve, or of two or three separate components, depending on the lens configuration \citep{schn_weiss86,erdl_schneider93}.

We illustrate the dependence of the point-source amplification $A_0$ on the position of the source by the contour plots in Figure~\ref{fig:point_mag}. We present nine different lens configurations, obtained by combining three lens-component mass ratios $q = \mu_A/\mu_B= \{1/9, 1/3, 1\}$ with three lens-component separations $d = |{\bf x}_B - {\bf x}_A| = \{0.65, 1.0, 2.05\}$. The choice of separations gives us three-component, continuous, and two-component caustics, respectively, for all three mass-ratio values. In binary microlensing terminology, these are known as the ``close", ``intermediate", and ``wide" binary configurations, respectively. In Figure~\ref{fig:point_mag}, as well as throughout the paper, we set the origin of the coordinate system at the center of mass of the lens and place the two point masses on the ${\rm y}_1$ axis, with the lighter mass (A) positioned to the left of the heavier mass (B).

\begin{figure*}[t]
\plottwo{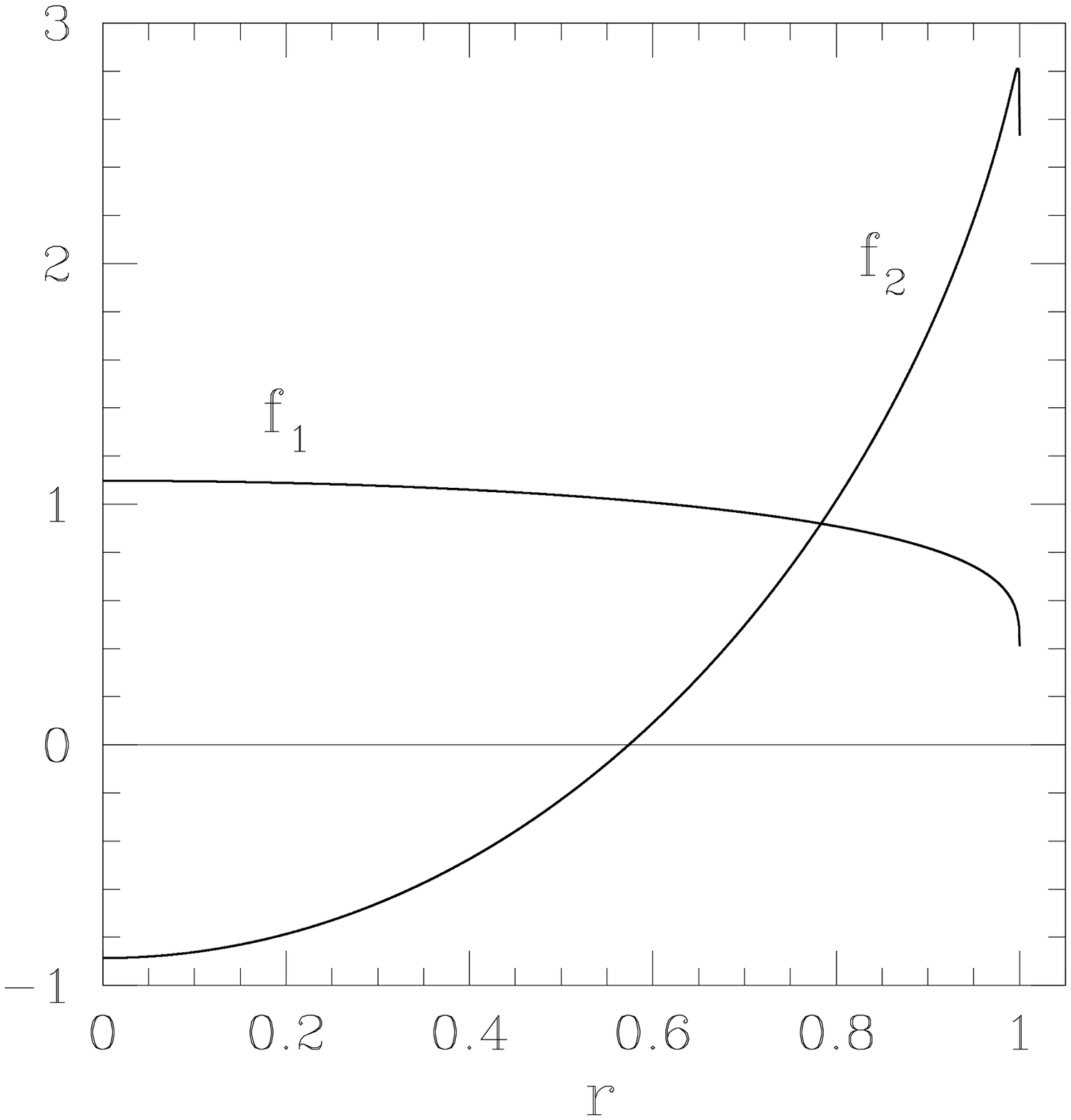}{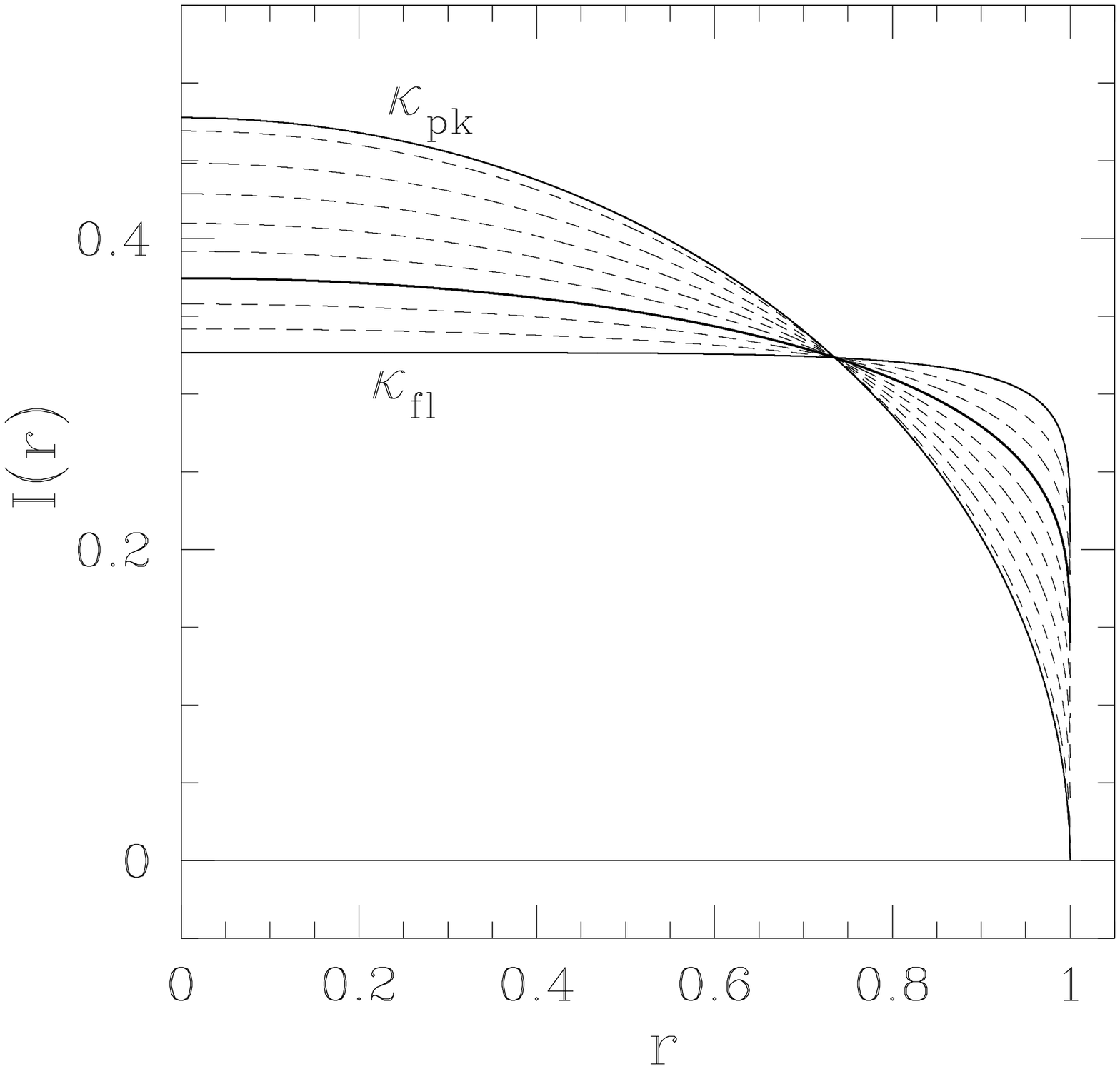}
\caption{Limb-darkening model used in this work. Left panel: the first two terms $f_1(r)$ and $f_2(r)$ of an orthonormal basis obtained by principal component analysis (PCA) of Kurucz's ATLAS stellar atmosphere models \citep{hey08}. Right panel: full range of limb-darkening profiles $I(r)$ described by the first two principal components following equation~(\ref{eq:pca_limb}). All profiles are normalized to unit total flux, $2\pi\int_0^1 I(r)r\,\rmd r=1$. The bold inner curve corresponds to $\kappa=0$, the outer solid curves are determined by limiting values $\kappa_{\rm pk}=-0.1620$ and $\kappa_{\rm fl}=0.0902$. The inner dashed curves correspond to $\kappa$ values spaced from $0$ by $0.03$, with positive values characterizing flatter profiles and negative values more peaked profiles.}
\label{fig:pca_ld}
\end{figure*}

In any of the lens configurations, contours far from the origin would eventually coincide with the circular contours of a single-point-mass lens with the same total mass. The point-source amplification grows asymmetrically from this asymptotic region and is divergent along the caustic. The increase to the divergence is continuous only when reaching the caustic at a cusp, elsewhere $A_0$ converges from outside the caustic to a finite value and discontinuously jumps to $\infty$. The highest amplification outside the caustic thus occurs at its cusps, with nearby contours ending abruptly at the caustic. Note the narrow higher-amplification regions connecting the facing cusps of the caustic components in the outer columns. Inside the caustic $A_0$ drops continuously to a high-amplification ``plateau'' with $A_0\geq 3$ \citep{witt_mao95}.

For actual astrophysical sources (i.e., stars, in our context) the point-source approximation breaks down in regions where $A_0$ varies substantially and nonlinearly on angular scales comparable to the source size, in particular close to the caustic. In such situations the amplification is affected by the extended nature and, more specifically, the surface-brightness distribution $B({\bf y}\,')$ of the source. The amplification $A_*$ of an extended source is obtained by integrating the point-source amplification $A_0$ weighted by $B$ over the source area $\Sigma_{\rm S}$ and dividing the result by the unamplified source flux. Alternatively, $A_*$ can also be computed as the ratio of the total flux in the images to the unamplified source flux. For a source with its center positioned at ${\bf y}_{\rm c}$ we thus have
\begin{eqnarray}
A_*({\bf y}_{\rm c}) = \frac{\int_{\Sigma_{\rm S}}A_0({\bf y}_{\rm c}+{\bf y}\,')B({\bf y}\,')\rmd^2 {\bf y}\,'}{\int_{\Sigma_{\rm S}} B({\bf y}\,')\rmd^2 {\bf y}\,'} =  \nonumber\\
= \frac{\int_{\Sigma_{\rm I}} B({\bf y}[{\bf x}]-{\bf y}_{\rm c})\rmd^2 {\bf x}}{\int_{\Sigma_{\rm S}} B({\bf y}\,')\rmd^2 {\bf y}\,'}\,.
\label{eq:amp_integral}
\end{eqnarray}
Here $\Sigma_{\rm I}$ stands for the total area covered by the images and ${\bf y}\,'$ is measured from the center of the source. In the numerator of the second expression we map the image position ${\bf x}$ to the corresponding position on the source ${\bf y}[{\bf x}]$ using equation~(\ref{eq:bin_lens}). For a source passing sufficiently close to the caustic of the lens, the spatial dependence of $B$ is thus encoded in the temporal dependence of $A_*$. In such cases the surface-brightness profile of the source can be measured, in principle, from the light curve of the lensing event.

\newpage
\section{COMPUTATIONAL METHOD}
\label{sec:method}

Evaluation of the extended-source amplification $A_*$ given by equation~(\ref{eq:amp_integral}) with an arbitrary surface-brightness distribution of the source is a challenging task. Direct integration using the first expression in equation~(\ref{eq:amp_integral}) is very inefficient, due to the root-finding necessary for evaluating $A_0$ for each point of the source and the need of special treatment of its divergence in caustic-crossing events. The inverse ray-shooting method and its variations \citep{schn_weiss86,rattetal02}, based on the second expression in equation~(\ref{eq:amp_integral}), require prohibitively many evaluations of equation~(\ref{eq:bin_lens}) to achieve a relative accuracy of at least $10^{-3}$. Such a minimum accuracy is necessary for analyzing light curves with good-quality photometry, and is essential for extracting the surface-brightness profile of the source from observational data. Computational methods based on Green's theorem \citep{gould_gaucherel97,dominik98,dominik07} are efficient for uniform sources, but have difficulties in the more realistic case of sources with general limb-darkening profiles. Hybrid methods, which use a combination of several techniques, present a more efficient approach. For example, \citet{dongetal06} use a combination of ray shooting and Green's theorem.  We employ an image-plane integration method suggested by \citet{bennett_rhie96} and \citet{verm00}, which combines elements of the direct and ray-shooting methods.

For a given source-center position we first find the positions of its images from the lens equation~(\ref{eq:bin_lens_complex}). The image positions then serve as starting points for a recursive scan-line flood-fill algorithm that checks points (pixels) in the neighborhood to see whether they ray-trace back to the source disk through equation~(\ref{eq:bin_lens_complex}). To obtain the numerator of the second expression in equation~(\ref{eq:amp_integral}), appropriate image pixel areas are then summed with a weight given by the surface brightness of their corresponding point on the source. The flood-fill procedure runs until image boundaries are reached in all directions. When evaluating the amplification for a given source position, we have to keep in mind that two new partial images appear whenever the edge of the source enters the caustic. These may be disconnected from the source-center images and thus could be potentially missed. In order to take them into account we calculate the intersections of the edge of the source and a polygon closely approximating the caustic, which is obtained as described in the text following equation~(\ref{eq:crit_converg}). For initializing the algorithm we then use a source point just interior of such an intersection instead of the source-center position, and keep track of the visited image pixels in order not to count any repeatedly. This method can easily yield the amplification of an extended source with an arbitrary surface-brightness profile, with very few redundant inversions of equation~(\ref{eq:bin_lens_complex}) and relatively few back-traced rays that do not hit the source.

We model the stellar source disk by a circle with angular radius $\theta_* = \rho_*\theta_{\rm E}$ and surface brightness described by a radially symmetric limb-darkening profile $B({\bf y}\,')=I(r)$, with the radial coordinate $r=|{\bf y}\,'|/\rho_*$ scaled to the source radius. We perform most of our computations for two different source radii $\rho_* = \{0.002, 0.02\}$ in units of the Einstein radius, roughly corresponding to typical bulge main sequence and red giant microlensed source stars, respectively. Limb darkening is usually described by the linear limb-darkening law,
\beq
I(r) = I_0 [1 - \upsilon (1-\sqrt{1-r^2})] \,.
\label{eq:lin_limb}
\eeq
However, the linear law often gives a poor approximation of limb-darkening profiles of model atmospheres. Such an inaccurate model would hinder the study of details of the caustic crossing. Instead, we model the limb darkening $I(r)$ using the first two terms, $f_1(r)$ and $f_2(r)$, of an orthonormal basis obtained by principal component analysis (PCA) of limb-darkening profiles of Kurucz's ATLAS model atmospheres \citep{hey03,hey08},
\beq
I(r) = a_1 f_1(r) + a_2 f_2(r) = a_1 [f_1(r) + \kappa f_2(r)] \,.
\label{eq:pca_limb}
\eeq
The first basis function $f_1(r)$ already gives a weighted-average limb-darkening shape of the set of model profiles, while the second term with parameter $\kappa$ provides the first correction for variations in the shape. The basis functions $f_1$ and $f_2$ are plotted in Figure~\ref{fig:pca_ld} together with the range of limb-darkening profiles $I(r)$ obtained by their linear combination. Such a model is much better suited for a realistic description of limb darkening, even though it has the same number of degrees of freedom as the linear law.

\section{SENSITIVITY TO AN EXTENDED SOURCE}
\label{sec:sensitivity}

In this section we explore the sensitivity of the microlensing amplification to the extended nature of the source. Our primary interest here is to find out how far from the caustic can the effect be felt. The variation of amplification with limb darkening, which is primarily important during caustic crossing but of secondary importance further off the caustic, is studied in \S~\ref{sec:chromaticity}. In most of this section we therefore set $\kappa=0$ and use only the first principal function $f_1(r)$ for the limb darkening\footnote{Such a profile is similar to that of various ATLAS models, for example the $B$-band profile of a model with effective temperature $T_{eff}=12500\,K$ and surface gravity $\log\,g=2.5$ or the $R$-band profile of a model with $T_{eff}=8000\,K$ and $\log\,g=1.5$ (both with solar metallicity and microturbulent velocity $v_t=2\,km\,s^{-1}$).}.

\begin{figure*}[t]
\begin{center}
\includegraphics[scale=.69]{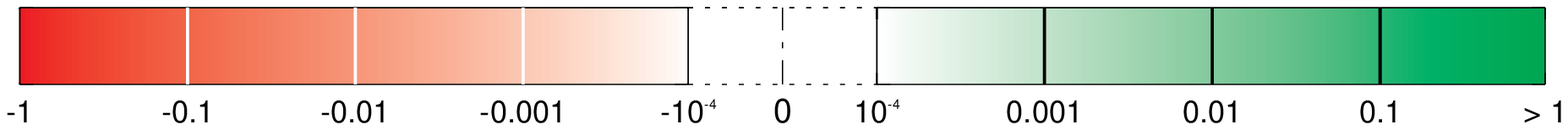}\vspace{5mm}\\
\includegraphics[scale=.31]{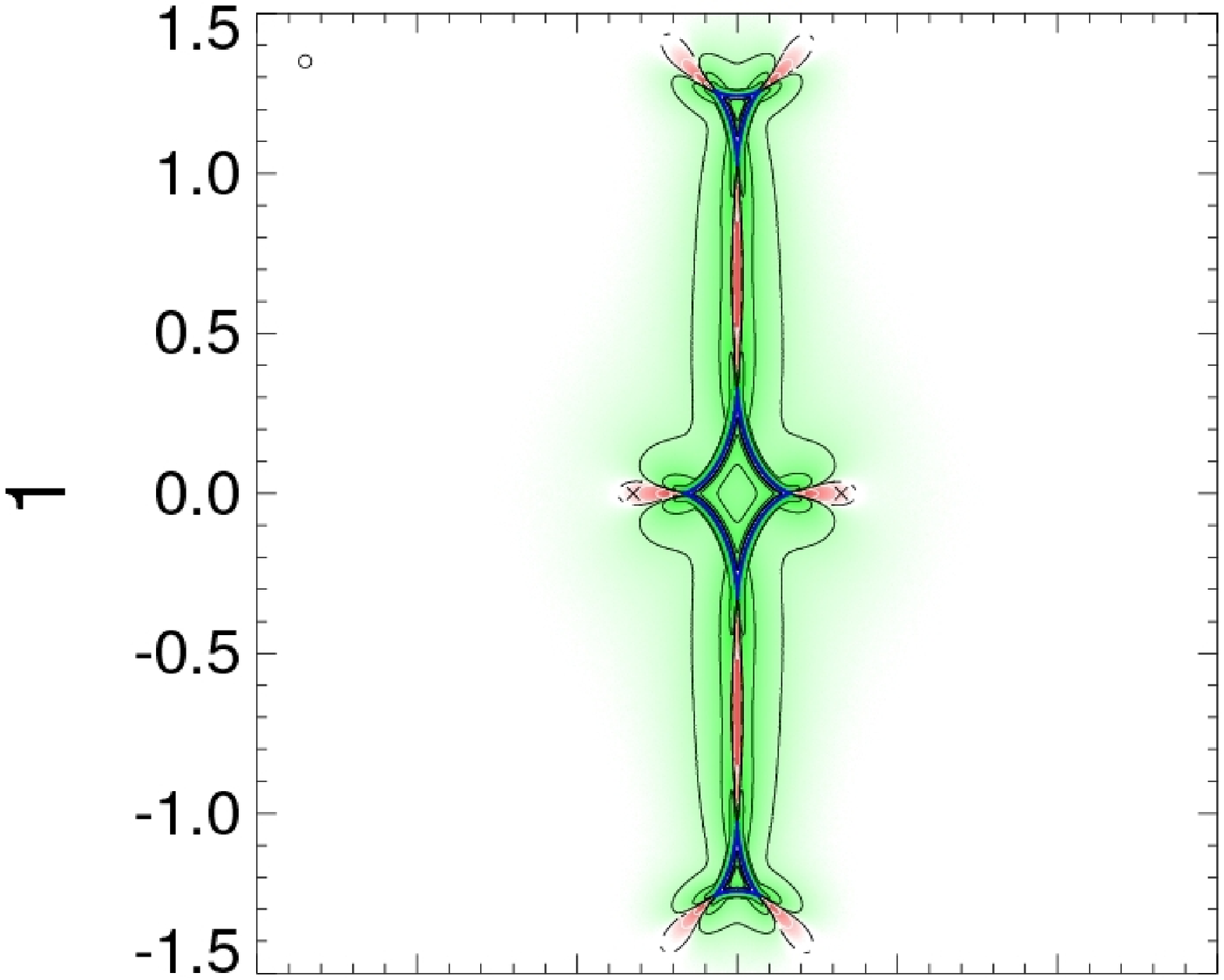}
\includegraphics[scale=.31]{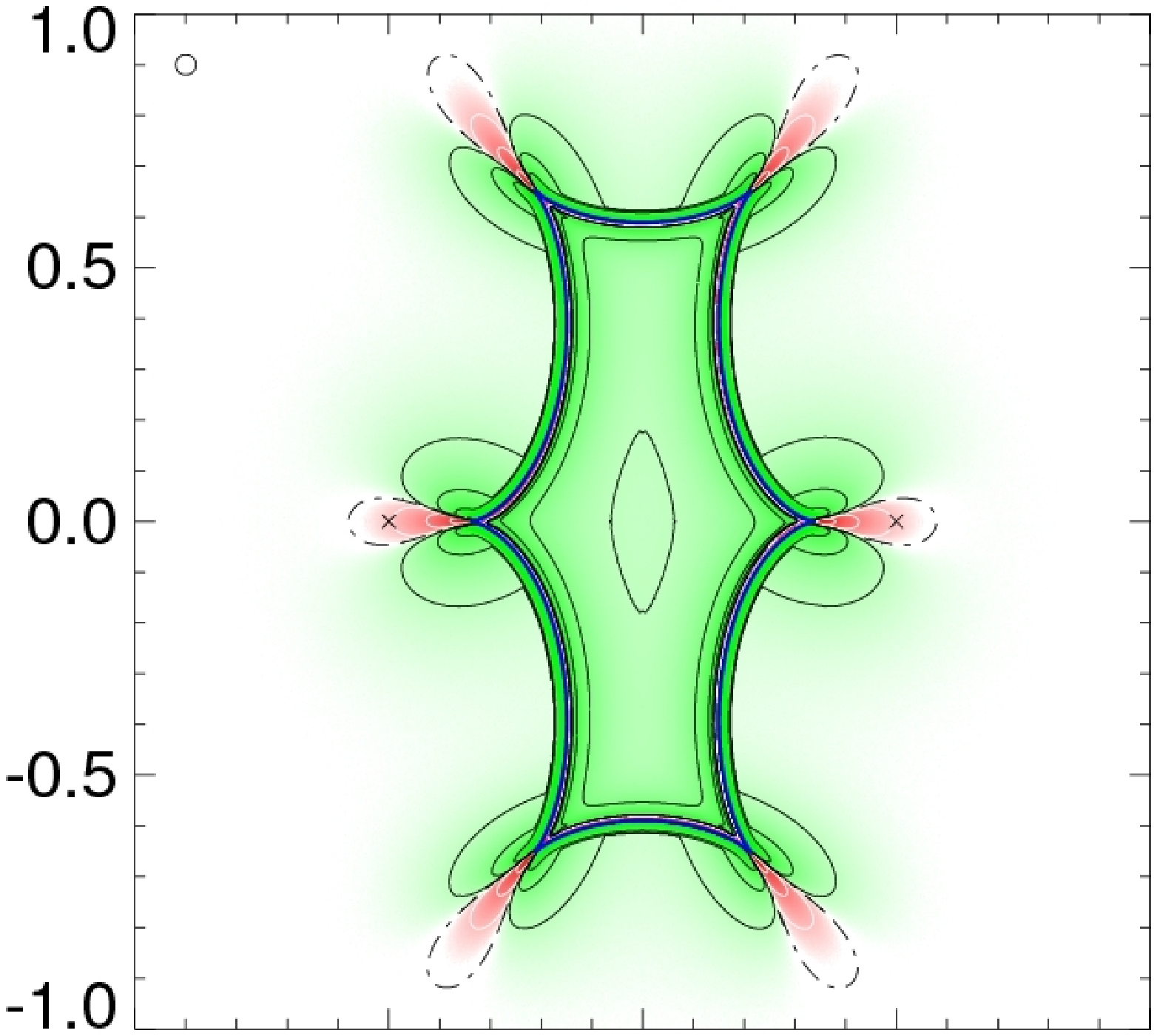}
\includegraphics[scale=.31]{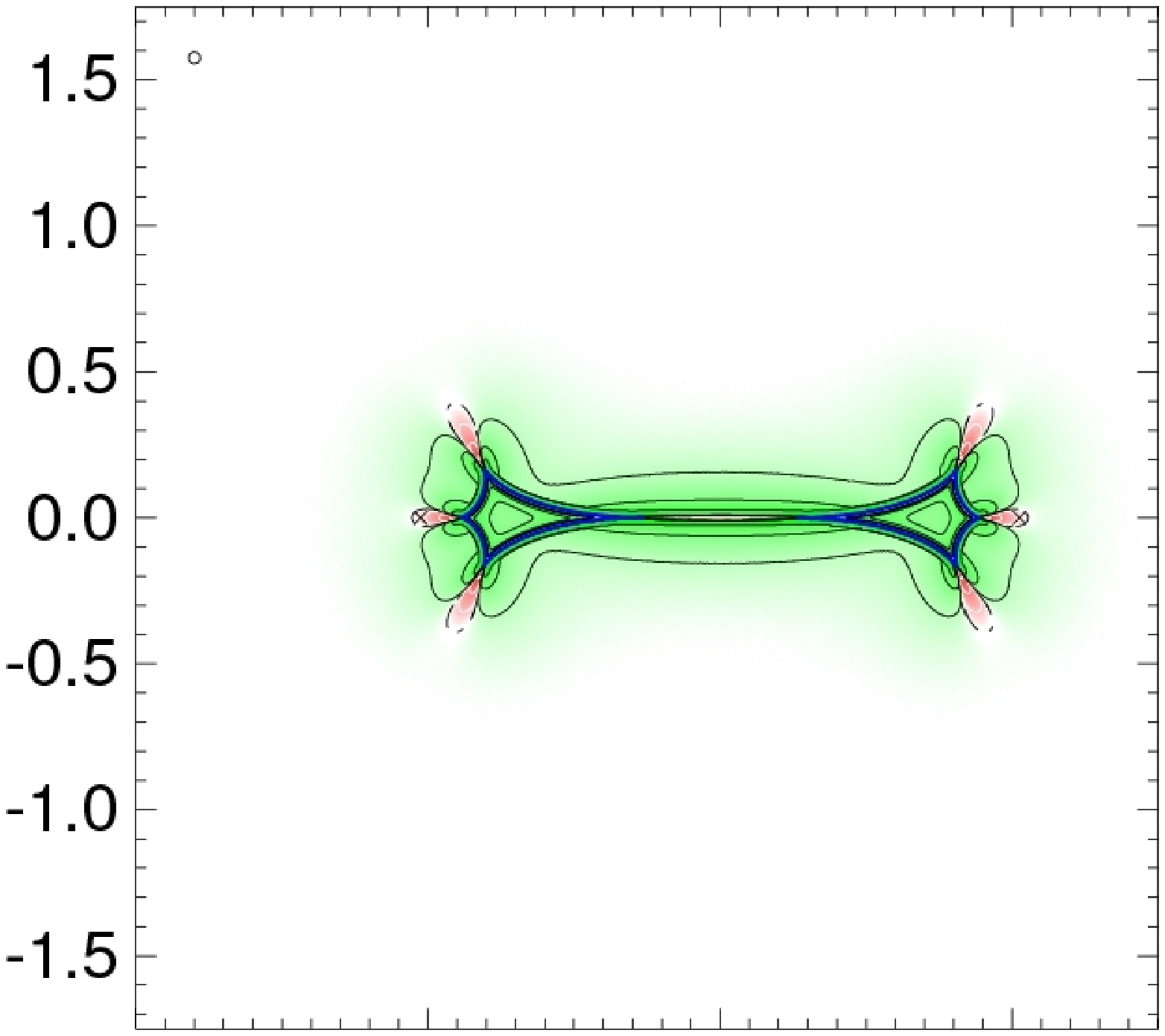}\\
\includegraphics[scale=.31]{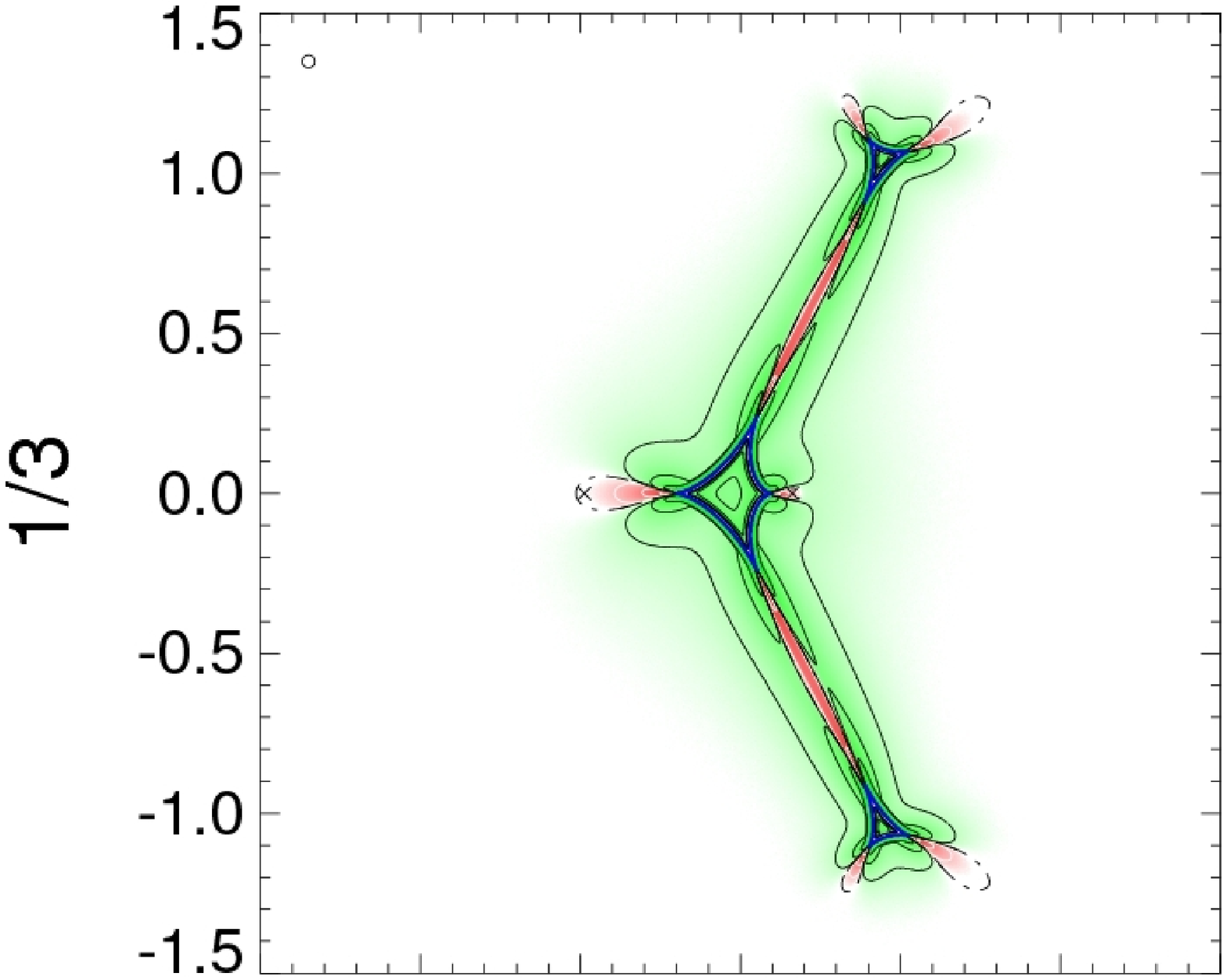}
\includegraphics[scale=.31]{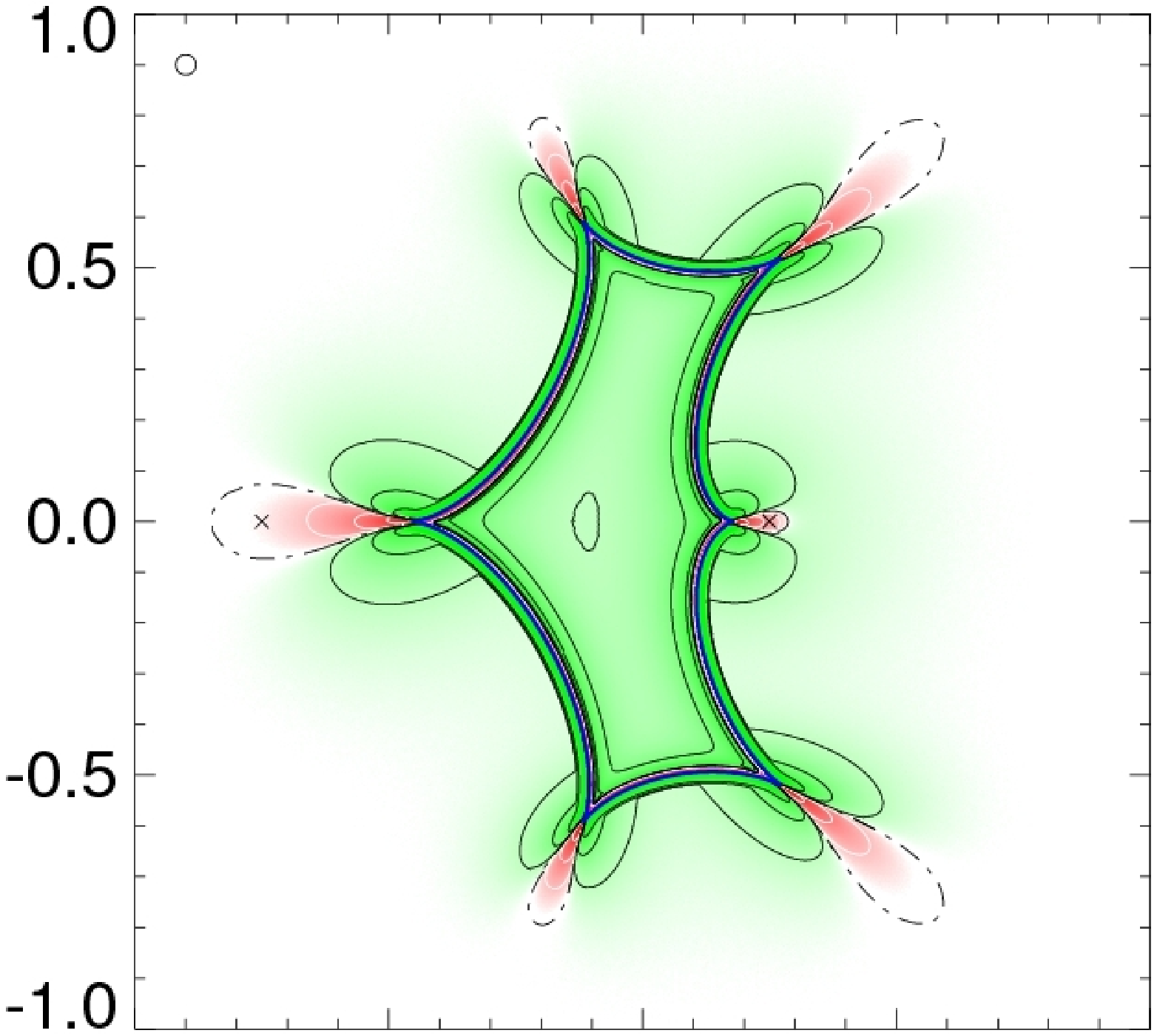}
\includegraphics[scale=.31]{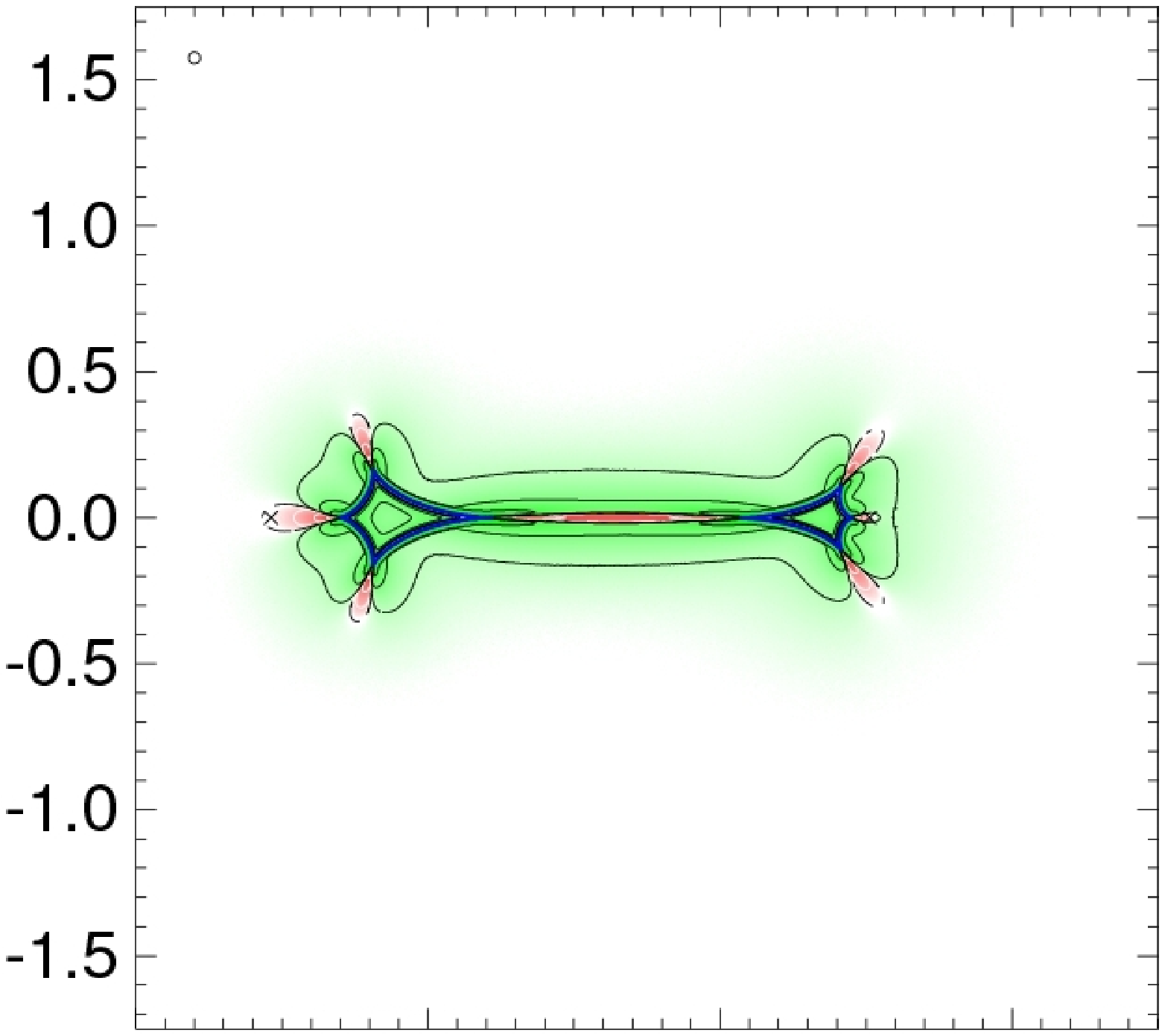}\\
\includegraphics[scale=.31]{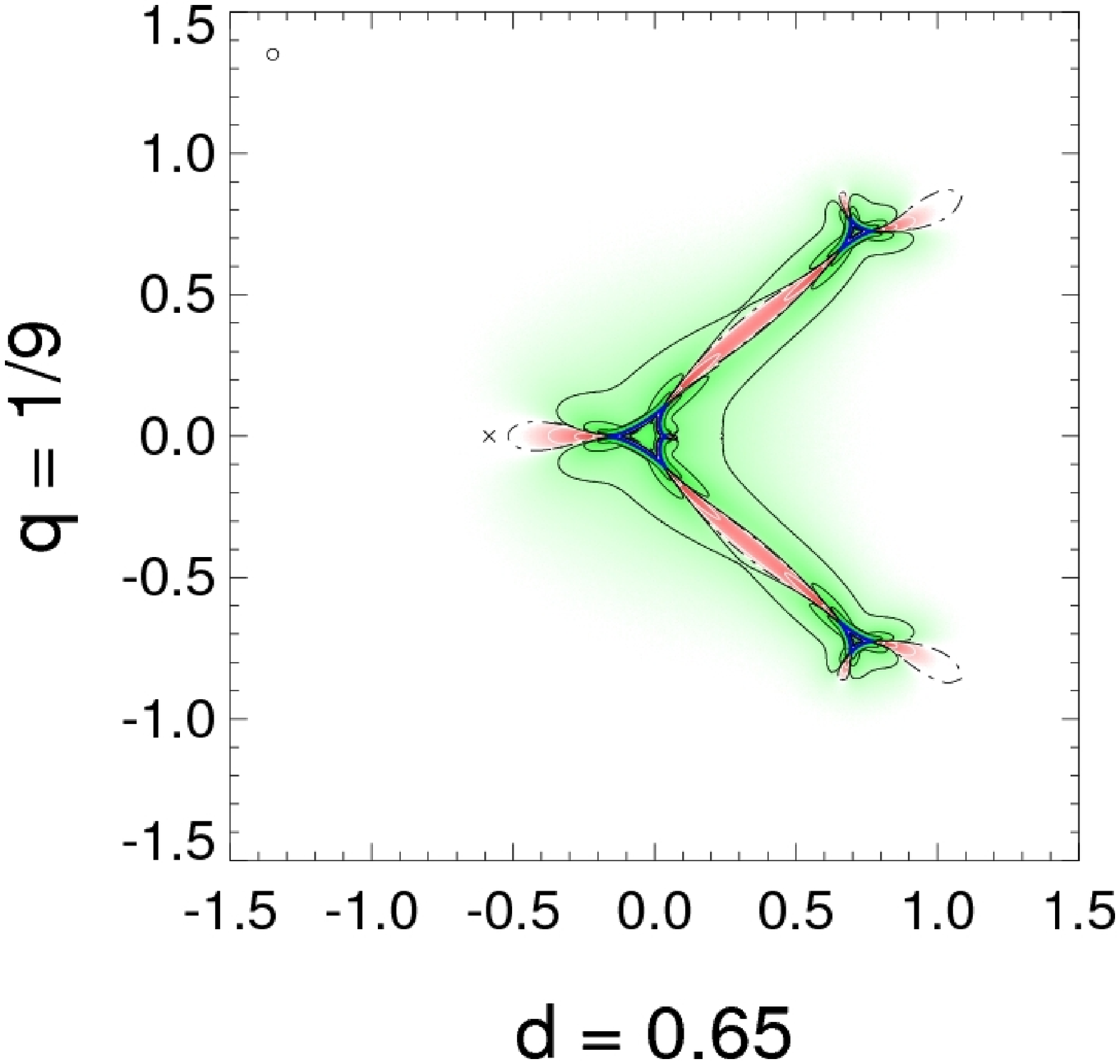}
\includegraphics[scale=.31]{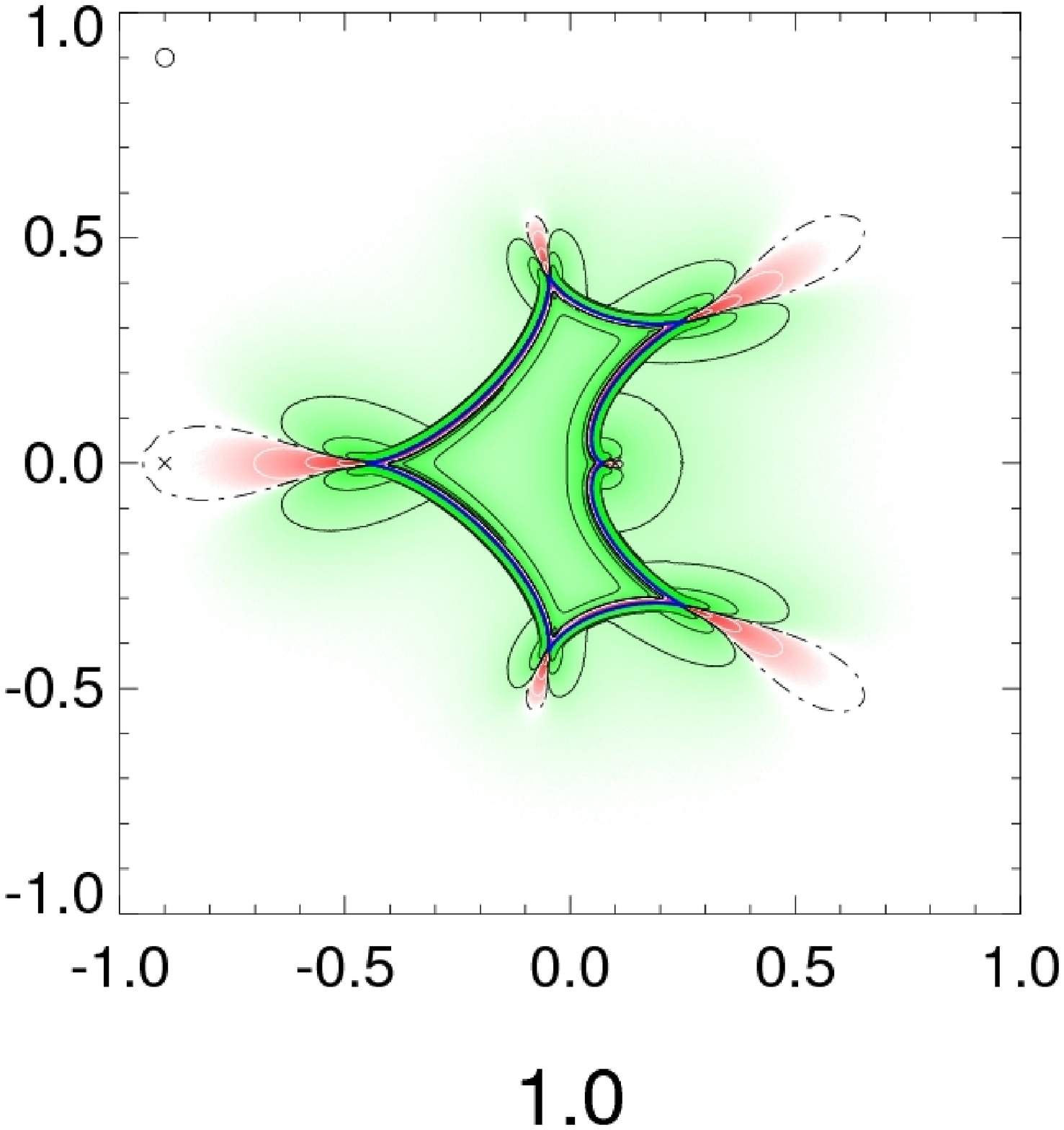}
\includegraphics[scale=.31]{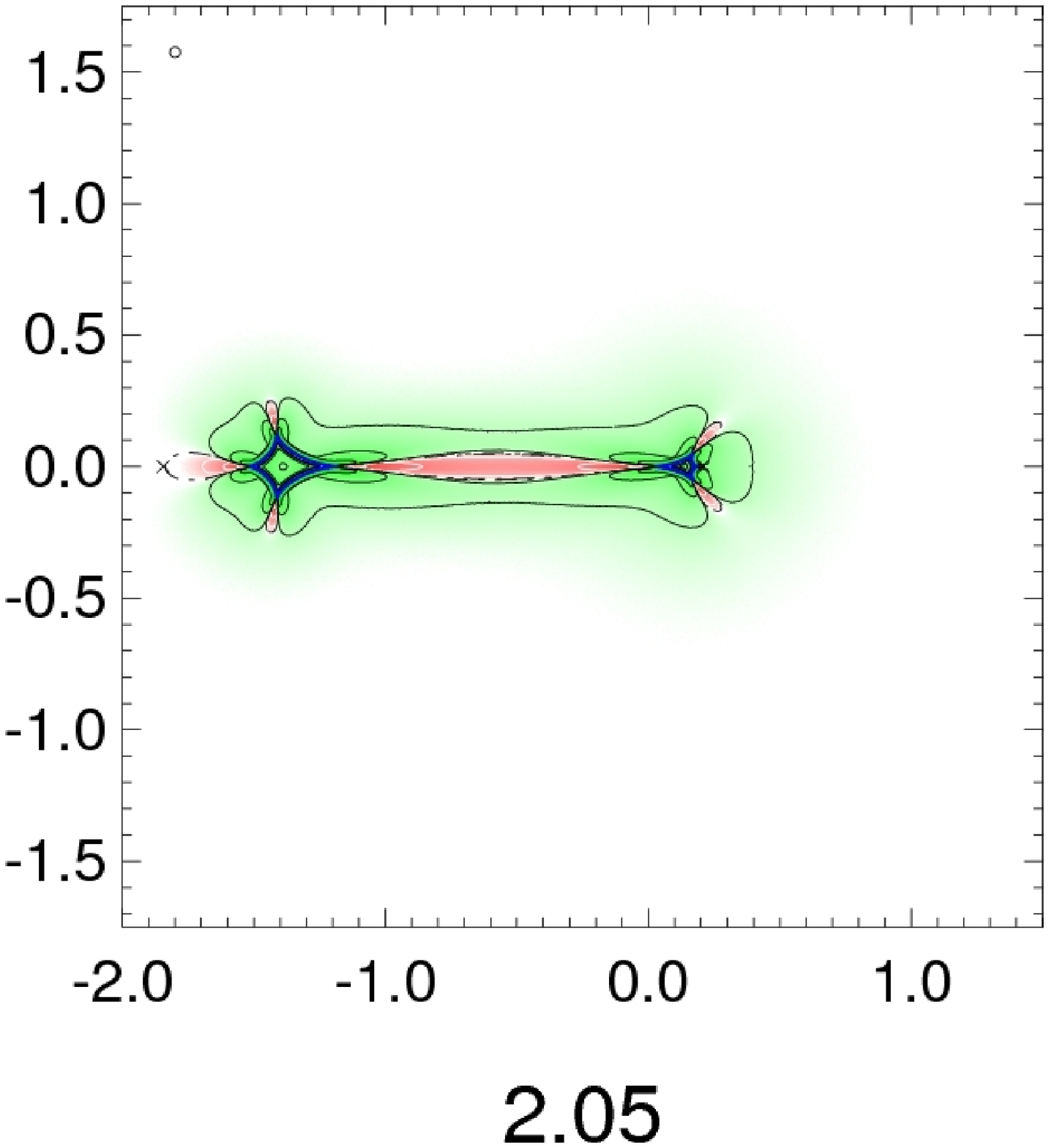}
\end{center}
\caption{Extended-source effect $\delex$ for a source with radius $\rho_* = 0.02$ and $\kappa=0$ limb darkening as a function of source-center position ${\bf y}_{\rm c}$. Panels correspond to the same lens configurations as in Figure~\ref{fig:point_mag}. Solid contours are plotted for $\delex=\pm 0.001$, $\pm 0.01$, and $\pm 0.1$; the dot-dashed contour corresponds to $\delex=0$. As marked in the color bar, positive values are mapped in shades of green with black contours, negative in shades of red with white contours. Areas with $|\delex|\leq10^{-4}$ are left white. The open circle in the upper left corner of each panel illustrates the size of the source (note the different scales in each column).}
\label{fig:delta_ex}
\end{figure*}

We first compute in \S~\ref{sec:region} maps of the sensitivity for different lens configurations and illustrate its general properties. In \S~\ref{sec:analytical} we derive an analytical estimate of the effect valid for any source not lying directly on the caustic. Finally, in \S~\ref{sec:probability} we compare the probability of events with an extended-source effect of a given amplitude with the probability of caustic-crossing events.

\subsection{Region of Sensitivity}
\label{sec:region}

For a source centered at ${\bf y}_{\rm c}$ we describe the sensitivity by the relative amplification excess over the point-source amplification,
\beq
\delex({\bf y}_{\rm c})= \frac{A_*({\bf y}_{\rm c}) - A_0({\bf y}_{\rm c})}{A_0({\bf y}_{\rm c})} \,.
\label{eq:delta_ex}
\eeq
Figure~\ref{fig:delta_ex} shows contour plots of $\delex$ for the lens geometries used in Figure~\ref{fig:point_mag} and a source star with radius $\rho_* = 0.02$. As anticipated, most of the area sensitive to the resolved source is concentrated near the caustic curves. However, the $|\delex|=0.001$ and 0.01 contours depart in some cases substantially from the caustic, as discussed further below and in \S~\ref{sec:probability}.

The dot-dashed contour extending from the cusps corresponds to a zero excess, with the extended-source amplification equal to the amplification of a point source positioned at its center. In the dominant positive green areas the amplification exceeds $A_0({\bf y}_{\rm c})$. This is to be expected, since for a source positioned fully outside or inside the caustic a large part of the limb lies closer to the caustic than the source center, and the point-source amplification increases convexly toward the caustic.

The limited negative red areas, in which the amplification is lower than $A_0({\bf y}_{\rm c})$, can be seen as compact lobes of varying extent along the outer axes of all cusps. Further away from any cusp along its axis $\delex$ changes back to positive. The more subtle reason for this effect is caused by the concave drop of amplification in both directions perpendicular to the axis, which is initially stronger than the outward convex decrease along the axis. The off-axis parts of the extended-source limb thus decrease the total amplification below the value corresponding to its center. Further from the caustic the relation reverses and the perpendicular drop becomes weaker than the radial drop. On a surface plot corresponding to the point-source amplification maps in Figure~\ref{fig:point_mag} these regions would appear as sharp ridges extending outward from the cusps and eventually merging with the surroundings. Neglecting the change of amplification perpendicular to the cusp axis can lead to misleading results, such as the positive extended-source effect obtained by \cite{gaudi_petters02b} along the full extent of the outer axis of a generic cusp. For details see \S~\ref{sec:analytical}.

\begin{figure*}[t]
\begin{center}
\hspace{5mm}\includegraphics[scale=.6]{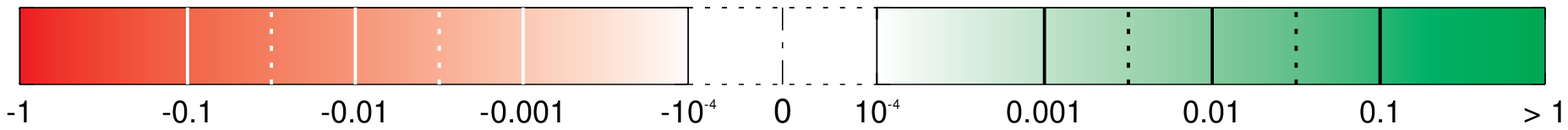}\vspace{5mm}\\
\includegraphics[scale=.7]{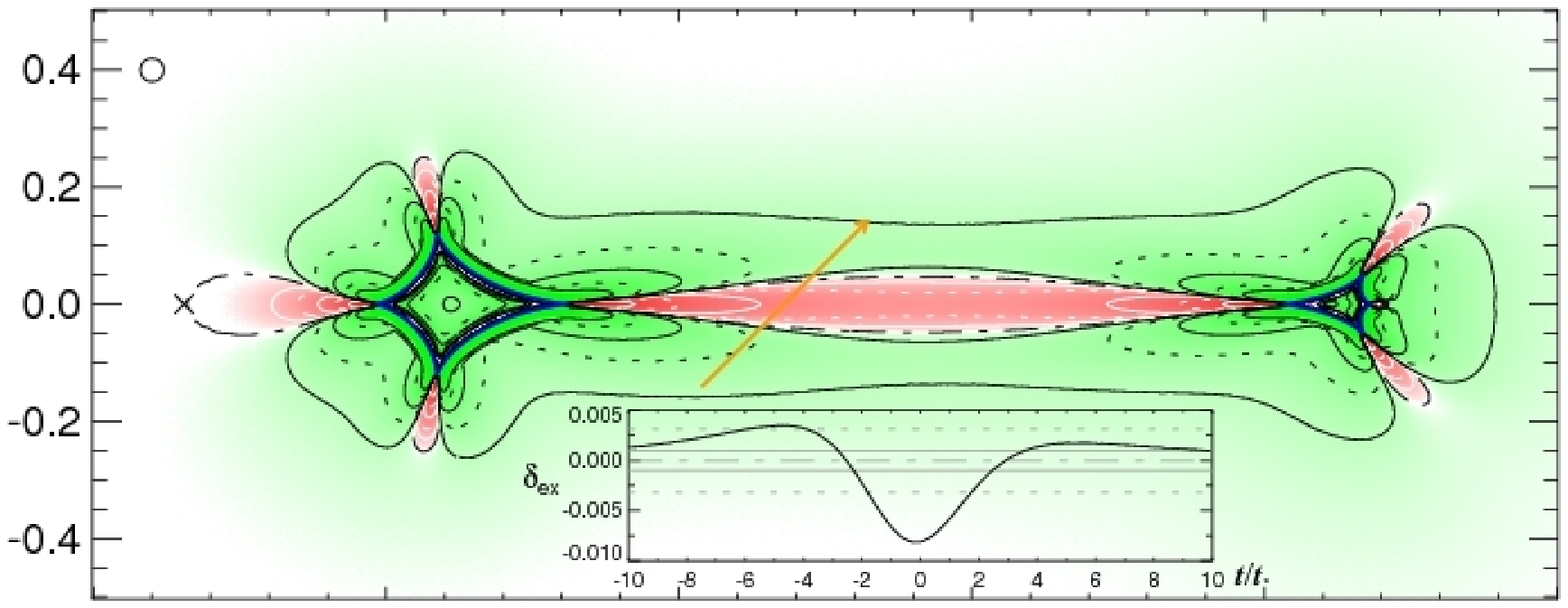}\\
\includegraphics[scale=.7]{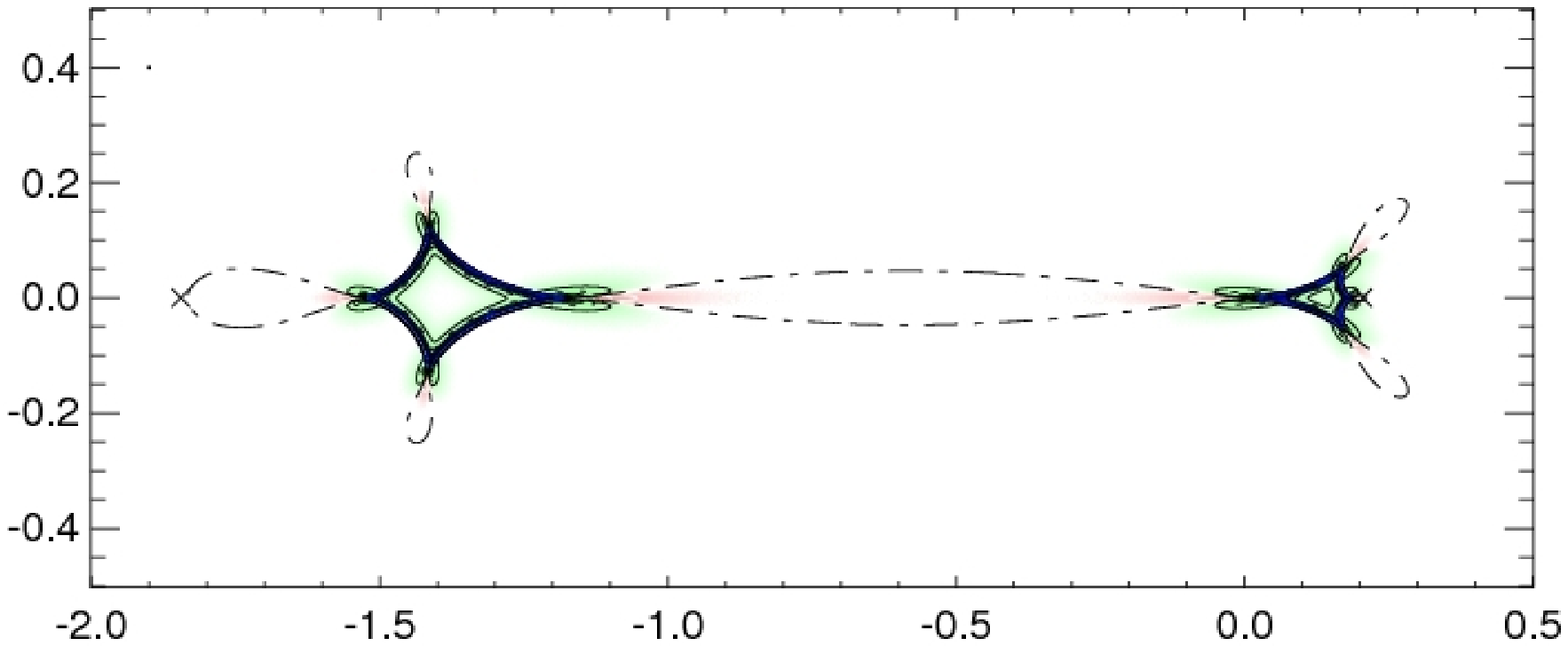}
\end{center}
\caption{Extended-source effect $\delex({\bf y}_{\rm c})$ for a lens with $d=2.05$ and $q=1/9$, and sources with $\rho_* = 0.02$ (top panel) and $\rho_* = 0.002$ (bottom panel). Symbols and lines have the same meaning as in Figure~\ref{fig:delta_ex} with four additional dotted contours at $\pm10^{-2.5}$ and $\pm10^{-1.5}$. Inset in top panel: cut along the source track marked with the orange arrow. The time axis is parameterized with $t$ expressed in units of source-radius crossing time $t_*$ and $t=0$ corresponding to the crossing of the caustic axis ${\rm y}_2 = 0$. Values corresponding to $\delex=0,\,\pm10^{-3},$ and $\pm10^{-2.5}$ contours are marked in the inset by gray lines.}
\label{fig:delta_ex_src}
\end{figure*}

The smallest negative lobes can be found along the lens axis: for non-equal-mass binaries ($q<1$) the smallest lies around the heavier lens-component; for an equal-mass binary ($q=1$) the two lobes closest to either component are the smallest. Interestingly, it is in this region that the increase of point-source amplification toward the caustic is the strongest. However, as indicated by the more circular contours in Figure~\ref{fig:point_mag}, in this case the extending ridge is more blunt, nearly radially symmetric in some cases. The perpendicular drop soon becomes comparable with the radial drop, which then dominates and leads to a positive excess.

The most interesting and unexpected feature can be seen in the outer columns with the compound caustics. The negative sensitive area connects facing cusps of the caustic components, extending many source radii along the axis from the caustic. These regions correspond to the narrow higher-$A_0$ features of Figure~\ref{fig:point_mag} noted earlier in \S~\ref{sec:extended}. The amplification excess can reach well over $1\%$ in these regions and is thus observationally significant. For a given lens separation $d$ the significance or strength is dependent on the mass ratio $q$. With decreasing $q$ the caustic components become more separated, the regions become more diffuse perpendicularly, and the maximum $|\delex|$ decreases. Similarly, for a fixed $q$ the importance of this region depends on $d$. For instance, for $q=1/3$ (middle row of Figure~\ref{fig:delta_ex}) the connecting area splits into opposite lobes for $d \lesssim 0.41$ and for $d \gtrsim 3.55$. Note that in the small-separation limit the two-point-mass lens converges to a single-point-mass lens with the same total mass, while the large-separation limit leads to two independent point-mass lenses. Many events in both regimes are hard to distinguish from single-point-mass microlensing.

In the top panel of Figure~\ref{fig:delta_ex_src} we present an enlargement from Figure~\ref{fig:delta_ex}, namely, the bottom right lens with $d=2.05$ and $q=1/9$. Details of the structure close to the caustic are better visible in this blowup. On the inner side of the caustic we may begin to discern a very narrow negative region, where the diverging point-source amplification dominates over the non-divergent extended-source amplification, bringing $\delex$ in the limit to $-1$ along the inner side of the caustic. Just outside the caustic, at a sufficient separation from cusps, the higher positive contours converge to a curve parallel to the caustic at a distance of one source radius. Here the extended-source limb enters the caustic while the point source remains outside, thus only the extended-source amplification starts to grow rapidly. The area between this curve and a similar parallel curve inside the caustic is the key region where extended-source effects have been considered seriously so far. From our results it is clear that at the $1\%$ level this region should be substantially expanded close to cusps, especially along the axis connecting the caustic components. In this example the region around the smaller caustic component on the right is fully dominated by cusp proximity, so that the one-source-radius parallel curve is relevant only at the $10\%$ level. At the level of $0.1\%$ the sensitive region broadly covers the entire caustic structure including the area between the components.

In order to demonstrate the large extent of the sensitive zone we included an inset plot illustrating the $\delex$ variation with time for a source traveling along the arrow marked in the main plot. The maximum effect reaches nearly $-1\%$ when the source is located along the axis 17.6 $\rho_*$ from the nearest caustic point ($-1\%$ would be reached 16.2 $\rho_*$ from the cusp). Hence, even in such entirely non-caustic-crossing events, neglecting extended-source effects may lead to inaccurate light-curve fits and biased event parameters.

\bef
\begin{center}
\plotone{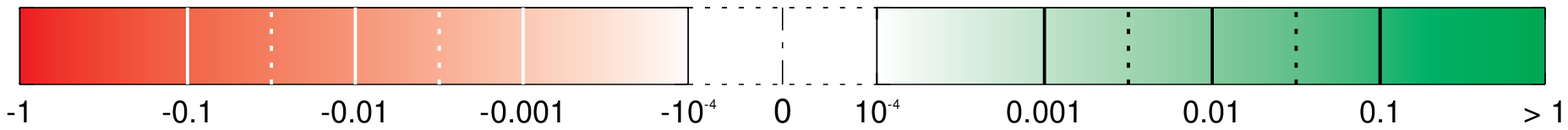}
\vspace{0.3cm}
\includegraphics[scale=0.22]{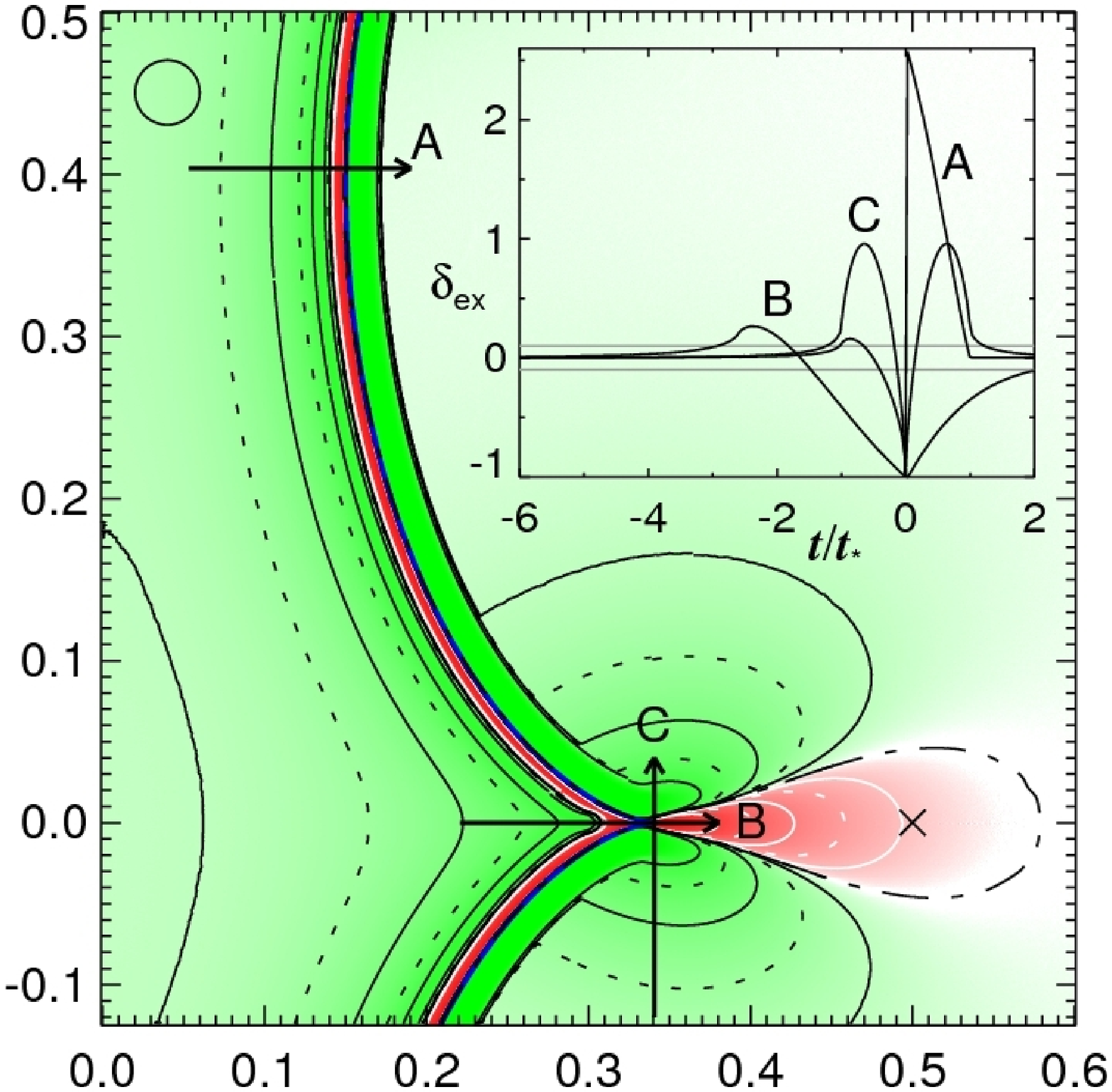}
\includegraphics[scale=0.22]{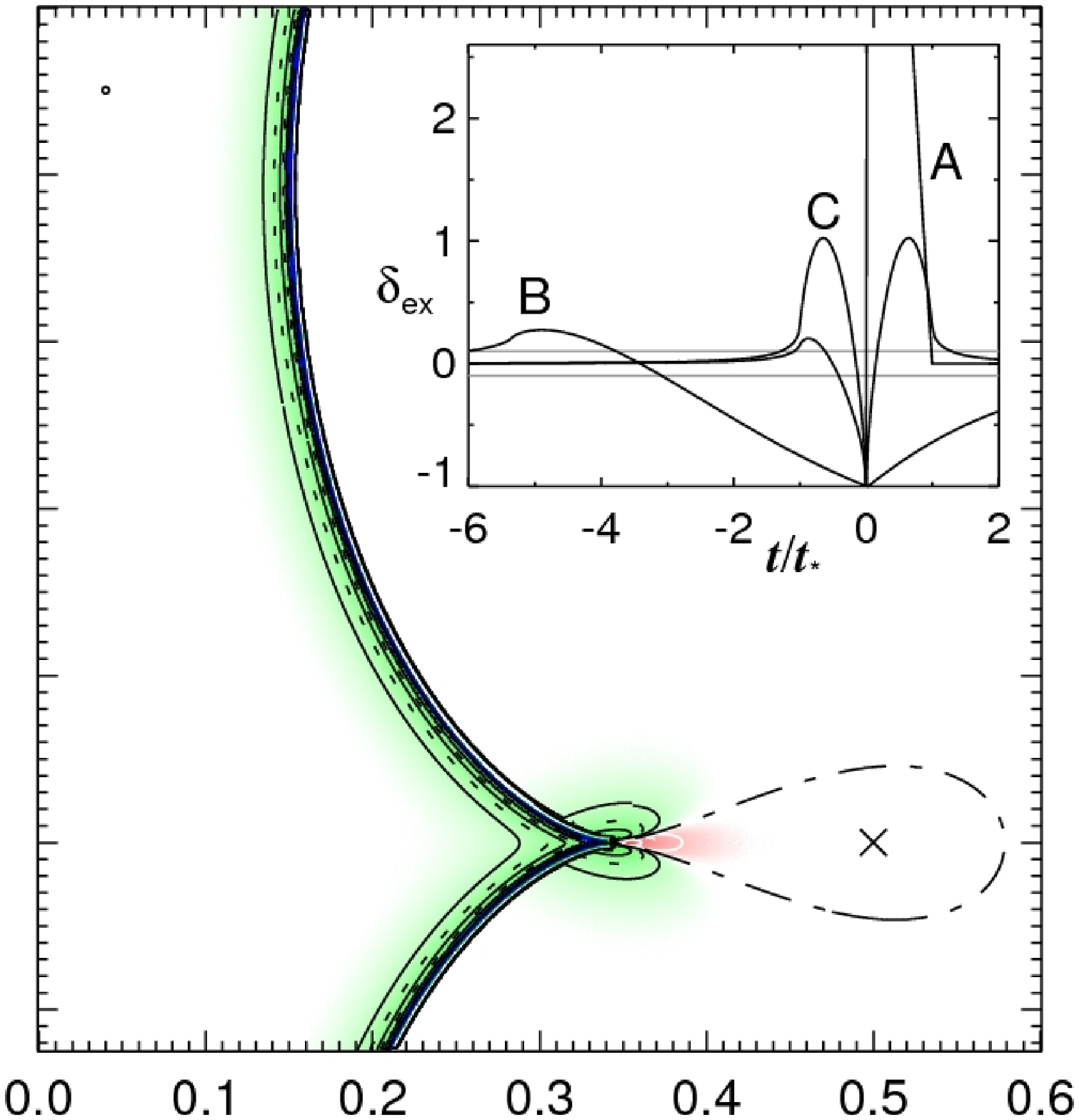}\\
\vspace{0.3cm}
\includegraphics[scale=0.44]{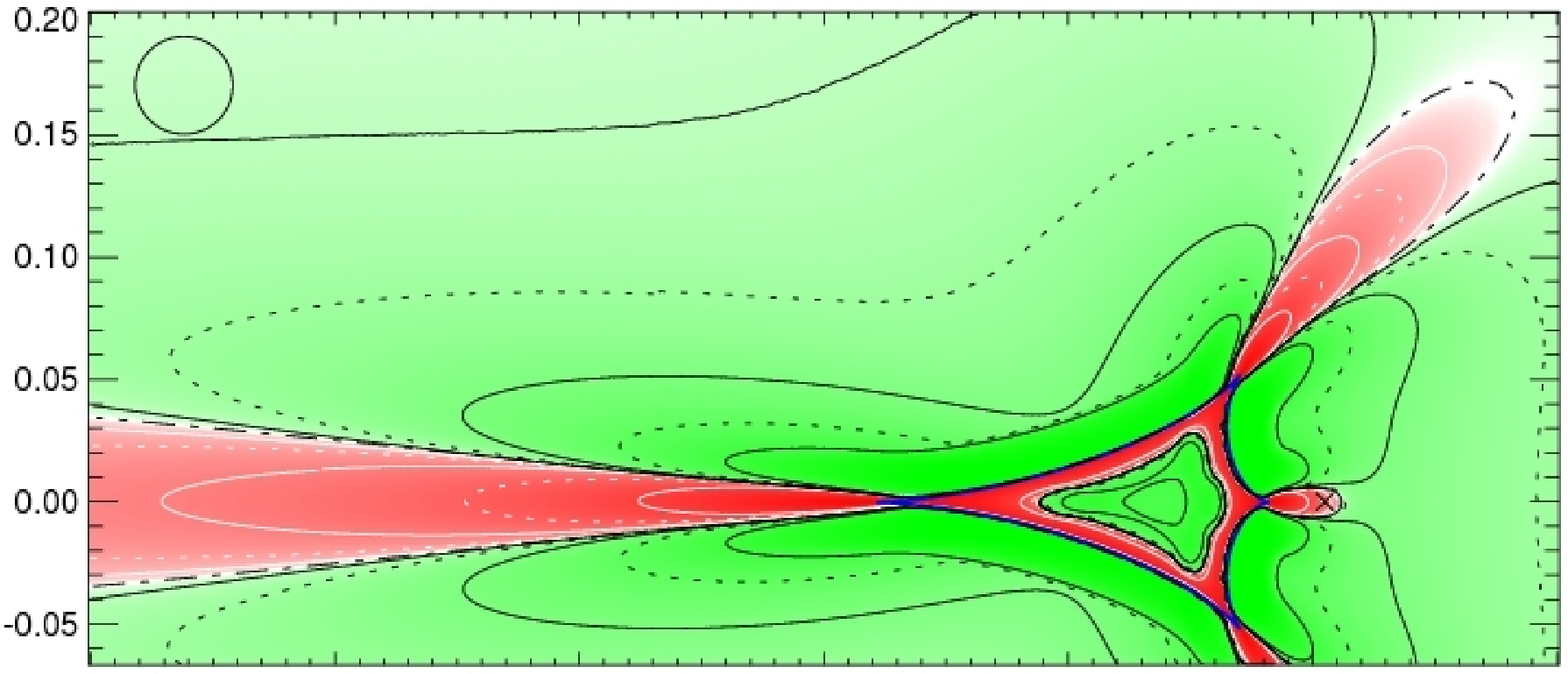}
\includegraphics[scale=0.44]{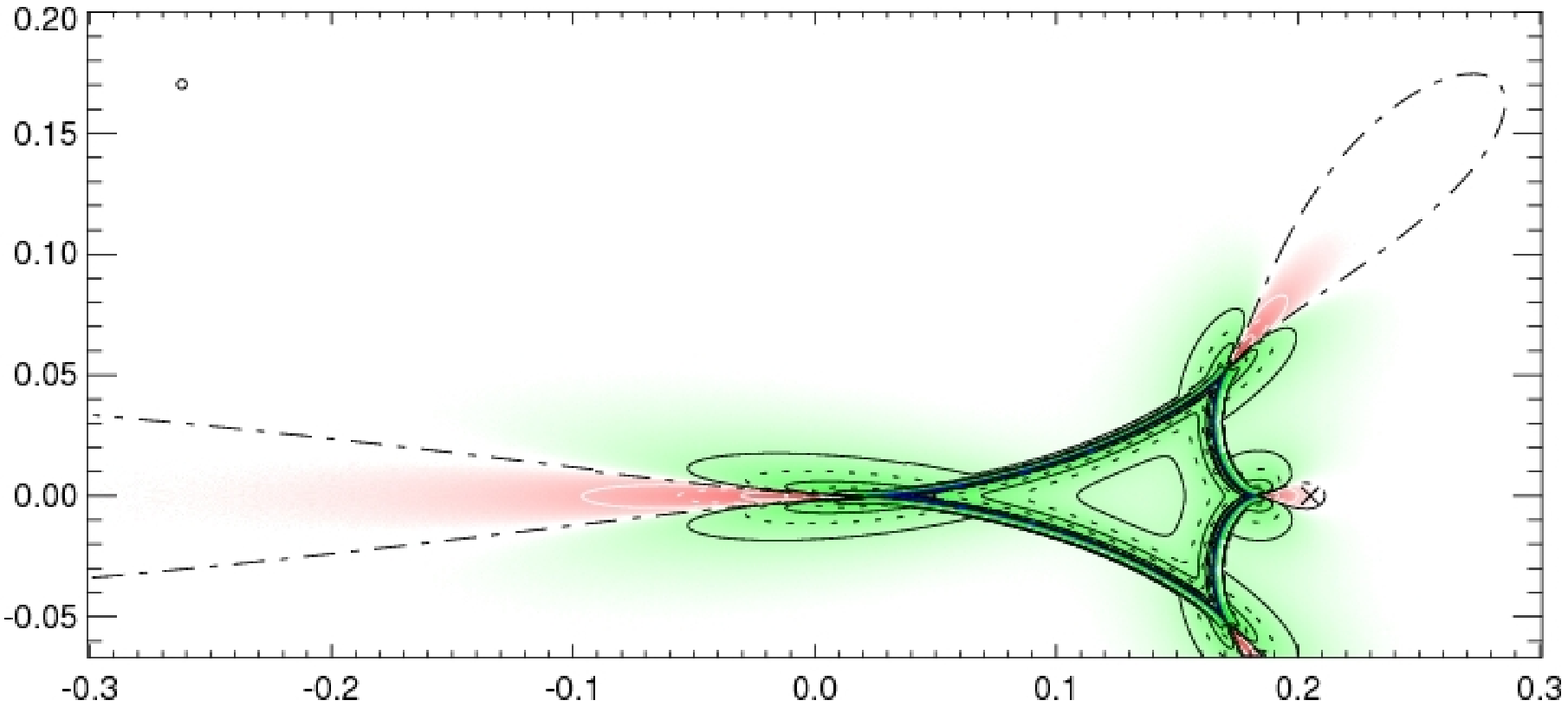}\\
\end{center}
\caption{Details of extended-source effect $\delex({\bf y}_{\rm c})$ for a $d=1,$ $q=1$ lens (top panels), and a $d=2.05$, $q=1/9$ lens (middle and bottom panels) for source radii $\rho_*=0.02$ (top left and middle panels) and $\rho_*=0.002$ (top right and bottom panels). Meaning of symbols and lines as in Figure~\ref{fig:delta_ex_src}. Top left inset: cuts for source tracks marked by arrows; top right inset: cuts for the same number of source radii along the same tracks. Caustic crossing of the source center occurs at time $t=0$; $t_*$ is the source-radius crossing time. Gray horizontal lines in the insets mark contour values $\delex=\pm0.1$.}
\label{fig:delta_ex_det}
\enf

In the bottom panel of Figure~\ref{fig:delta_ex_src} we illustrate the scaling of $\delex$ with source size by using the same lens parameters and a ten times smaller source, $\rho_* = 0.002$. Here even the outer $0.1\%$ contour closely follows the caustic. As expected, in comparison with the $1\%$ and higher contours from the top panel, the distance of the contours from the caustic generally scales with the source size. Nevertheless, the scaling is not uniform, as can be noticed from the somewhat more elongate non-zero contours along the cusp axes. Indeed, for a source crossing the caustic axis a $-1\%$ effect occurs as far as 28 $\rho_*$ to the right of the cusp of the larger caustic component.

This can be seen also in Figure~\ref{fig:delta_ex_det}, which includes a detail of the top center panel of Figure~\ref{fig:delta_ex} (with $q=1$ and $d=1$) and a further blow-up of Figure~\ref{fig:delta_ex_src}, each for source radii $0.02$ and $0.002$. In the top left panel we include cuts through the plot along three marked source trajectories crossing the caustic. In the top right panel we include cuts for the smaller source along the same straight lines extending the same number of source radii from the caustic (i.e., in this panel the arrows would be shrunk ten times towards the caustic). The A trajectories correspond to the well-studied fold-crossing regime. As the source approaches the caustic from inside, the positive effect peaks just after the limb touches the caustic and drops to $\delex=-1$ at the caustic. The peak value is only slightly higher for the smaller source. Once the source center is outside the caustic, the corresponding point-source amplification jumps to a low value and the extended-source amplification dominates by a factor roughly proportional $\rho_*^{-1/2}$. Nevertheless, $\delex$ drops rapidly practically to zero when the source limb fully exits the caustic. The region with $|\delex|\gtrsim 0.01$ is thus limited to a band from $2.4\,\rho_*$ inside to $1\,\rho_*$ outside the caustic. For the region with $|\delex|\gtrsim 0.001$ the outer limit remains the same, but the character of the inner region in this particular situation depends on the source size. While for the smaller source the inner limit of the band lies at $7.8\,\rho_*$, for the larger source the width of the entire caustic is less than double (only $15\,\rho_*$) along the horizontal line extended from A. The $\delex$ value thus remains larger than $0.001$ along the line inside the caustic. The validity of the fold-caustic approximation is discussed further in \S~\ref{sec:fold}.

Sources moving along trajectories B directly exit the caustic through the cusp along its axis. The character inside the caustic is similar to A with a few important differences. The peaks are substantially higher than those of the corresponding A trajectories. The positive peak occurs already $2.4\,\rho_*$ from the cusp for the larger source, when the limb of the source touches the inclined parts of the caustic above and below the axis. Due to the curvature of the caustic, for the smaller source the peak occurs as far as $4.9\,\rho_*$ from the cusp. Outside the caustic the curves are entirely different from A, slowly rising from $-1$ towards zero, the larger source passing $\delex=-0.1$ at $2\,\rho_*$, $\delex=-0.01$ at $4.4\,\rho_*$, and $\delex=-0.001$ at $7.6\,\rho_*$ from the cusp. The rise is even slower for the smaller source, which passes $\delex=-0.1$ at $4.6\,\rho_*$, $\delex=-0.01$ at $11\,\rho_*$, and $\delex=-0.001$ as far as $23\,\rho_*$ from the cusp. Inside the caustic the larger source crosses $\delex=0.01$ at $5.9\,\rho_*$ and $\delex=0.001$ at $14\,\rho_*$ from the cusp, the smaller source at $12\,\rho_*$ and $26\,\rho_*$, respectively. The important result that the smaller source size leads to a larger extent of the sensitive region in terms of source radii on both sides of the cusp contradicts one of the analytical findings of \cite{gaudi_petters02b}, who concluded that the extent was independent of source size for sufficiently small sources. We discuss this point further in \S~\ref{sec:analytical}.

Source-center trajectories C pass through the cusp perpendicular to its axis. Except for the immediate vicinity of the cusp, the values are positive, peaking at $\delex\simeq 1$ at positions $0.65\,\rho_*$ from the cusp, dropping to $0.1$ at $1.2\,\rho_*$ from the cusp, to $0.01$ at $3.1\,\rho_*$, and to $0.001$ at $8.2\,\rho_*$ from the cusp. The decay rate is thus intermediate between cases A and B. In this case the curves for the two source sizes are nearly identical, differing only slightly in the peak height.

\begin{figure*}
\begin{center}
\includegraphics[scale=0.3]{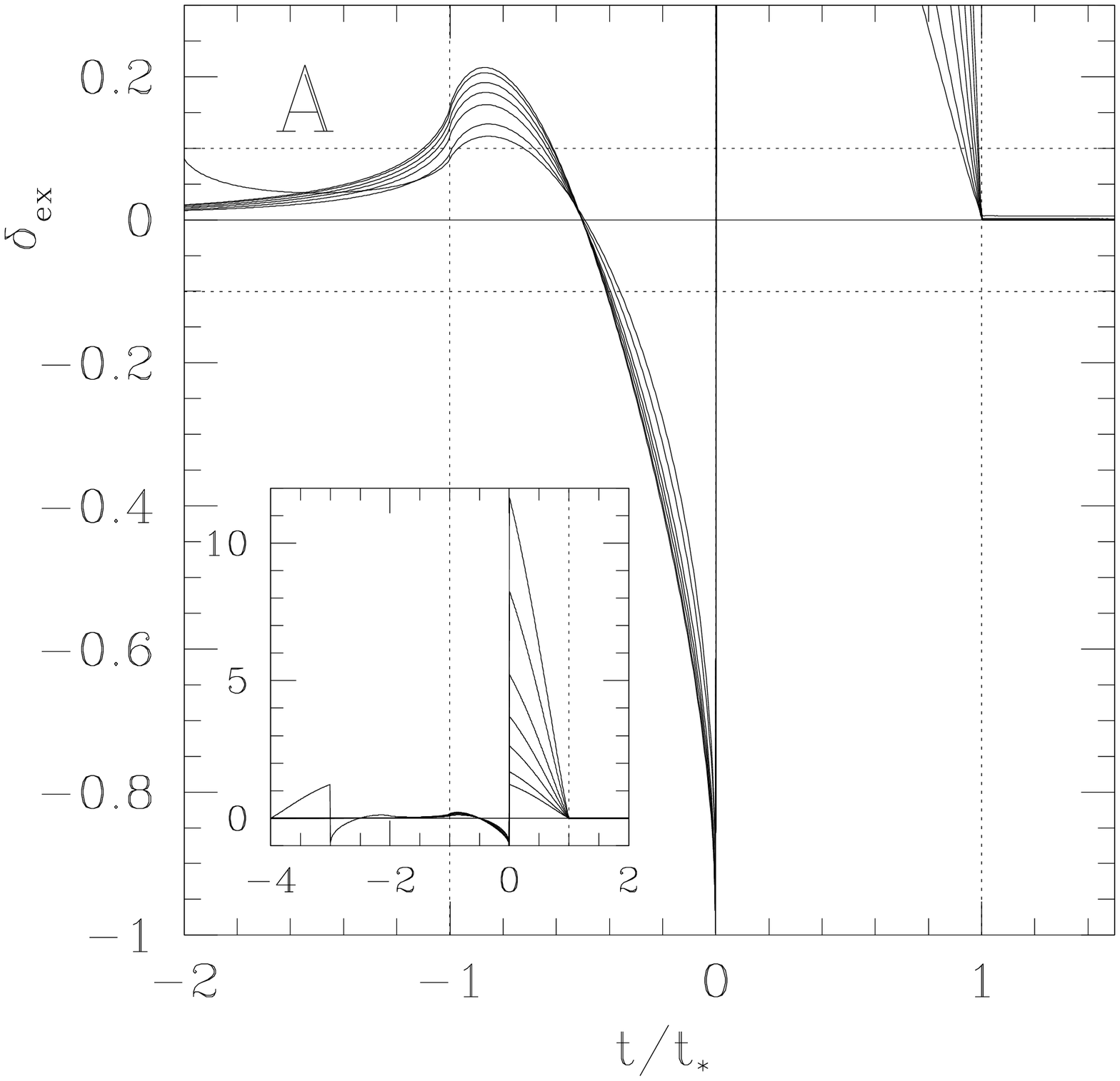}
\includegraphics[scale=0.3]{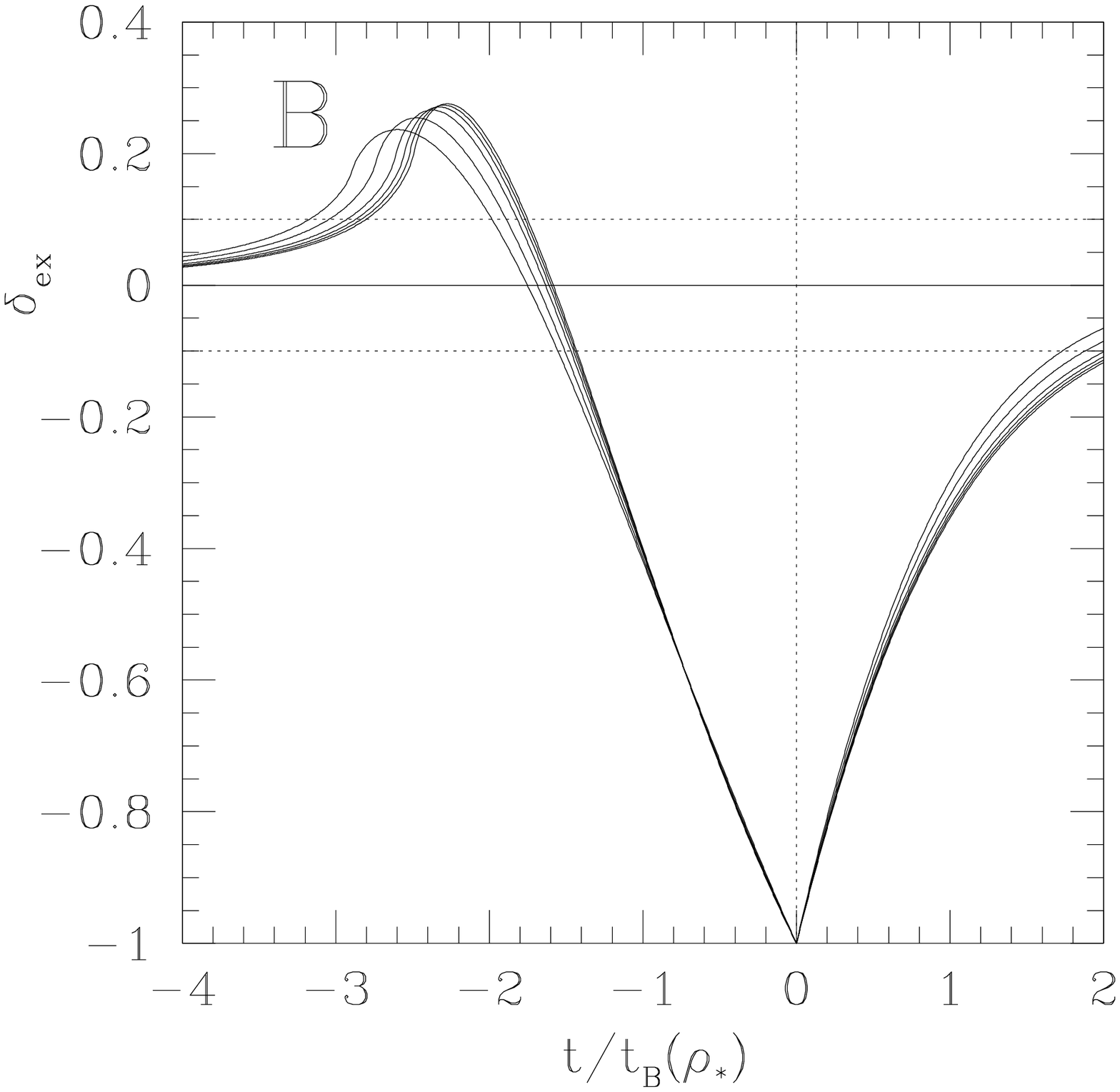}\\
\includegraphics[scale=0.3]{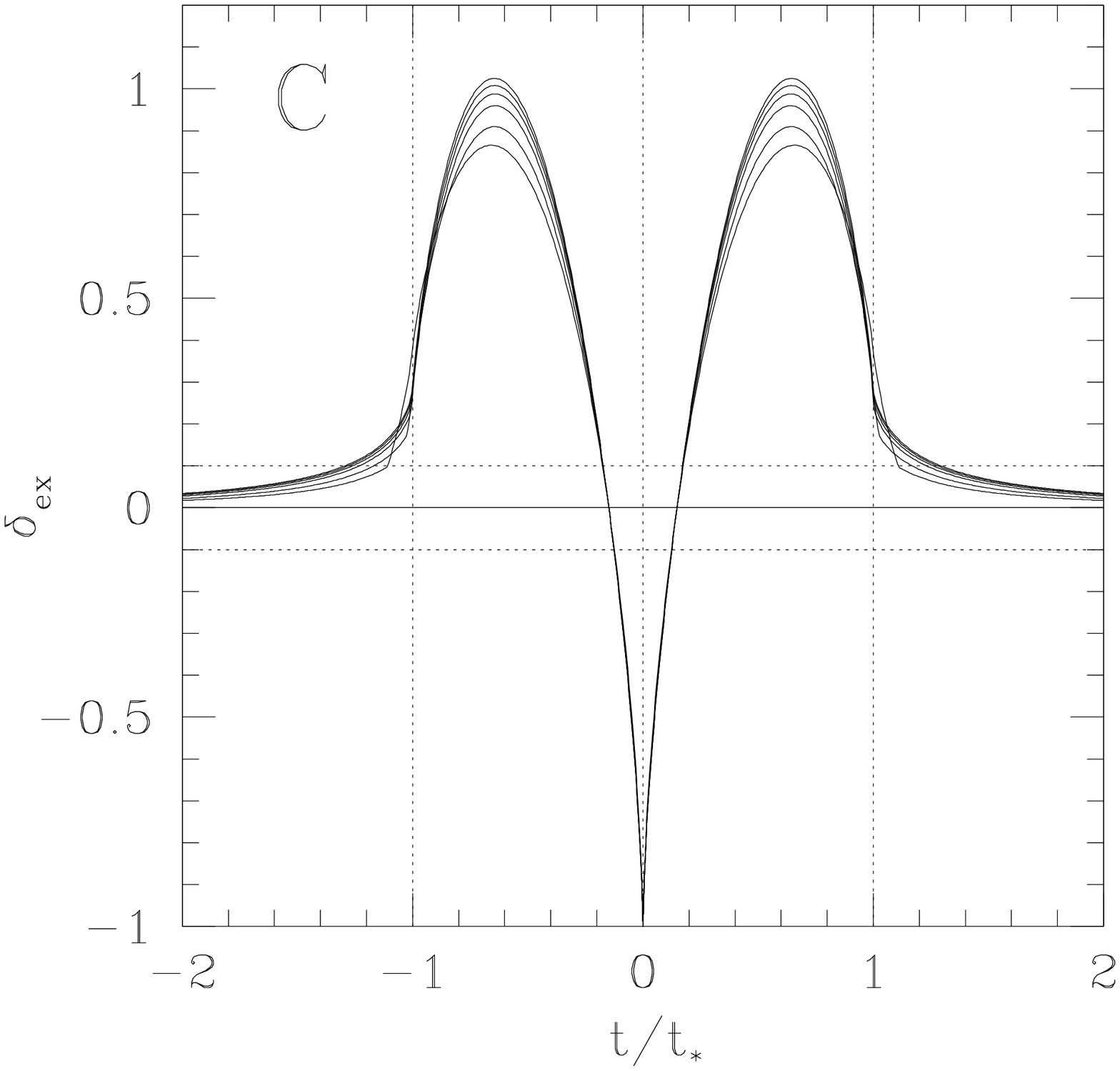}
\includegraphics[scale=0.3]{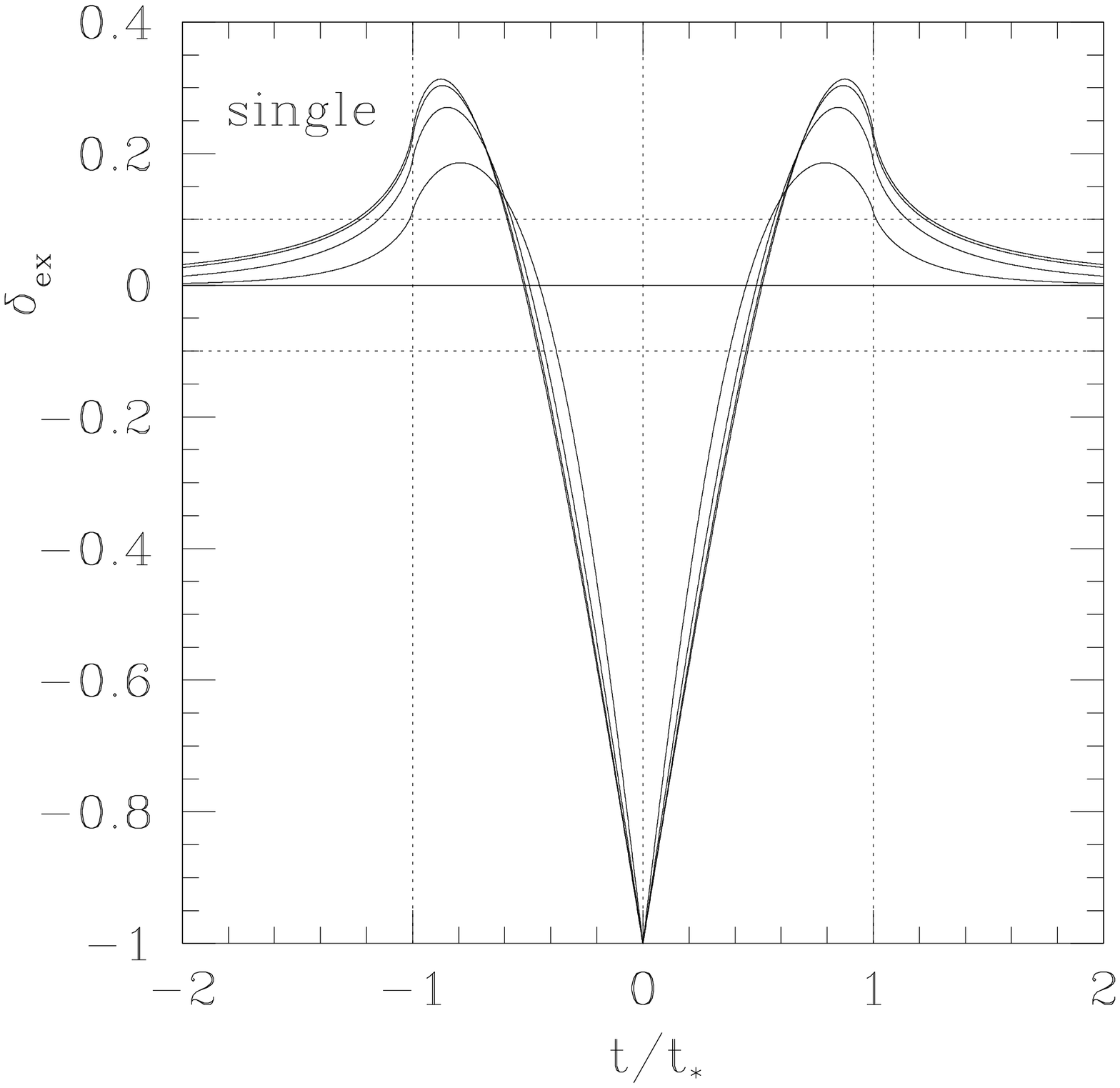}\\
\end{center}
\caption{Source-size dependence of extended-source effect $\delex$. Panels correspond to the source tracks marked in Figure~\ref{fig:delta_ex_det} as noted in the top left corner, and to a source-center-crossing single-point-mass lens (bottom right) for comparison. Panel A inset shows an expanded view of the plot. Curves with increasing peak height correspond to source radii $\rho_*=\{0.1,0.05,0.02,0.01,0.005,0.002,0.001\}$ for panel A, $\rho_*=\{0.1,0.05,0.02,0.01,0.005,0.002\}$ for panels B and C, and $\rho_*=\{2,1,0.5,0.2\}$ for the single lens. Time is marked in units of $\rho_*$-crossing time $t_*$ except in panel B, where it is marked in units of the rescaled $t_{\rm B} (\rho_*)=(\rho_*/0.02)^{-1/3}\,t_*$ to account for the changing characteristic scale along the cusp axis. Vertical dotted lines mark the times of first contact of the source with the crossing point of the caustic, source-center crossing, and last contact. In panel B the first and last contact with the cusp occurs at $t/t_{\rm B}(\rho_*)=\mp(\rho_*/0.02)^{1/3}$. Horizontal dotted lines mark $\delex=\pm0.1$.
\label{fig:delex_size}}
\end{figure*}

The source-size dependence of the extended-source effect is illustrated in more detail in Figure~\ref{fig:delex_size} for the three cuts from Figure~\ref{fig:delta_ex_det} and a source-center-crossing single-point-mass lens for comparison. The graphs for cuts A, C, and for the single lens are plotted as a function of time in units of the source-radius crossing time $t_*$. For cut B time is shown in units of $t_{\rm B}(\rho_*)=(\rho_*/0.02)^{-1/3}\,t_*$, in order to account for the changing characteristic scale along the cusp axis discussed above. In each panel the curves with increasing peak height correspond to decreasing source size. For cut A, the values of $\rho_*=\{0.1,0.05,0.02,0.01,0.005,0.002,0.001\}$, with the previous caustic entrance visible in the inset for $\rho_*=0.1$; for cuts B and C, $\rho_*=\{0.1,0.05,0.02,0.01,0.005,0.002\}$; for the single lens, $\rho_*=\{2,1,0.5,0.2\}$. The $\delex$ curves converge for $\rho_*\rightarrow 0$ nearly everywhere to limit curves, so that curves for smaller source radii would be indistinguishable in the plots. The only obvious exception occurs for a source centered within one radius outside the caustic in cut A, where $A_0(t/t_*)$ converges to a constant while $A_*(t/t_*)$ and thus also $\delex(t/t_*)$ diverge as $\rho_*^{-1/2}$. It is interesting to note the rapidity of the convergence in the case of the single-point-mass lens, in which the limit curve is an analog of the uniform-source $B(z)$ curve of \cite{gould94}. In the case of the three presented binary-lens cuts the limit curves are valid for sources at least two orders of magnitude smaller than in the single-lens case. For instance, along the fold-crossing cut A, the $t<0$ part of the $\delex$ curve is independent of the source radius only for $\rho_*\lesssim 0.001$, and the {\em shape} of the $t<t_*$ part of the corresponding $A_*$ light curve is independent of $\rho_*$ only for $\rho_*\lesssim 0.005$ (not shown here\footnote{The faster convergence of the light-curve shape is caused by the partial compensation of the steeper drop of $\delex$ at $t\rightarrow 0^-$ for a smaller source by the steeper rise of $A_0$ closer to the caustic.}).

Even though we explored the source-size dependence only for three specific caustic crossings of a specific lens, due to the universality of the local lensing behavior the cut A limit curve is valid for any fold crossing, provided we replace $t_*$ by the source-radius {\em caustic}-crossing time in non-perpendicular crossings. Similarly, the cut B and C limit curves are valid for any parallel and perpendicular cusp crossings, respectively. However, in all cases the rate of convergence as $\rho_*\rightarrow 0$ {\em does} depend on the geometry of the particular crossing, primarily on the local curvature of the caustic, cusp shape, and vicinity of (other) cusps.

The opposite large-source regime is illustrated in the middle panel of Figure~\ref{fig:delta_ex_det}, where the source size is comparable with the size of the caustic component. Hence, the proximity of all cusps has an influence on any caustic-crossing event, as demonstrated by the smooth deformed contours, very different from those in the top left panel. Along the lens axis the $\delex=-0.01$ contour extends $14.6\,\rho_*$ out from the leftmost cusp, but only $1.29\,\rho_*$ out from the rightmost cusp. The structure for the smaller source in the bottom panel looks more familiar. Here the ratio between the $\delex=-0.01$ left and right contour extent is relatively less extreme but still large: $26.2\,\rho_*$ from the leftmost cusp and $4.8\,\rho_*$ from the rightmost cusp. Note that the only significant region inside the caustic with a negligible extended-source effect $|\delex|<0.001$ lies within the innermost contour.

To summarize the influence of source size, a smaller source reaches higher amplification close to the caustic. The amplification excess $|\delex|$ typically also increases, but for sufficiently small sources on the inner side of folds and in any direction from cusps it remains constant. While the extended-source effect for a smaller source is relevant in smaller regions around the caustic in terms of Einstein radii, it extends substantially further outwards from cusps in terms of source radii $\rho_*$. Along the axis connecting caustic components a $1\%$ effect may occur tens of $\rho_*$ from cusps for the smaller source, and significantly over $10\,\rho_*$ even for the larger source.

\subsection{Analytical Estimate of Extended-Source Effect}
\label{sec:analytical}

\subsubsection{General Results}
\label{sec:analytical_general}

For any sufficiently small source we may study the amplification excess $\delex$ analytically by expanding the point-source amplification appearing in equation~(\ref{eq:amp_integral}) around ${\bf y}_{\rm_c}$. In principle, such an approach should give us good results for any region where $A_0({\bf y})$ is analytic within a source radius $\rho_*$ of point ${\bf y}_{\rm_c}$, which means anywhere except within a source radius of the caustic. The extended-source amplification can then be written as
\begin{eqnarray}
\lefteqn{\!\!\!\!\!\!A_*({\bf y}_{\rm c}) = A_0({\bf y}_{\rm c})+\sum_{j=1}^2 \frac{\partial A_0}{\partial {\rm y}_j}({\bf y}_{\rm c})\frac{\int_{\Sigma_{\rm S}} {\rm y}'_j B({\bf y}\,')\rmd^2 {\bf y}\,'}{\int_{\Sigma_{\rm S}} B({\bf y}\,')\rmd^2 {\bf y}\,'} +} \nonumber \nopagebreak[4]\\
& & \!\!\!\!\!\!\!\!\!\!\!\!+ \frac{1}{2}\sum_{j,k=1}^2 \frac{\partial^2 A_0}{\partial {\rm y}_j\, \partial {\rm y}_k}({\bf y}_{\rm c})\frac{\int_{\Sigma_{\rm S}}{\rm y}'_j{\rm y}'_k B({\bf y}\,')\rmd^2 {\bf y}\,'}{\int_{\Sigma_{\rm S}} B({\bf y}\,')\rmd^2 {\bf y}\,'} + O (\rho^3_*)\, .
\label{eq:exp_gen}
\end{eqnarray}
The first correction to the point-source amplification is given by the dot product of the point-source amplification gradient and the position of the center of brightness of the source measured from source center,
\beq
{\bf y}_{\rm bc}=\frac{\int_{\Sigma_{\rm S}} {\bf y}\,' B({\bf y}\,')\rmd^2 {\bf y}\,'}{\int_{\Sigma_{\rm S}} B({\bf y}\,')\rmd^2 {\bf y}\,'}={\bf y}_{\rm b}-{\bf y}_{\rm c}\, ,
\label{eq:ctr_o_br}
\eeq
where ${\bf y}_{\rm b}$ is the position of the center of brightness measured from the origin of the global coordinates. The following terms in equation~(\ref{eq:exp_gen}) involve higher-order derivatives of the point-source amplification and higher-order moments of the brightness distribution.

In the special case of a circularly symmetric brightness distribution $B({\bf y}\,')=I(r)$, which is of primary interest here, we have ${\bf y}_{\rm b}={\bf y}_{\rm c}$ and all terms of odd order drop out and the first three terms of the resulting expansion are
\begin{eqnarray}
A_*({\bf y}_{\rm c}) = A_0({\bf y}_{\rm c})+\frac{1}{4}\,\Delta A_0({\bf y}_{\rm c})\rho^2_*\,\frac{\int_0^1 I(r)r^3\rmd r}{\int_0^1 I(r)r\rmd r} + \nonumber \nopagebreak[0]\\
\!\!\!\!\!\! +\frac{1}{64}\,\Delta^2 A_0({\bf y}_{\rm c})\rho^4_*\,\frac{\int_0^1 I(r)r^5\rmd r}{\int_0^1 I(r)r\rmd r} + O (\rho^6_*)\, ,
\label{eq:exp_sph}
\end{eqnarray}
where $\Delta$ is the Laplacian and $\Delta^2=\Delta\Delta$ is the biharmonic operator. For a symmetric source the leading-order approximation of  equation~(\ref{eq:delta_ex}) therefore is
\beq
\delex({\bf y}_{\rm c}) \approx \frac{\Delta A_0({\bf y}_{\rm c})}{4\,A_0({\bf y}_{\rm c})}\;\rho_*^2\; \frac{\int_0^1 I(r)r^3\rmd r}{\int_0^1 I(r)r\rmd r}\,.
\label{eq:delta_ex_theo}
\eeq
This expression provides an easy way to estimate the sensitivity to an extended source merely from the knowledge of the point-source amplification\footnote{Its Laplacian can be easily computed numerically from a map of $A_0(\bf y)$ such as those in Figure~\ref{fig:point_mag}.}.

Except for the immediate vicinity of the caustic, the relative amplification excess can thus be approximated by a product of three terms, each depending on different parameters. Only the first term depends on the source-center position and the lens parameters, therefore it fully determines the contour geometry. The contour shape away from the caustic is thus independent of the source size and limb darkening. The second term defines the quadratic scaling with source radius, i.e., doubling the source size at a given position will increase $\delex$ four times. The last term depends purely on the limb darkening of the source and varies only weakly. For the linear law given by equation~(\ref{eq:lin_limb}) the term equals $(1.5-0.7\upsilon)/(3-\upsilon)$, for our PCA model given by equation~(\ref{eq:pca_limb}) we get $(0.2155+0.3097\kappa)/(0.4665+0.3272\kappa)$. For either model these values range from 0.4 to 0.5 from the most peaked to the flattest possible profile. Such a weak dependence on the limb darkening away from the caustic justifies our usage of a fixed limb-darkening profile in this section. Its parameter $\kappa=0$ gives us $0.462$ as the value of the last term. For orientation, the most peaked profile would have $|\delex|$ lower by $13\%$, the flattest would have $|\delex|$ higher by $8\%$ away from the caustic.

\bef
\plotone{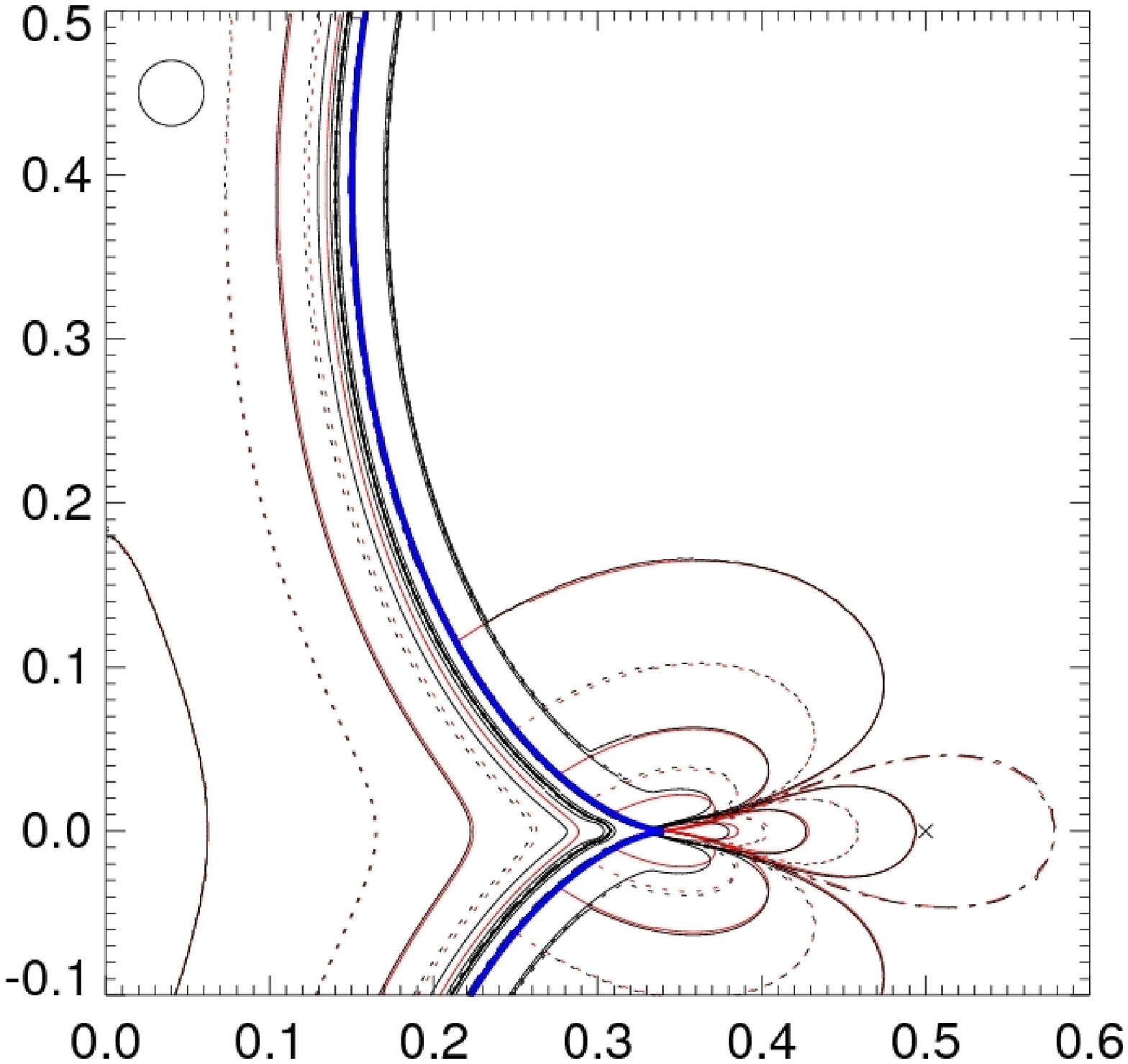}
\caption{Comparison of extended-source effect $\delex({\bf y}_{\rm c})$ and its approximation given by equation~(\ref{eq:delta_ex_theo}) for the lens and source from the top left panel of Figure~\ref{fig:delta_ex_det}. The caustic is marked by the bold blue line, exact contours are plotted in black and approximate contours in red; contour levels are the same as in Figure~\ref{fig:delta_ex_det}.}
\label{fig:laplace}
\enf

In Figure~\ref{fig:laplace} we compare this approximation to the true $\delex$ using the detail in the top left panel of Figure~\ref{fig:delta_ex_det}. The agreement is remarkable anywhere more than a source diameter from the caustic and very good up to a source radius from the caustic except near the cusp. In the outer vicinity of the caustic the exact contours bend and run parallel to the caustic at a distance of $\rho_*$, while the approximate contours extend directly to the caustic as if $\delex$ continued analytically. The slight discrepancy closer to the cusp corresponds to the change of contour geometry with source size in this region, as noted earlier in the discussion of Figures~\ref{fig:delta_ex_src}--\ref{fig:delex_size}. Adding the higher-order term from equation~(\ref{eq:exp_sph}) would improve the agreement here\footnote{This was demonstrated after the manuscript submission by \cite{gould08}, whose hexadecapole approximation is a numerical implementation of equation~(\ref{eq:exp_sph}) including the $O (\rho^4_*)$ term.}. Finally, note that all outer approximate contours extend directly to the cusp, while all exact contours extend to points slightly offset to both sides of the cusp.

The zero-effect contour clearly has a special significance. Equation~(\ref{eq:delta_ex_theo}) shows that away from the caustic $\delex=0$ corresponds to points with $\Delta A_0({\bf y}_{\rm c})=0$, independent of the source properties. This explains the similarity of the zero contours for both values of $\rho_*$ in Figure~\ref{fig:delta_ex_src} or Figure~\ref{fig:delta_ex_det}. In fact, the similarity allowed us to use the zero-Laplacian curve instead of the zero-$\delex$ curve in several of the previous plots further from the caustic, as it is less sensitive to low-level numerical noise. We also point out specifically that away from the caustic the zero-effect contour remains the same for any limb-darkening profile $I(r)$, as discussed further in \S~\ref{sec:chromaticity}.

For completeness, we briefly turn to the case of an asymmetric brightness profile, relevant in our case for example for a spotted source star. We can simplify equation~(\ref{eq:exp_gen}) similarly by expanding $A_0$ around the center of brightness ${\bf y}_{\rm b}$ instead of the source center ${\bf y}_{\rm c}$. This approach again leads to the elimination of the first correction and we are left with
\begin{eqnarray}
\lefteqn{\!\!\!\!\!\!\!\!A_*({\bf y}_{\rm c}) = A_0({\bf y}_{\rm b})+ \frac{1}{2}\sum_{j,k=1}^2 \frac{\partial^2 A_0}{\partial {\rm y}_j \partial {\rm y}_k}({\bf y}_{\rm b})\times} \nonumber \\
& \times \left[\frac{\int_{\Sigma_{\rm S}}{\rm y}'_j{\rm y}'_k B({\bf y}\,')\rmd^2 {\bf y}\,'}{\int_{\Sigma_{\rm S}} B({\bf y}\,')\rmd^2 {\bf y}\,'}\,-{\rm y}_{{\rm bc}_j}{\rm y}_{{\rm bc}_k} \right]+ O (\rho^3_*)\, ,
\label{eq:exp_asph}
\end{eqnarray}
where ${\bf y}_{\rm bc}$ is given by equation~(\ref{eq:ctr_o_br}). The amplification of an asymmetrically bright extended source lying off the caustic is thus equal to the amplification of a point source placed at its center of brightness plus a correction of second order in source radius.

We emphasize here that the expansions in equation~(\ref{eq:exp_gen}) and equation~(\ref{eq:exp_asph}) are valid for any gravitational lens and a small source of any shape and brightness distribution, not just a binary lens and a circular source. Similarly, equation~(\ref{eq:exp_sph}) and equation~(\ref{eq:delta_ex_theo}) are valid for any lens and any small source with a circularly symmetric brightness distribution. For instance, in the special case of a single-point-mass lens and a circular or elliptical source equation~(\ref{eq:exp_gen}) yields the analytical result obtained by \citet{heylo97}. Equation~(\ref{eq:exp_sph}) applied to a single-point-mass lens and a circular symmetric source reproduces the analytical result given by \citet{heysalo00}.

\subsubsection{Extended Source Near a Cusp}
\label{sec:analytical_cusps}

In order to examine the peculiarities of extended-source effects near cusps, we start from the analytical expressions for the amplification of images formed by a generic cusp, in the form derived by \cite{zakharov95,zakharov99} and presented here in the Appendix. For a sufficiently small source lying off the cusp we may then utilize equation~(\ref{eq:delta_ex_theo}) derived above to obtain $\delex$. We express the results here in terms of the source position $({\rm y_\parallel},{\rm y_\perp})$ measured in a local coordinate system centered on a given cusp, with the first axis identified with the cusp axis and the second perpendicular to it. As noted in the Appendix, without loss of generality we rotate the coordinates so that the cusp is pointed towards negative ${\rm y_\parallel}$ and opened towards positive ${\rm y_\parallel}$.

We are primarily interested in getting expressions for $\delex$ of a small source placed along the cusp axis $({\rm y_\perp}=0)$ outside $({\rm y_\parallel}<0)$ or inside $({\rm y_\parallel}>0)$ the cusp, or perpendicular to the cusp axis (with ${\rm y_\parallel}=0$). For this purpose we compute the Laplacian of the total amplification $A^{\rm tot}$, express $\delex$ using equation~(\ref{eq:delta_ex_theo}), and get our results by setting ${\rm y_\perp}=0$ or ${\rm y_\parallel}=0$, as required.

Outside the caustic the generic cusp approximation yields a single image and $A^{\rm tot}$ is given by equation~(\ref{eq:cusp_outer_total_ampl}). Starting with the most interesting region along the outer cusp axis, we arrive at the exact expression
\beq
\frac{\Delta A^{\rm tot}}{A^{\rm tot}}\,({\rm y_\perp}=0,{\rm y_\parallel}<0) =\frac{8\,K}{9\,{\rm y^3_\parallel}}+\frac{2}{{\rm y^2_\parallel}}\,,
\label{eq:cusp_outer axis_Lapl}
\eeq
where the cusp parameter $K$ defined in equation~(\ref{eq:cusp_parameter}) is always positive in our coordinate system. Note that the first term, which dominates in the vicinity of the cusp and originates from the second derivative in the perpendicular direction, is negative, because ${\rm y_\parallel}<0$ along the outer cusp axis. From the leading order term we readily get
\beq
\delex ({\rm y_\perp}=0) \approx \frac{2\,K}{9\,{\rm y^3_\parallel}}\;\rho_*^2\; \frac{\int_0^1 I(r)r^3\rmd r}{\int_0^1 I(r)r\rmd r}\,,
\label{eq:cusp_outer axis_delex}
\eeq
a negative extended-source effect along the axis dropping off with the inverse cube of the distance from the cusp. The effect is directly proportional to the cusp parameter $K$, hence $\delex$ is suppressed near broad cusps with small $K$ (such as the cusp close to the heavier component of the binary) and boosted near narrower cusps with large $K$, which explains the large extent of the sensitive zone between the narrow facing cusps of multi-component caustics. Expressing the distance in source-radius units, we see that at a given number of source radii from the cusp $\delex$ is inversely proportional to the source radius $\rho_*$. By inverting equation~(\ref{eq:cusp_outer axis_delex}) we get the scaled contour extent
\beq
\frac{{\rm y_\parallel}}{\rho_*}\,({\rm y_\perp}=0) \approx \left[ \frac{2\,K}{9\,\rho_*\delex}\;\frac{\int_0^1 I(r)r^3\rmd r}{\int_0^1 I(r)r\rmd r} \right]^{1/3}\,.
\label{eq:cusp_outer axis_extent}
\eeq
Hence, even though for smaller sources the extent of a given $\delex$ contour along the outer cusp axis shrinks as $\rho_*^{2/3}$ when measured in Einstein radii, it grows as $\rho_*^{-1/3}$ when measured in source radii, explaining our observations from \S~\ref{sec:region}. Measured either way, the extent also scales as $\delex^{-1/3}$.

The expressions derived above are in good agreement with the numerical results presented in \S~\ref{sec:region}. Their disagreement with the results obtained along the outer cusp axis by \cite{gaudi_petters02b} and \cite{gould08} comes from our accounting for the change of the amplification in the direction perpendicular to the axis in its immediate vicinity. As mentioned earlier in \S~\ref{sec:region}, close to the cusp this change dominates over the relatively slow drop along the axis. The approach of \cite{gaudi_petters02b} and \cite{gould08} takes into account only the radial drop and thus yields only the smaller second term in equation~(\ref{eq:cusp_outer axis_Lapl}), which in turn incorrectly implies a positive extended-source effect that is independent of cusp properties. The obtained contour-extent scaling would be the same as at a fold, i.e., linear with source radius if measured in units of Einstein radii, or independent of source radius if measured in units of source radii.

Staying outside the cusp and studying the effects of a small source displaced from the cusp perpendicularly to its axis, we arrive at the exact result
\beq
\frac{\Delta A^{\rm tot}}{A^{\rm tot}}\,({\rm y_\parallel}=0)=\frac{10\,}{9\,{\rm y^2_\perp}}\;.
\label{eq:cusp_perp_axis_Lapl}
\eeq
For the extended-source effect we get
\beq
\delex ({\rm y_\parallel}=0) \approx \frac{5\,}{18\,{\rm y^2_\perp}}\;\rho_*^2\; \frac{\int_0^1 I(r)r^3\rmd r}{\int_0^1 I(r)r\rmd r}\,,
\label{eq:cusp_perp axis_delex}
\eeq
a positive effect dropping off as the inverse square of the perpendicular separation from the cusp, in agreement with \cite{gaudi_petters02b}. Expressing the scaled perpendicular contour extent
\beq
\frac{|{\rm y_\perp}|}{\rho_*}\,({\rm y_\parallel}=0) \approx \left[ \frac{5\,}{18\,\delex}\;\frac{\int_0^1 I(r)r^3\rmd r}{\int_0^1 I(r)r\rmd r} \right]^{1/2}\,,
\label{eq:cusp_perp axis_extent}
\eeq
we see it is independent of the source radius, in agreement with our C cuts in Figure~\ref{fig:delta_ex_det}. Unlike the inverse-cube-root scaling with $\delex$ along the axis, perpendicularly from the cusp the extent scales as $\delex^{-1/2}$.

Inside the cusp there are three images with total amplification given by equation~(\ref{eq:cusp_inner_total_ampl}). Repeating a similar procedure as above, we get exactly
\beq
\frac{\Delta A^{\rm tot}}{A^{\rm tot}}\,({\rm y_\perp}=0,{\rm y_\parallel}>0) =\frac{8\,K}{9\,{\rm y^3_\parallel}}+\frac{2}{{\rm y^2_\parallel}}\,,
\label{eq:cusp_inner axis_Lapl}
\eeq
the same as the result in equation~(\ref{eq:cusp_outer axis_Lapl}) for the outer axis. However, along the inner part of the cusp axis ${\rm y_\parallel}>0$, thus both terms are positive. The corresponding result of \cite{gaudi_petters02b} includes only the smaller second term.

Not surprisingly, equations~(\ref{eq:cusp_outer axis_delex}) and (\ref{eq:cusp_outer axis_extent}) are valid even inside the cusp, the only difference being the positive values of $\delex$. Sources positioned along the axis symmetrically on opposite sides of a cusp thus have an opposite value of $\delex$, provided they are far enough to lie off the caustic. The required distance can be estimated by computing the position on the axis of a small source tangent to the cusp from inside,
\beq
\frac{{\rm y_{tan}}}{\rho_*} \approx \left[ \frac{K\,}{\rho_*}\right]^{1/3}\,.
\label{eq:cusp_inner_tangent_pos}
\eeq
The tangent position of the source thus also scales as $\rho_*^{-1/3}$ when expressed in source-radius units. This explains another observation from \S~\ref{sec:region}, namely, the change with radius of the position of the first peak along cut B in Figure~\ref{fig:delta_ex_det}. Even though the peak doesn't lie exactly at the tangent point, it lies only slightly closer to the cusp where a larger part of the limb is aligned with the caustic. We note that the requirement ${\rm y_\parallel}\gtrsim{\rm y_{tan}}$ implies that inside the cusp equation~(\ref{eq:cusp_outer axis_extent}) holds for values $\delex\lesssim 0.08$.

Usage of the formulae derived above in actual gravitational lensing scenarios requires more than just satisfying the conditions of equation~(\ref{eq:delta_ex_theo}), i.e., having a small source lying off the caustic. Additional constraints are placed by the applicability of the generic isolated cusp model, primarily putting limitations on the distance from the cusp. For instance, in binary lensing there are three images of a source positioned outside the caustic and five of a source inside, as opposed to one and three images, respectively, in generic cusp lensing. At a greater distance from the cusp the additional images cannot be neglected. The case of facing cusps also places obvious distance limits.

\subsection{Probability of Extended-Source Effect}
\label{sec:probability}

Having shown that the region sensitive to the source size may be considerably larger than the region occupied by the caustic, we now quantify the increase in detection probability. As in all similar cases, rigorous estimation of the probability of detecting an extended-source effect would require a light curve by light curve analysis, based on measuring the residuals from a best-fit point-source binary-lens light curve. Nevertheless, understanding that the actual probability will tend to be lower, we may estimate the sought probability increase simply by computing the number of random straight-line source trajectories that cross a given $|\delex|$ contour. For trajectories oriented at a given angle $\alpha$ from the lens axis $({\rm y}_1)$, the number is proportional to the perpendicularly projected extent of the contour. Specifically, we define $l_w(\alpha)$ as the total projected width\footnote{Sum of projected widths of all components of the contour with any possible regions of overlap counted only once} perpendicular to trajectory orientation $\alpha$ of the contour $|\delex| = w$. Similarly, to estimate the frequency of caustic-crossing events, we define $l\cacr(\alpha)$ as the total projected width perpendicular to trajectory orientation $\alpha$ of the outer parallel curve of the caustic at a distance $\rho_*$.

\bef
\plotone{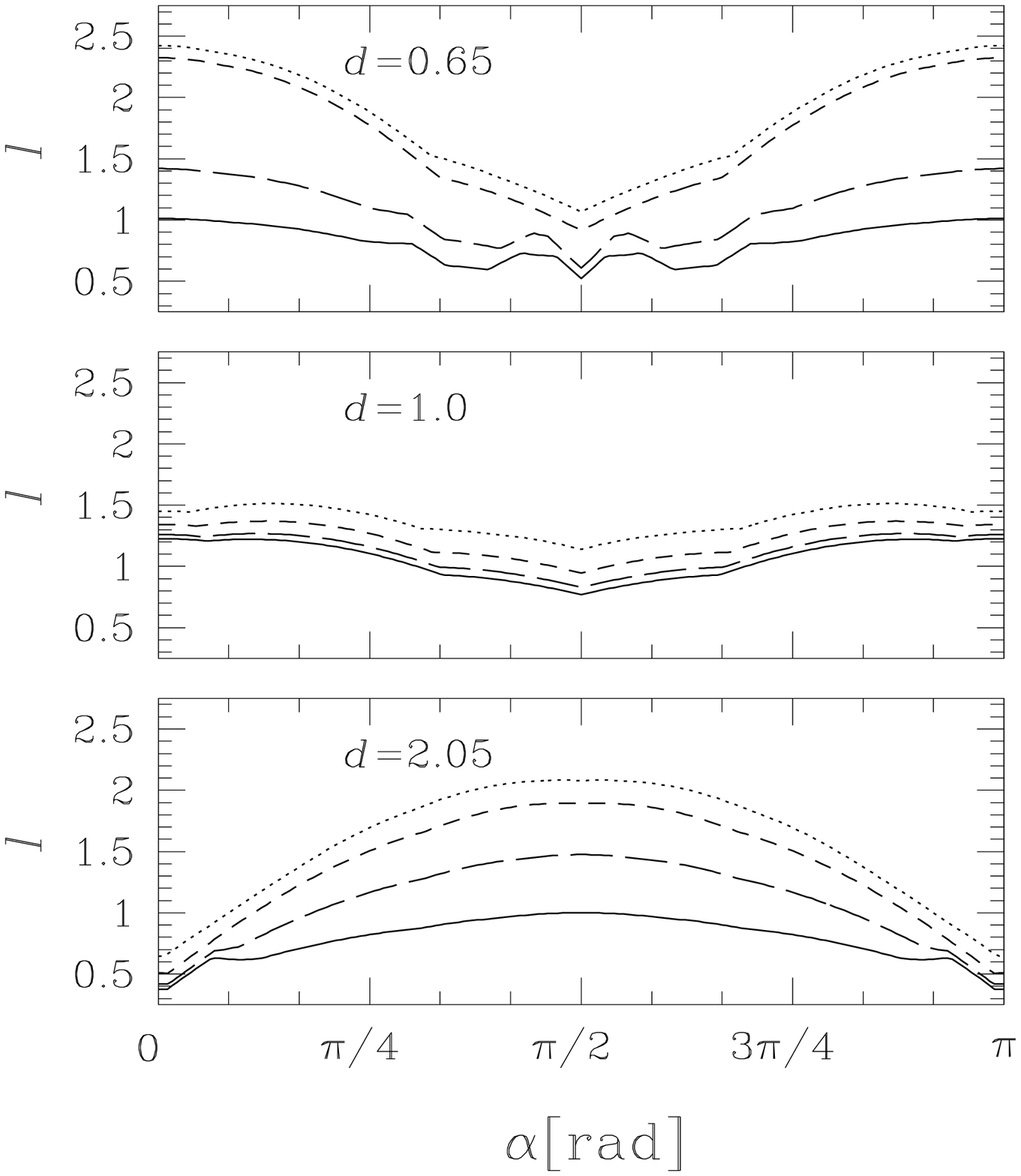}
\caption{Projected widths $l_w(\alpha)$ in Einstein radii of regions with absolute extended-source effect $|\delex|\geq w$ for source trajectories oriented at an angle $\alpha$ from the lens axis. Lens geometries correspond to the middle row of Figure~\ref{fig:delta_ex} ($q = 1/3$, $d$ marked in each panel) with source radius $\rho_* = 0.02$. Curves for $w=0.001$, $0.01$, and $0.1$ are marked by dotted, short-dashed, and long-dashed lines, respectively. Solid lines mark the projected width of the caustic-crossing region $l\cacr(\alpha)$. The symmetry around $\alpha=\pi/2$ arises from the axial symmetry of the lens.
\label{fig:cross_section}}
\enf

\begin{deluxetable*}{cccccccccccccccc}
\tablecolumns{16}
\tablewidth{0pt}
\tablecaption{Probability $P_w$ of occurrence of extended-source effect of amplitude $|\delex|\geq w$ in multiples of caustic-crossing probability
\label{tab:cross}}
\tablehead{
\colhead{} & \multicolumn{15}{c}{$d$} \\
\colhead{} & \multicolumn{3}{c}{0.65} & \colhead{} & \multicolumn{3}{c}{1.0} & \colhead{} & \multicolumn{3}{c}{2.05} & \colhead{} & \multicolumn{3}{c}{$[ 0.5,3.0]$}\\
\cline{2-4} \cline{6-8} \cline{10-12} \cline{14-16}\\
\colhead{$q$} & \colhead{$P_{0.001}$} & \colhead{$P_{0.01}$} & \colhead{$P_{0.1}$} & \colhead{} &
\colhead{$P_{0.001}$} & \colhead{$P_{0.01}$} & \colhead{$P_{0.1}$} & \colhead{} &
\colhead{$P_{0.001}$} & \colhead{$P_{0.01}$} & \colhead{$P_{0.1}$} & \colhead{} &
\colhead{$\bar{P}_{0.001}$} & \colhead{$\bar{P}_{0.01}$} & \colhead{$\bar{P}_{0.1}$}\\
\hline
\multicolumn{16}{c}{$\rho_* = 0.02$} }
\startdata
   1  &   2.01 & 1.89 & 1.40 &   & 1.27 & 1.14 & 1.04  &    &  1.57  & 1.41 & 1.32 &   & 1.94 & 1.43 & 1.15\\
  1/3 &   2.24 & 2.09 & 1.33 &   & 1.30 & 1.15 & 1.05  &    &  2.00  & 1.77 & 1.38 &   & 2.09 & 1.50 & 1.17\\
  1/9 &   2.78 & 2.08 & 1.27 &   & 1.40 & 1.19 & 1.06  &    &  3.13  & 1.95 & 1.25 &   & 2.54 & 1.63 & 1.19\\
\cutinhead{$\rho_* = 0.002$}
   1  &   1.76 & 1.27 & 1.07 &   & 1.09 & 1.04 & 1.01  &    &  1.42  & 1.37 & 1.11 &   & 1.30 & 1.14 & 1.04\\
  1/3 &   1.65 & 1.27 & 1.09 &   & 1.10 & 1.05 & 1.02  &    &  1.68  & 1.23 & 1.08 &   & 1.35 & 1.17 & 1.06\\
  1/9 &   1.67 & 1.30 & 1.10 &   & 1.13 & 1.06 & 1.02  &    &  1.55  & 1.24 & 1.08 &   & 1.47 & 1.24 & 1.08\\
\enddata
\end{deluxetable*}

Sample $l_w(\alpha)$ and $l\cacr(\alpha)$ functions are displayed in Figure~\ref{fig:cross_section} for lenses with $q=1/3$ and source radius $\rho_*=0.02$ (middle row of Figure~\ref{fig:delta_ex}). Clearly, the greatest increase over caustic-crossing events occurs when contours connecting multiple caustic parts are crossed perpendicularly to their maximum elongation. For instance, for horizontal ($\alpha=0,\pi$) trajectories and the three-component $d=0.65$ caustic, the $1\%$ effect region is 2.3 times wider than the caustic-crossing region. For vertical ($\alpha=\pi/2)$ trajectories and the two-component $d=2.05$ caustic, the $1\%$ region is 1.9 times wider than the caustic-crossing region.

Note that there is a slight difference between the quantities used here and similar ones found in the literature. The width $\Delta p$ used by \citet{mao_paczynski91} and the equivalent $s\cacr$ used by \citet{nightetal07} are similar to $l\cacr$. However, unlike $l\cacr$ these quantities purely measure the size of the caustic, thus including only trajectories for which the {\em source center} crosses the caustic instead of all caustic-crossing trajectories. For tiny sources the difference is negligible, but even for the $\rho_*=0.02$ source used in Figure~\ref{fig:cross_section}, $s\cacr(\pi/2)$ in the top panel is $16\%$ (four source radii) lower than $l\cacr(\pi/2)$. In addition, when the caustic shrinks to isolated points in the very wide $d\gg 1$ or very close $d\ll 1$ binary limits, $s\cacr$ drops to zero while $l\cacr$ converges to two or three source diameters, respectively\footnote{For any orientation except for angles perfectly aligned with lines connecting the caustic components.}.

\begin{figure*}
\includegraphics[scale=0.46]{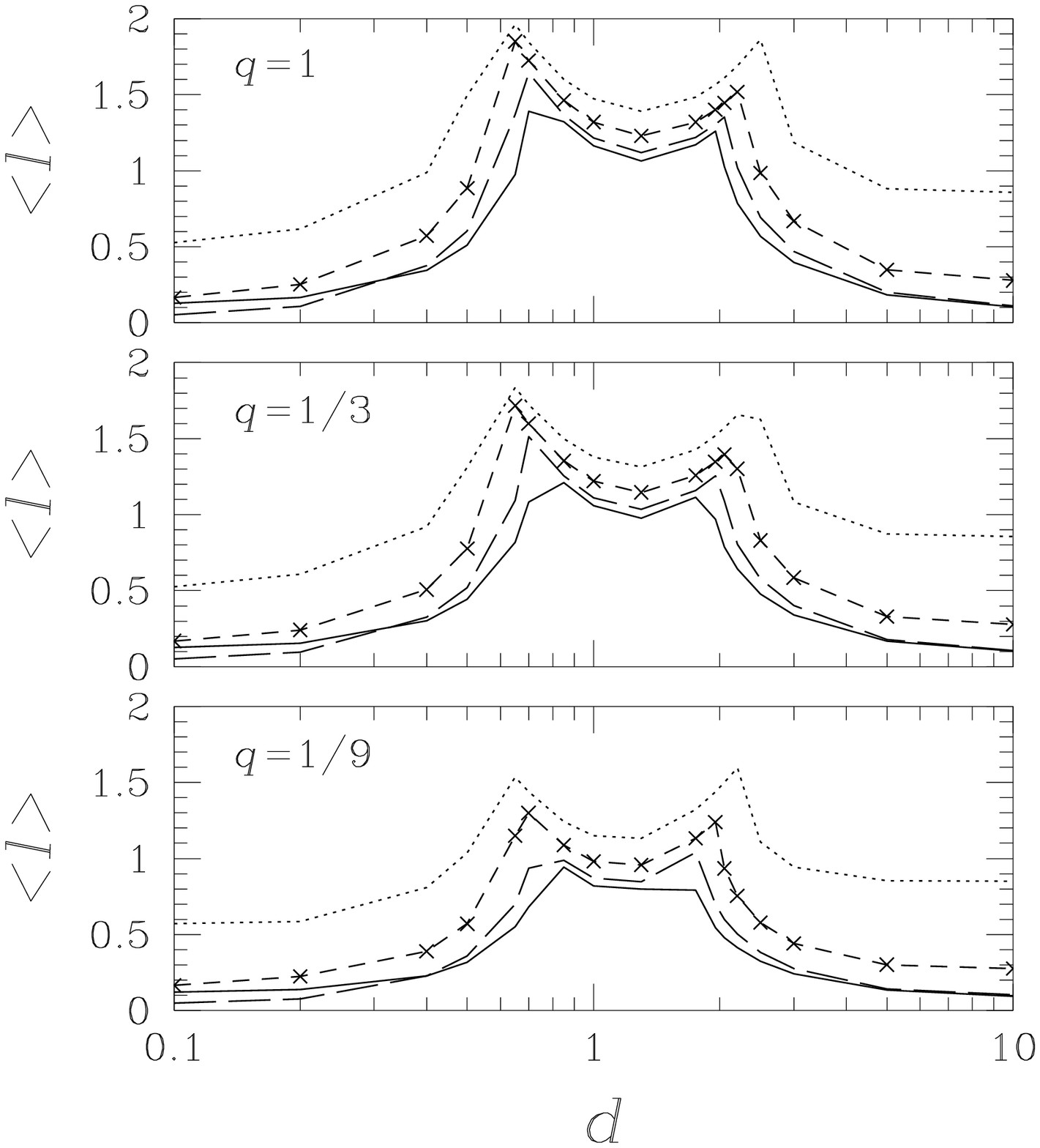}
\includegraphics[scale=0.46]{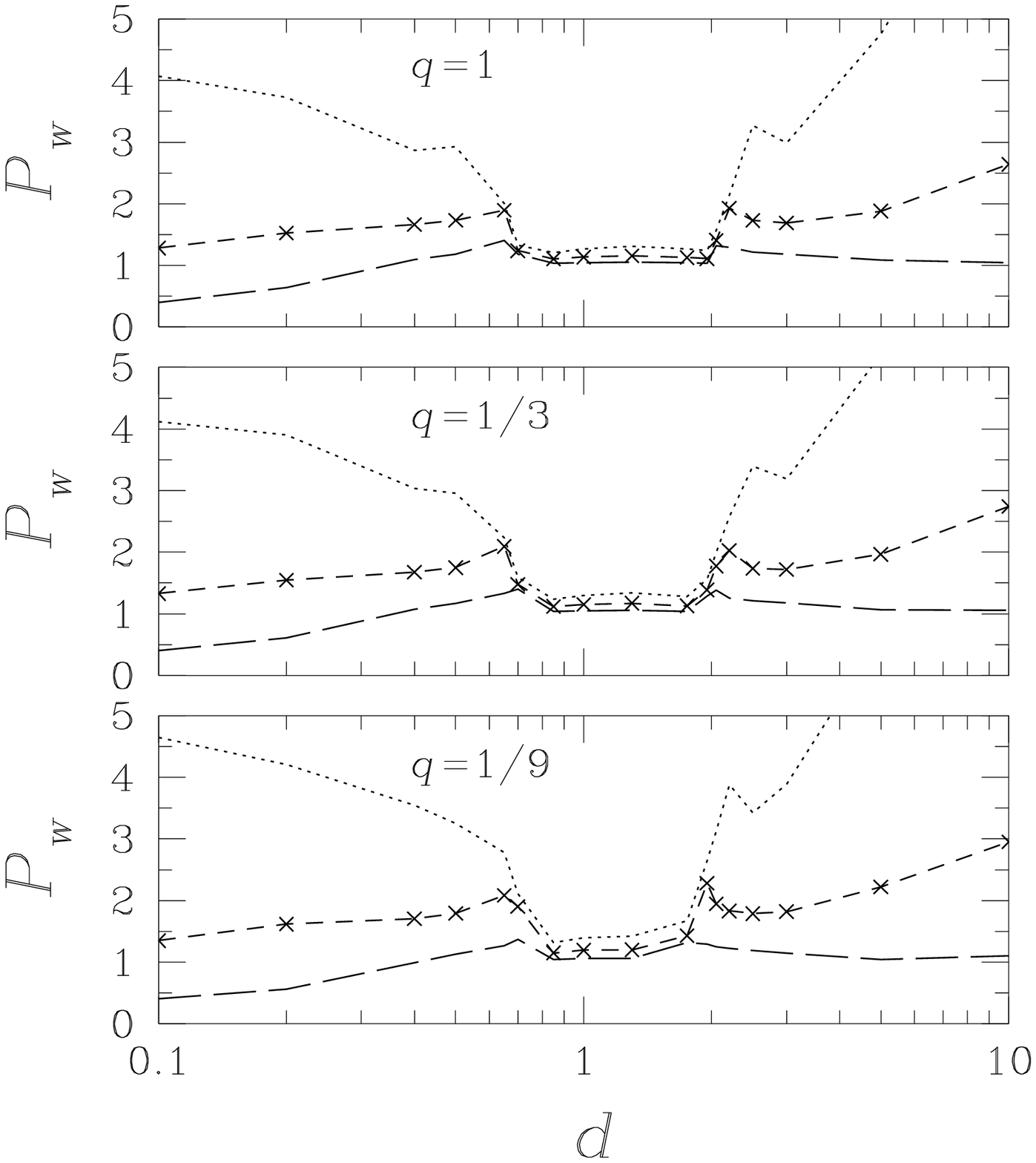}
\caption{Angle-averaged projected widths $\mean{l_w}$ of regions with absolute extended-source effect $|\delex|\geq w$ (left column) and relative probabilities $P_w$ (right column) of achieving an $|\delex|\geq w$ effect as a function of lens-component separation $d$ for three mass ratios (as marked in each panel) and a $\rho_*=0.02$ source. Curves for $w=0.001$, $0.01$, and $0.1$ are marked by dotted, short-dashed, and long-dashed lines, respectively. Solid lines in the left column mark the angle-averaged projected width of the caustic-crossing region $\mean{l\cacr}$. The relative probabilities $P_w$ are computed in comparison with the probability of caustic crossing, $P_w=\mean{l_w}/\mean{l\cacr}$. The values of $d$ used in the computations are marked by crosses along the $w=0.01$ curves.
\label{fig:probability}}
\end{figure*}

The relative probability of achieving at least an $|\delex|=w$ extended-source effect in comparison with the probability of caustic crossing is given by the ratio of the angle-averaged projected widths,
\beq
P_w = \frac{\mean{l_w}}{\mean{l\cacr}} = \frac{\int_0^{2\pi}l_w(\alpha)\,\rmd \alpha}{\int_0^{2\pi}l\cacr(\alpha)\,\rmd \alpha}\,.
\label{eq:total_cross}
\eeq
We present the values of $P_w$ for the nine lens geometries of Figure~\ref{fig:delta_ex} in the left part of Table~\ref{tab:cross} for source radii $\rho_*=0.02$ and 0.002. The obtained values demonstrate that the probability of an extended-source effect can be much higher than the probability of caustic crossing. In the case of the larger source the probability ratio for a $1\%$ effect exceeds 2 in two of the nine studied geometries. For the smaller source $P_{0.01}\geq 1.3$ also in two of the nine. Apart from the obvious decrease of $P_w$ with decreasing $\rho_*$ and increasing effect amplitude $w$, few trends in the values can be easily generalized. From the tabulated data we can at least see that for a given combination of $q$, $\rho_*$, and $w$ the value of $P_w$ is lowest for the $d=1$ single large caustic case. Such a caustic also has the lowest number of protruding cusps when averaged over projection angles.

To get a better understanding of the dependence on lens geometry, we include in Figure~\ref{fig:probability} plots of the angle-averaged widths $\mean{l_w}$, $\mean{l\cacr}$, and the relative probability $P_w$ as a function of lens-component separation $d$ for the three mass ratios $q$ and the larger source ($\rho_*=0.02$). The angle-averaged caustic-crossing-region width $\mean{l\cacr}$ is highest in the single-caustic regime, where it approaches or exceeds 1 Einstein radius, and drops to the limits mentioned above after the caustic splits into several components (see also \citealt{mao_paczynski91}; \citealt{nightetal07}). The contour widths $\mean{l_w}$ for decreasing $w$ are progressively higher and peak later after the splitting, but generally follow a similar trend. For small $d$ these widths converge to the corresponding contour widths of a single-point-mass lens, because the contours surrounding the small three-cusp caustics shrink to a point together with the caustics. For large $d$ the widths converge to double the respective value.

The right panels in Figure~\ref{fig:probability} illustrate the change of the relative probability $P_w$ with $d$, thus clarifying the interpretation of Table~\ref{tab:cross}. We first note that the character of the curves is the same for all three values of $q$. In the single-caustic regime the values of $P_w$ are low. Still, for example $P_{0.01}$ does not drop below $1.1$ for any of the mass ratios. $P_w$ increases rapidly to a peak when $d$ moves out of the single-caustic interval, and progresses to asymptotic values in the $d\gg 1$ and $d\ll 1$ regimes. In the wide-binary limit the value of $P_w$ is simply given by the corresponding single-point-mass contour width expressed in units of source {\em diameter}. For the three included contours and $\kappa=0$ limb darkening we get $P_{0.1}=1.2$, $P_{0.01}=3.4$, and $P_{0.001}=11$. In the close-binary limit the relative probabilities are one third of the corresponding wide-limit values, i.e., $P_{0.1}=0.42$, $P_{0.01}=1.2$, and $P_{0.001}=3.6$. The quoted values are valid for all three mass ratios and both source sizes\footnote{The curves for $w=0.001$ can be expected to turn over for $d<0.1$, the lowest computed separation.}. In fact, they remain constant for any source sufficiently smaller than the Einstein radius of the lower-mass component of the binary. More specifically, from single-point-mass extended-source computations we find that for the $1\%$ effect this requires $\rho_*\lesssim 0.2 \sqrt{q/(q+1)}$, while for the $0.1\%$ and $10\%$ effects the factor before the square root is approximately $0.05$ and $0.5$, respectively.

The character of the $\mean{l_w}(d)$ and $P_w(d)$ curves for the small source ($\rho_* = 0.002$) is similar to the large-source results, except for reaching generally lower values. It is worth noting here that the scaling of equation~(\ref{eq:delta_ex_theo}) with $\rho_*$ implies that for instance $\mean{l_{0.01}}$ for a source with $\rho_* = 0.02$ should be the same as $\mean{l_{0.0001}}$ for a source with $\rho_* = 0.002$, keeping $q$ and $d$ fixed. This can also be noticed in Figures~\ref{fig:delta_ex_src} and \ref{fig:delta_ex_det}. The scaling doesn't hold for $\mean{l_{0.1}}$, because the corresponding contour lies closer to the caustic than the domain of validity of equation~(\ref{eq:delta_ex_theo}). In either case $P_w$ will include an additional variation due to the increase of $\mean{l\cacr}$ with $\rho_*$.

In order to condense the information from Figure~\ref{fig:probability}, we average the value of the relative probability over the main range of separations in observed two-point-mass microlensing events, $d\in [0.5, 3.0]$. The obtained values of $\bar{P}_w$ for both source sizes are listed in the last three columns of Table~\ref{tab:cross}. For the larger source we see that encountering a $\geq1\%$ extended-source effect is on average 1.43--1.63 times more probable than crossing the caustic, the values increasing from $q=1$ to $q=1/9$. For the smaller source we get a factor of 1.14--1.24 increase in probability. It is clear from Figure~\ref{fig:probability} that these average values are particularly sensitive to setting the upper endpoint of the averaging interval in $d$, at which $P_{0.01}$ keeps on increasing.

\section{CHROMATICITY}
\label{sec:chromaticity}

Equation~(\ref{eq:delta_ex_theo}) demonstrated that the amplification depends on the limb-darkening profile even away from the caustic, albeit fairly weakly. In general, the variation of stellar limb-darkening profiles with wavelength introduces a chromatic effect in microlensing. In this section we explore the dependence of the microlensing amplification on the limb-darkening profile using the two-term PCA model given by equation~(\ref{eq:pca_limb}).

By introducing a model of this form in equation~(\ref{eq:amp_integral}) we obtain a limb-darkening-parameter $\kappa$ dependence of the extended-source amplification $A_*({\bf y}_{\rm c}, \kappa)$. The rate of its change with $\kappa$ is given by the derivative
\beq
\frac{\partial A_*({\bf y}_{\rm c}, \kappa)}{\partial \kappa} = \frac{\int_0^1 f_1(r)r\rmd r \int_0^1 f_2(r)r\rmd r}{\{\int_0^1 [f_1(r)+\kappa f_2(r)] r\rmd r\}^2}\;D_A({\bf y}_{\rm c}, \rho_*)\,,
\label{eq:amp_derivative}
\eeq
where we introduced the geometry-dependent function
\begin{eqnarray}
D_A({\bf y}_{\rm c}, \rho_*)=\frac{\int_{\Sigma_{S'}} A_0({\bf y}_{\rm c}+\rho_* {\bf r}) f_2(r) \rmd^2 {\bf r}}{2\pi\int_0^1 f_2(r)r\rmd r} - - \nonumber \\
 \mbox{} - \frac{\int_{\Sigma_{S'}} A_0({\bf y}_{\rm c}+\rho_* {\bf r}) f_1(r) \rmd^2 {\bf r}}{2\pi\int_0^1 f_1(r)r\rmd r}\,.
\label{eq:chrom_geometry}
\end{eqnarray}
The integrals in the numerators of the expression are computed over the unit disk of the source $\Sigma_{S'}$. The function $D_A({\bf y}_{\rm c}, \rho_*)$ can be readily interpreted as the amplification difference between hypothetical sources with limb darkening given by $f_2(r)$ and $f_1(r)$, respectively. Of course, the function $f_2$ alone does not describe the limb-darkening profile of any realistic source.

More interestingly, for a given two-term limb-darkening basis the dependence on individual parameters is separated, with the factor preceding $D_A({\bf y}_{\rm c}, \rho_*)$ in equation~(\ref{eq:amp_derivative}) depending purely on the limb-darkening parameter $\kappa$, and $D_A({\bf y}_{\rm c}, \rho_*)$ depending on the source position ${\bf y}_{\rm c}$ and source radius $\rho_*$. Equation~(\ref{eq:amp_derivative}) implies that at any source position ${\bf y}_{\rm c}$ the amplification changes monotonously with the limb-darkening parameter $\kappa$. The sign of the derivative changes along a curve ${\bf y}_{\rm c}={\bf y}_{\rm ac}(\tau)$ implicitly defined by setting $D_A({\bf y}_{\rm ac}, \rho_*)=0$. The parameter separation obviously persists in higher-order derivatives, and is retained even when computing the amplification difference between sources with different limb darkening (or a source observed at different wavelengths),
\beq
A_*({\bf y}_{\rm c}, \kappa_1)-A_*({\bf y}_{\rm c}, \kappa_2)= h(\kappa_1,\kappa_2)\,D_A({\bf y}_{\rm c}, \rho_*)\, ,
\label{eq:amp_difference}
\eeq
with the $\kappa$-dependent factor
\begin{eqnarray}
& \!\!\!\!\!\!\!\!\!\!\!\!h(\kappa_1,\kappa_2)&=(\kappa_1-\kappa_2)\times\nonumber \\
&\!\!\!\!\!\! \!\!\!\!\!\!\!\!\!\!\!\! \times& \!\!\!\!\!\!\!\!\!\!\!\! \frac{\int_0^1 f_1(r)r\rmd r \int_0^1 f_2(r)r\rmd r}{\int_0^1 [f_1(r)+\kappa_1 f_2(r)] r\rmd r \int_0^1 [f_1(r)+\kappa_2 f_2(r)] r\rmd r}\, .
\label{eq:kappa_factor}
\end{eqnarray}
We conclude that for a given source size and a given two-term limb-darkening model of the same form as the law in equation~(\ref{eq:pca_limb}) there exists a unique achromatic curve ${\bf y}_{\rm ac}(\tau)$ obtained by setting $D_A({\bf y}_{\rm ac}, \rho_*)=0$. The amplification of a source located anywhere along this curve is independent of the specific value of its limb-darkening parameter. Note that the curve does depend on the source size and on the specific functions $f_1(r), f_2(r)$ of the limb-darkening model. For example, it is somewhat different for linear limb darkening and for the PCA model. The $\rho_*$ dependence disappears away from the caustic, as shown below.

For the PCA model we use, the variation of $h(\kappa_1,\kappa_2)$ with $\kappa_{1,2}$ is dominated by the $(\kappa_1-\kappa_2)$ proportionality of the numerator of equation~(\ref{eq:kappa_factor}), while the $\kappa$-dependence of the denominator has only a weak influence. Hence, the largest value is achieved for the largest difference in limb-darkening parameters. As shown in Figure~\ref{fig:pca_ld}, the extreme physically permitted $\kappa$ values for our model are $\kappa_{\rm pk}=-0.1620$ for the most centrally peaked limb-darkening profile\footnote{ Closely resembling the $B$-band profile of a range of cool K and M giants, e.g., the solar-metallicity ATLAS model with $T_{eff}=3500\,K$, $\log\,g=1.5$, $v_t=2\,km\,s^{-1}$.}, and $\kappa_{\rm fl}=0.0902$ for the flattest profile\footnote{Resembling the $I$-band limb darkening of a range of O dwarfs, e.g., the solar-metallicity ATLAS model with $T_{eff}=48000\,K$, $\log\,g=5$, $v_t=2\,km\,s^{-1}$.}.

Based on the preceding we use these extreme values of the limb-darkening parameter $\kappa$ to define the microlensing chromaticity
\beq
\delchr({\bf y}_{\rm c}) = \frac{A_*({\bf y}_{\rm c},\kappa_{\rm pk}) - A_*({\bf y}_{\rm c},\kappa_{\rm fl})}{A_*({\bf y}_{\rm c},0)}\,,
\label{eq:delta_chr}
\eeq
where we divided equation~(\ref{eq:amp_difference}) by the amplification of a source with $\kappa=0$ used in \S~\ref{sec:sensitivity}. Thus defined, the microlensing chromaticity determines the maximum relative variation of the microlensing amplification with the parameter of a given limb-darkening model. In other words, $\delchr({\bf y}_{\rm c})$ is the minimum photometric precision necessary for distinguishing between the steepest and the flattest limb-darkening profiles of a source located at ${\bf y}_{\rm c}$. Note that $\delchr$ depends on non-divergent extended-source amplifications, and its behavior close to the caustic is therefore quite different from $\delex$. In particular, the change of $\delchr$ across the caustic is continuous, and its value varies even along the caustic.

In order to understand the variation of $\delchr$ away from the caustic analytically, we expand equation~(\ref{eq:delta_chr}) for a small source in regions where $A_0$ is analytic, similarly as in \S~\ref{sec:analytical}. We obtain, to first order in $\rho_*^2$,
\begin{eqnarray}
&\delchr({\bf y}_{\rm c})&\approx h(\kappa_{\rm pk},\kappa_{\rm fl}) \times \nonumber \\
& \times&\!\!\!\!\!\!\left[\frac{\int_0^1 f_2(r)r^3\rmd r\int_0^1 f_1(r)r\rmd r}{\int_0^1 f_2(r)r\rmd r\int_0^1 f_1(r)r^3\rmd r}-1 \right]\delex({\bf y}_{\rm c})\,,
\label{eq:delta_chr_delex}
\end{eqnarray}
where we utilized the approximate expression for the extended-source effect $\delex$ from equation~(\ref{eq:delta_ex_theo}) for a source with a brightness distribution $I(r)=f_1(r)$. The entire expression preceding $\delex$ in equation~(\ref{eq:delta_chr_delex}) depends only on the limb darkening, hence the source-position dependence and source-size scaling is exactly the same as for $\delex$. At a distance from the caustic (and its cusps in particular), the geometry of $\delchr({\bf y}_{\rm c})$ contours is thus exactly the same as that of the extended-source effect $\delex({\bf y}_{\rm c})$ contours. This also means that the achromatic curve ${\bf y}_{\rm ac}(\tau)$ coincides in this region with the $\delex=0$ contour, as implied earlier in \S~\ref{sec:analytical}, and thus it also becomes independent of $\rho_*$. Using our PCA basis and its $\kappa_{\rm pk}$ and $\kappa_{\rm fl}$ parameters, we obtain $-0.197$ as the value of the factor preceding $\delex$ in equation~(\ref{eq:delta_chr_delex}). The contours of $\delex$ away from the caustic thus coincide with contours of a five times lower value of chromaticity, $\delchr({\bf y}_{\rm c})\approx -0.2\,\delex({\bf y}_{\rm c})$.

It is worth noting here that when defining $\delchr$ in equation~(\ref{eq:delta_chr}), one could just as well divide the amplification difference for example by the amplification corresponding to $\kappa_{\rm pk}$, $\kappa_{\rm fl}$, or the average of these two amplifications. In fact, away from the caustic any of these choices would lead exactly to the same approximation given by equation~(\ref{eq:delta_chr_delex}). The advantage of the choice we made is apparent after rewriting equation~(\ref{eq:delta_chr}) in the form
\begin{eqnarray}
&\delchr({\bf y}_{\rm c})& = h(\kappa_{\rm pk},\kappa_{\rm fl})\left[\frac{\int_{\Sigma_{S'}} A_0({\bf y}_{\rm c}+\rho_* {\bf r}) f_2(r) \rmd^2 {\bf r}}{\int_{\Sigma_{S'}} f_2(r) \rmd^2 {\bf r}}\,\times \right. \nonumber \\
& & \qquad \times \left.\frac{\int_{\Sigma_{S'}} f_1(r) \rmd^2 {\bf r}}{\int_{\Sigma_{S'}} A_0({\bf y}_{\rm c}+\rho_* {\bf r}) f_1(r) \rmd^2 {\bf r}}-1\right]\,.
\label{eq:delta_chr_separated}
\end{eqnarray}
Such a separation of the limb-darkening-parameter dependence from the position dependence cannot be achieved by the other possible choices of the denominator of equation~(\ref{eq:delta_chr}). Only with the present choice does the full geometry of the contours remain unchanged even when comparing the amplification of sources with two arbitrary values of $\kappa$. This reason aside, from the preceding analysis we see that the computational differences obtained when using other choices would be limited to minor variations of effect amplitude in the vicinity of the caustic rather than causing substantial changes.

The relative amplification difference $\delta_{12}$ for two arbitrary profiles can be trivially computed from $\delchr$ by a simple rescaling,
\begin{eqnarray}
&\!\!\!\!\!\!\!\!\!\!\!\!\delta_{12}({\bf y}_{\rm c})& = \frac{A_*({\bf y_c}, \kappa_1) - A_*({\bf y_c}, \kappa_2)}{A_*({\bf y_c}, 0)} =\nonumber \\
& & \qquad\qquad =  \frac{h(\kappa_1,\kappa_2)}{h(\kappa_{\rm pk},\kappa_{\rm fl})}\;\delchr({\bf y}_{\rm c})\,,
\label{eq:delta_12_chr}
\end{eqnarray}
in view of equations~(\ref{eq:amp_difference}) and (\ref{eq:delta_chr}). With values ranging from -1 to 1, the $h$--factor ratio can be used to recompute contour values in the subsequent chromaticity plots, as deemed necessary. More specifically, for our PCA basis the ratio is given by
\beq
\frac{h(\kappa_1,\kappa_2)}{h(\kappa_{\rm pk},\kappa_{\rm fl})}=\frac{-(\kappa_1-\kappa_2)}{[0.2676+0.1876(\kappa_1+\kappa_2)+0.1316\,\kappa_1\kappa_2]}
\label{eq:h-ratio}
\eeq
and is plotted in Figure~\ref{fig:chrom_rescaling}.

\bef
\plotone{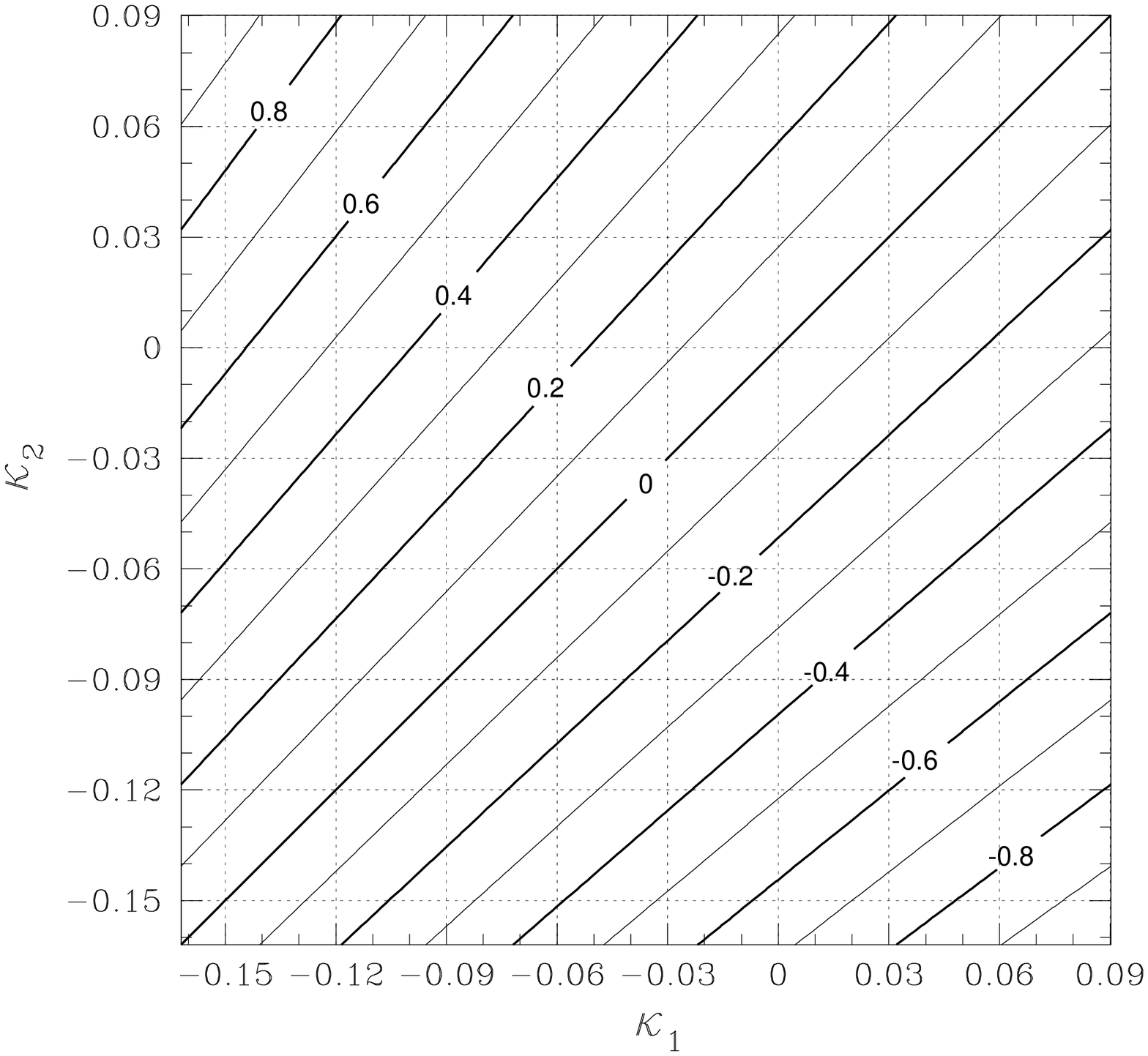}
\caption{Contour plot of the $h(\kappa_1,\kappa_2)/h(\kappa_{\rm pk},\kappa_{\rm fl})$ factor for multiplying $\delchr$ according to equation~(\ref{eq:delta_12_chr}) in order to get the relative amplification difference $\delta_{12}$ of two arbitrary PCA limb-darkening profiles with parameters $\kappa_1$ and $\kappa_2$. More specifically, all chromaticity values in the following three figures can be rescaled as required in this way. The factor ranges from 1 in the upper left corner to -1 in the lower right, and is numerically given by equation~(\ref{eq:h-ratio}). }
\label{fig:chrom_rescaling}
\enf

\begin{figure*}
\begin{center}
\includegraphics[scale=.69]{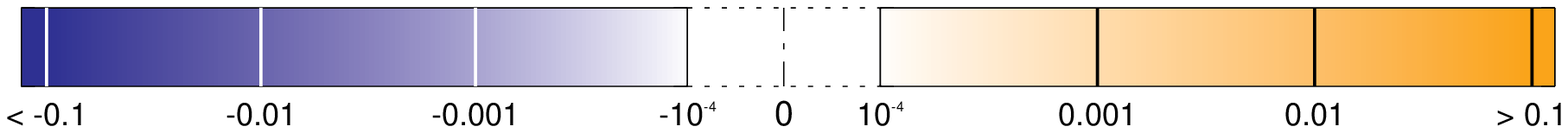}\\
\includegraphics[scale=.31]{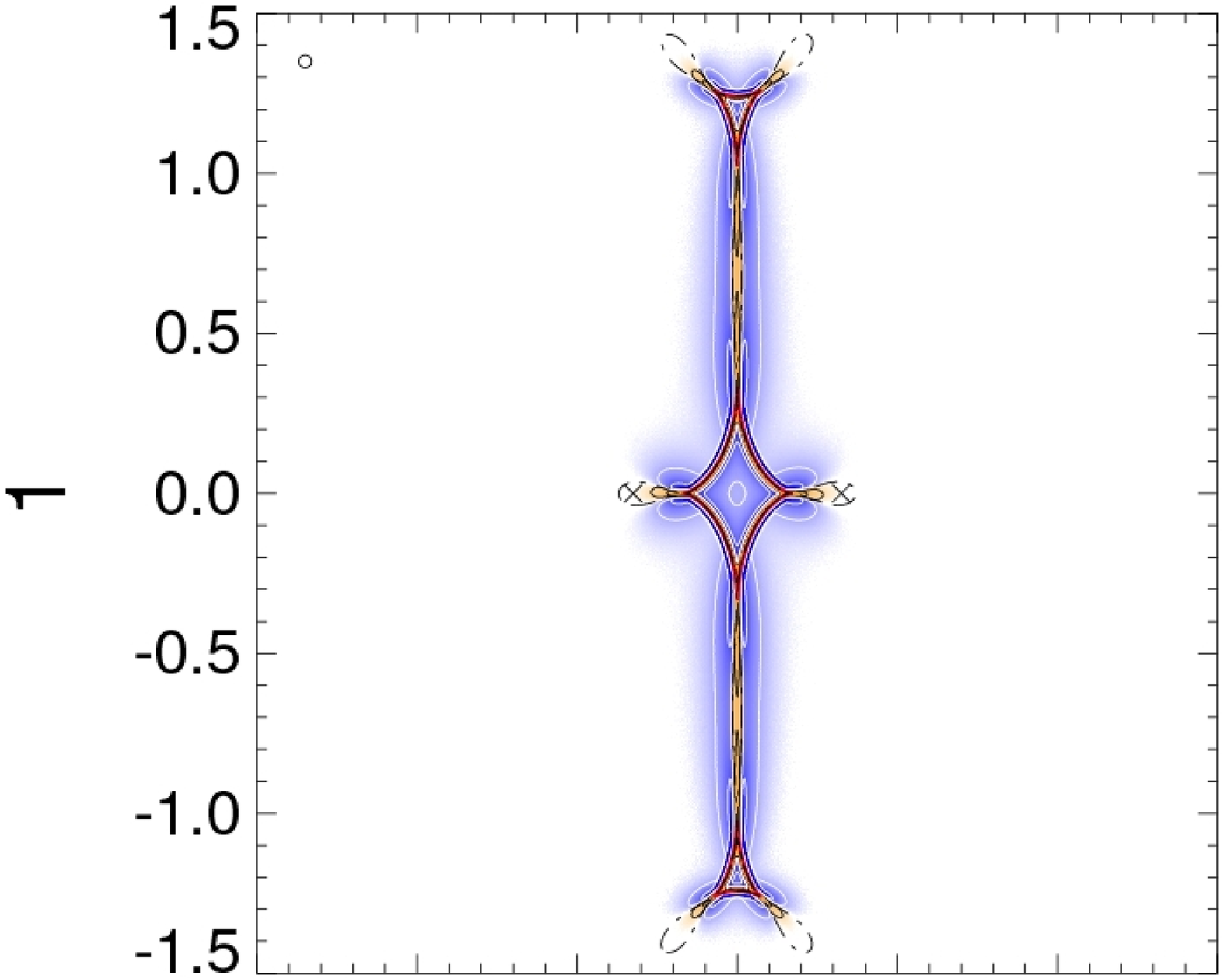}
\includegraphics[scale=.31]{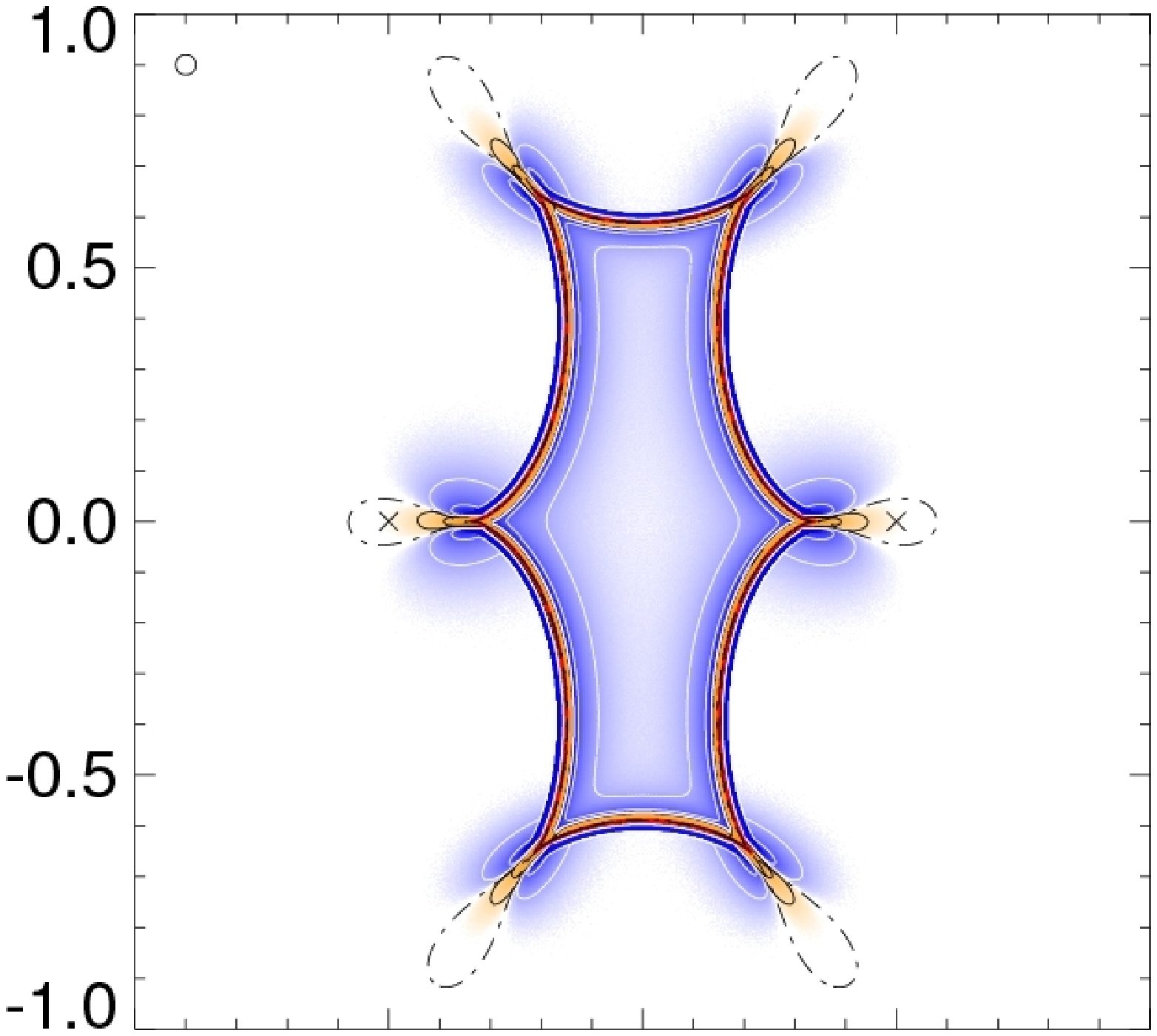}
\includegraphics[scale=.31]{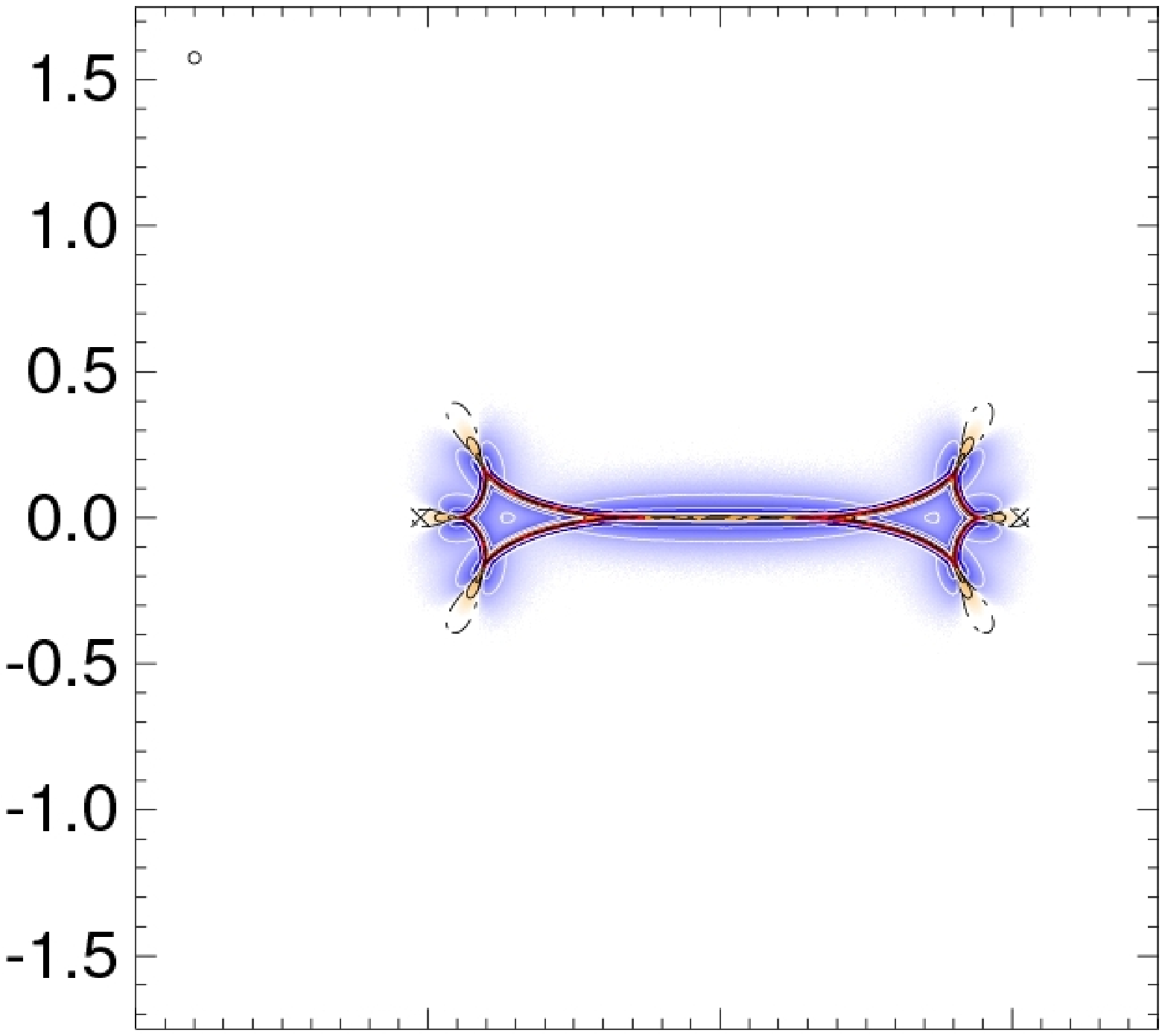}\\
\includegraphics[scale=.31]{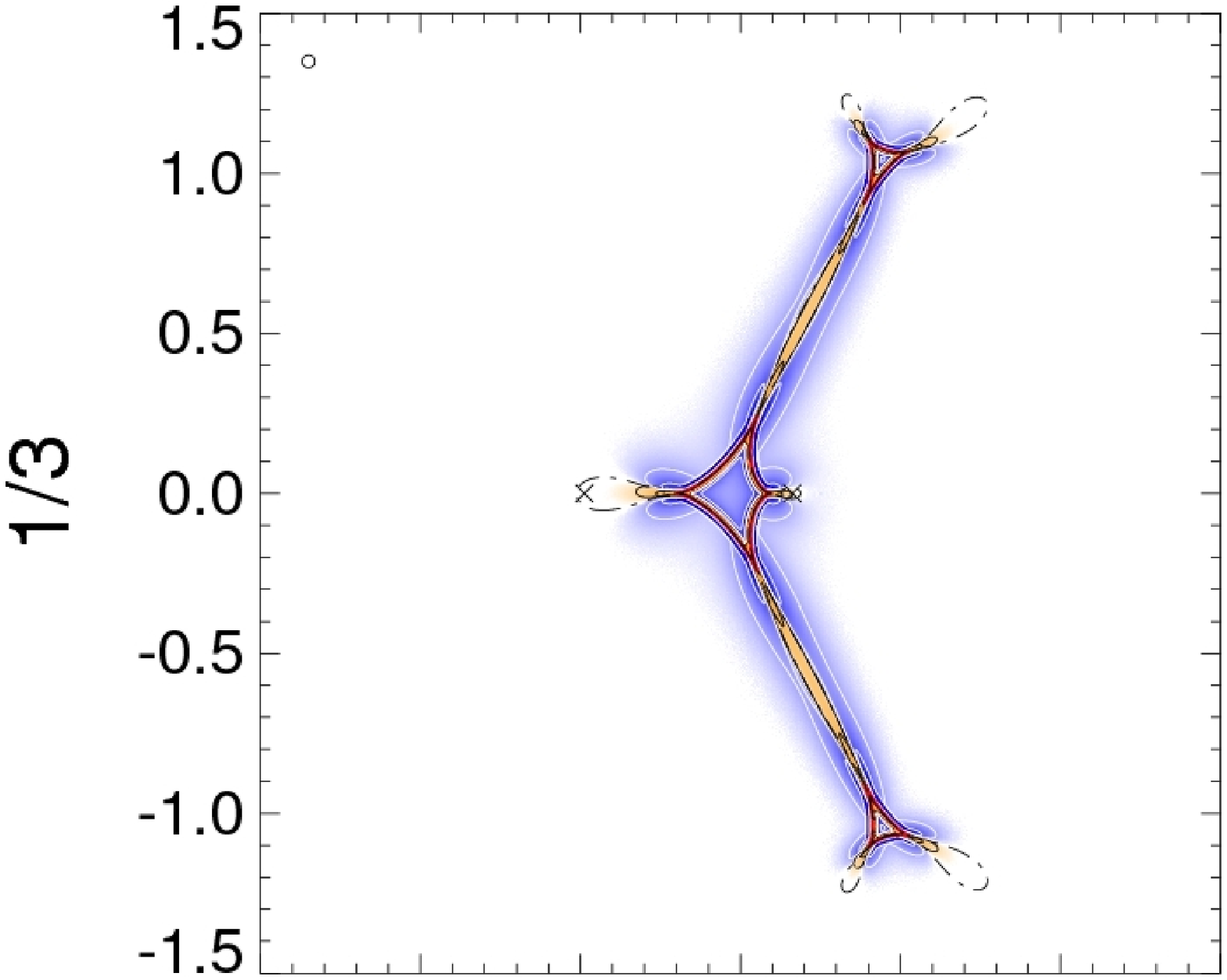}
\includegraphics[scale=.31]{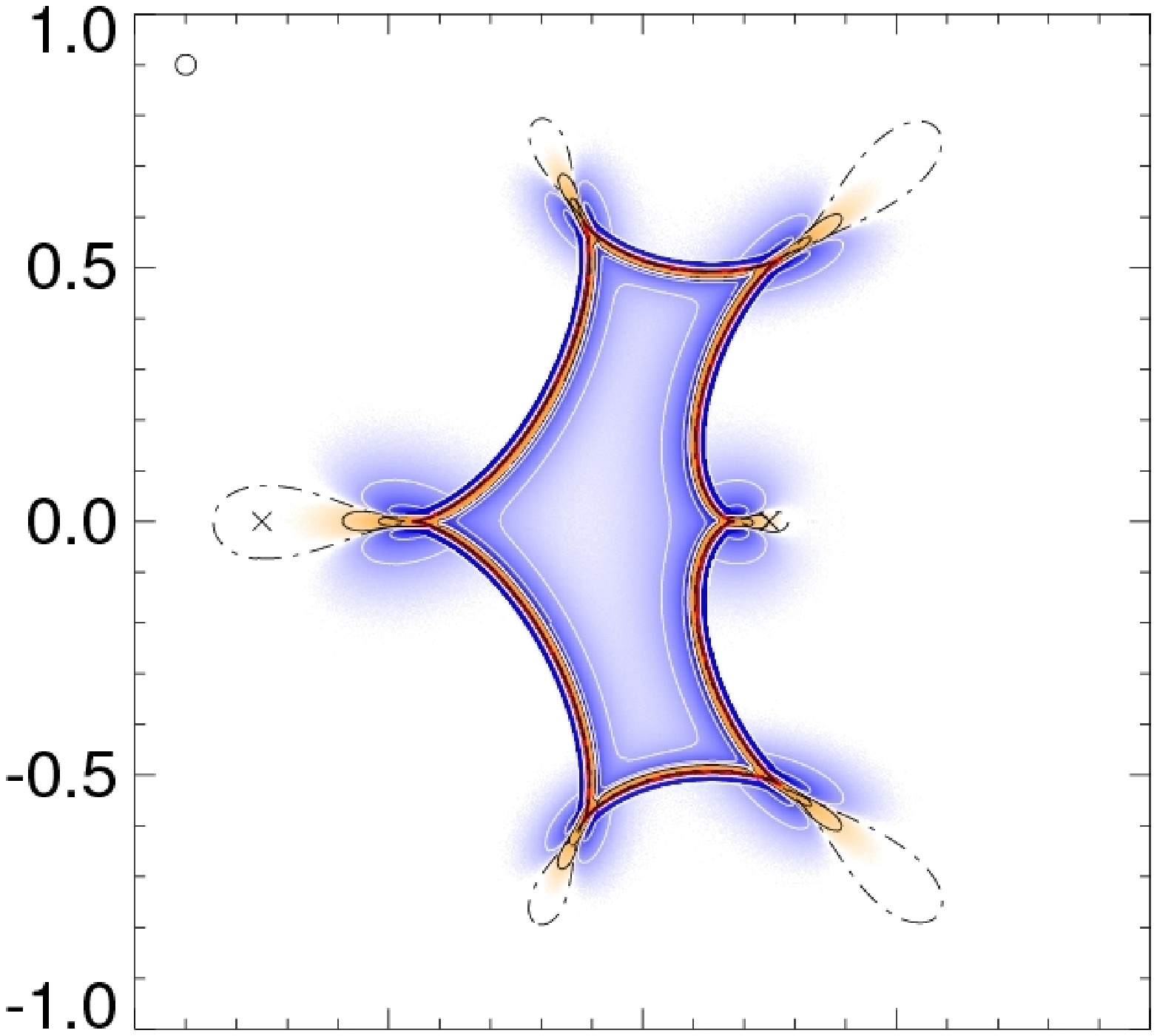}
\includegraphics[scale=.31]{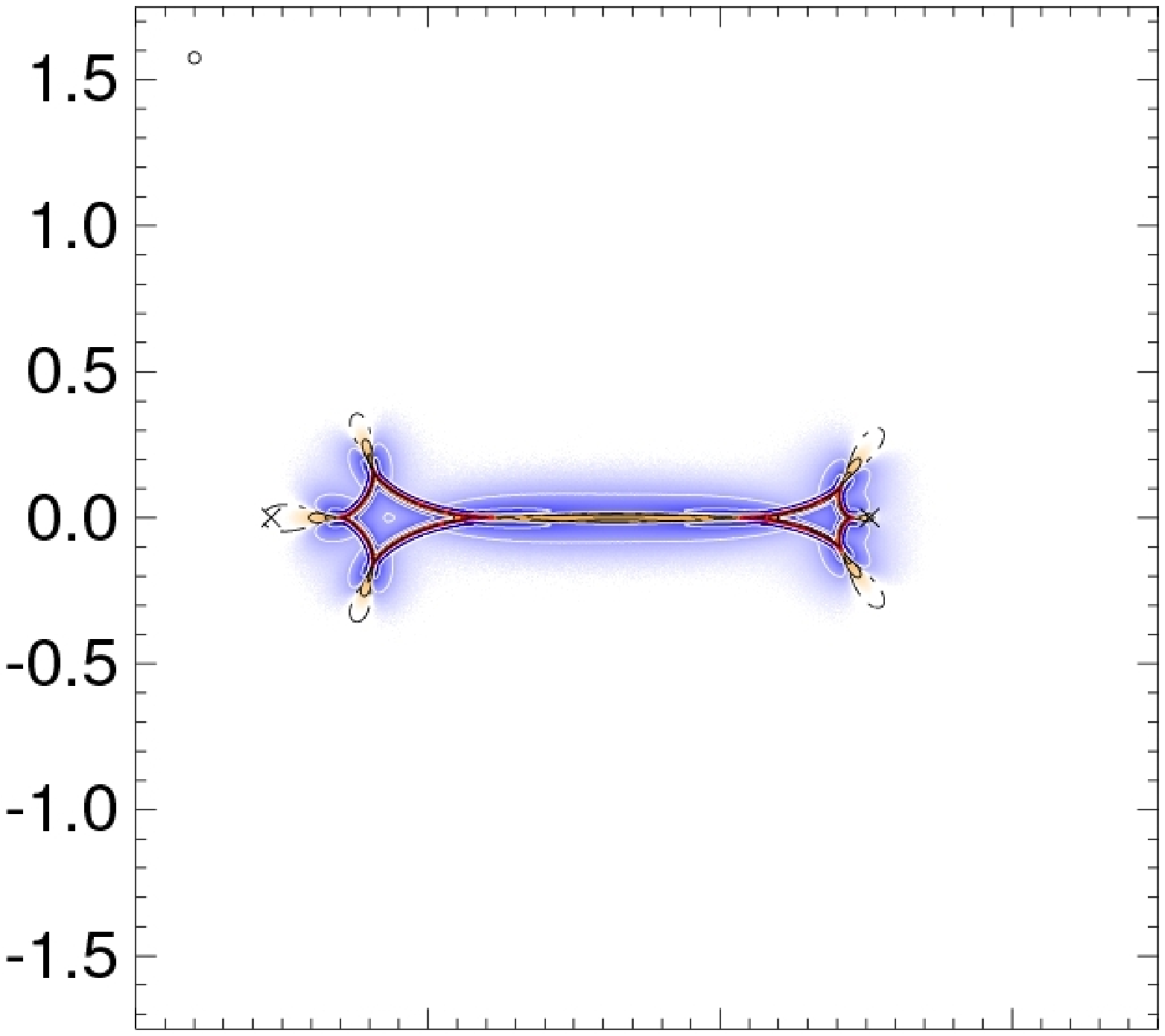}\\
\includegraphics[scale=.31]{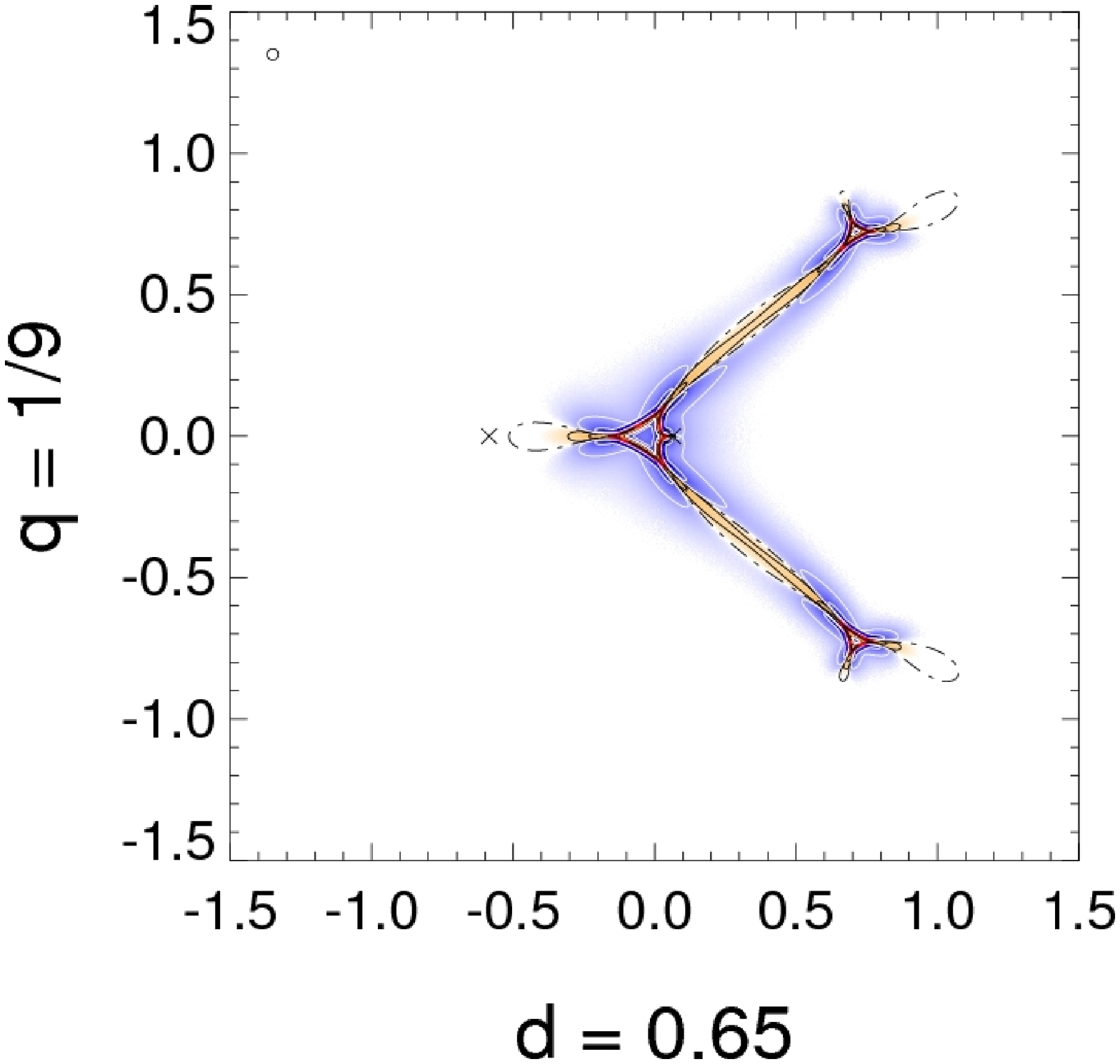}
\includegraphics[scale=.31]{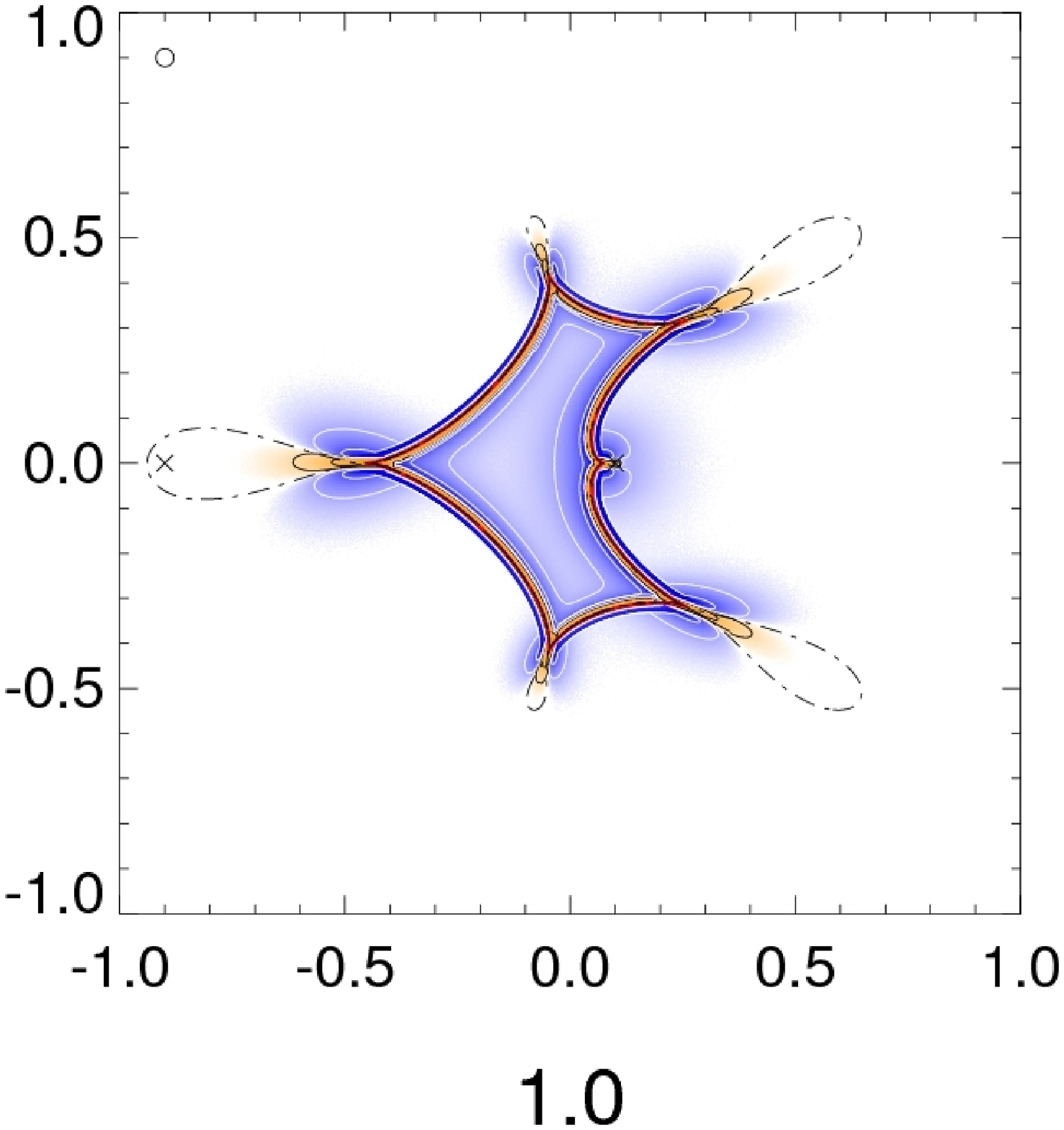}
\includegraphics[scale=.31]{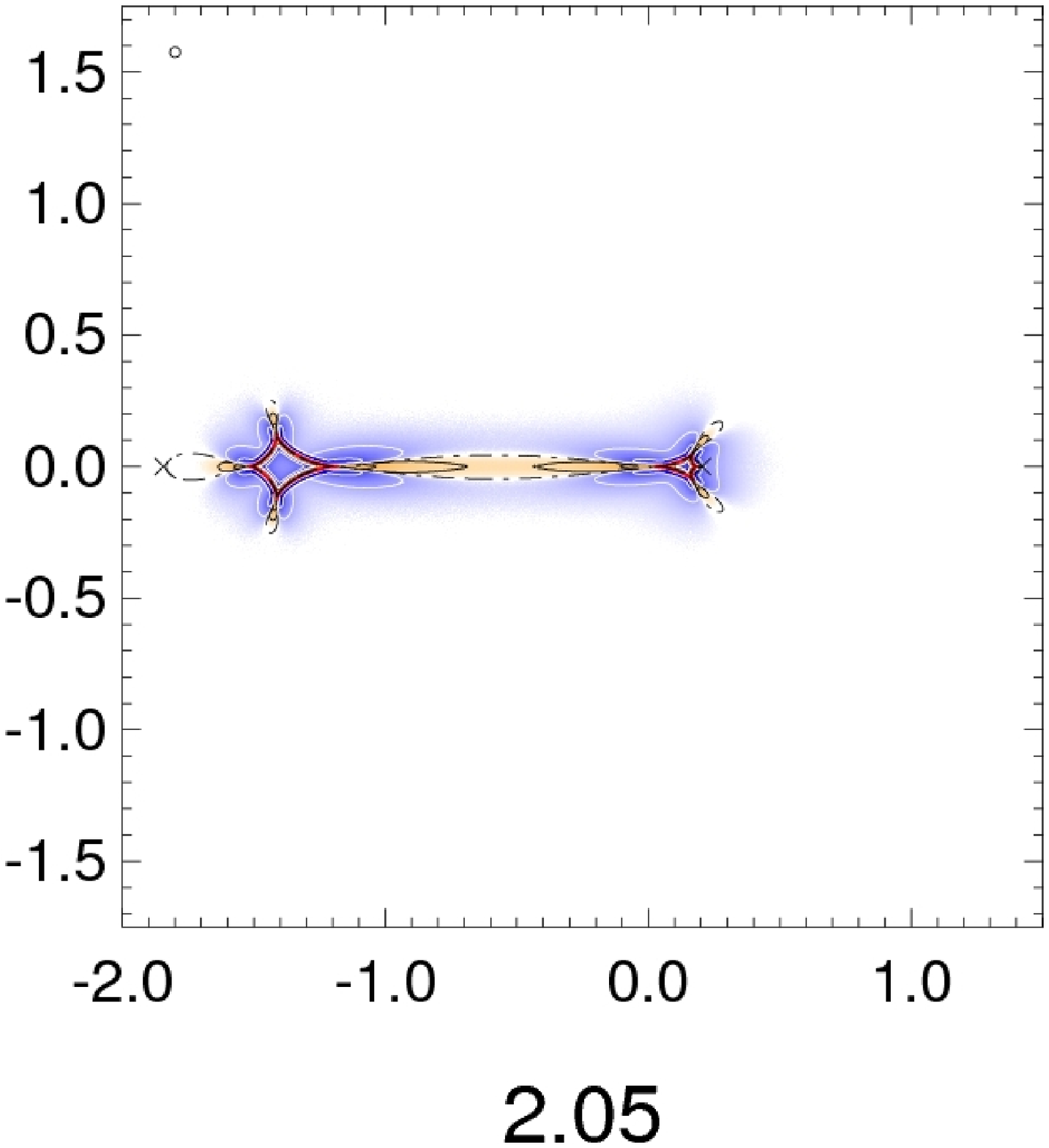}
\end{center}
\caption{Chromaticity $\delchr({\bf y}_{\rm c})$ for a source with radius $\rho_* = 0.02$ and PCA limb darkening. Panels correspond to the same lens configurations as in Figures~\ref{fig:point_mag} and \ref{fig:delta_ex}. Solid contours are plotted for $\delchr=\pm 0.001$, $\pm 0.01$, and $\pm 0.1$ (the latter localized close to the caustic and hardly visible here); the dot-dashed contour corresponds to the achromatic curve $\delchr({\bf y}_{\rm ac})=0$. As marked in the color bar, positive values are mapped in shades of orange with black contours, negative in shades of blue with white contours. Areas with $|\delchr|<10^{-4}$ are left white. Circles in panel corners illustrate the source size and the scale difference between columns.}
\label{fig:delta_chr}
\end{figure*}

Before proceeding to numerical results, we emphasize that all formulae and equations in this section except equation (\ref{eq:h-ratio}) are valid not just for the two-point-mass lens, but more generally for any lens and any small source with a circularly symmetric brightness distribution of the form given by equation~(\ref{eq:pca_limb}). In the case of quasar microlensing, however, one should keep in mind that much of the chromaticity is caused by the different extent of the emitting region at different wavelengths \citep{wambsganss_paczynski91}.

\bef
\begin{center}
\plotone{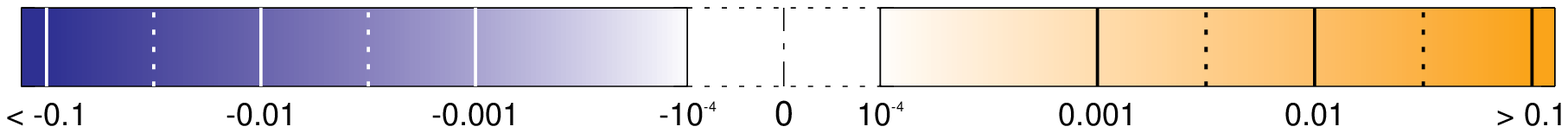}\\
\vspace{0.3cm}
\includegraphics[scale=0.22]{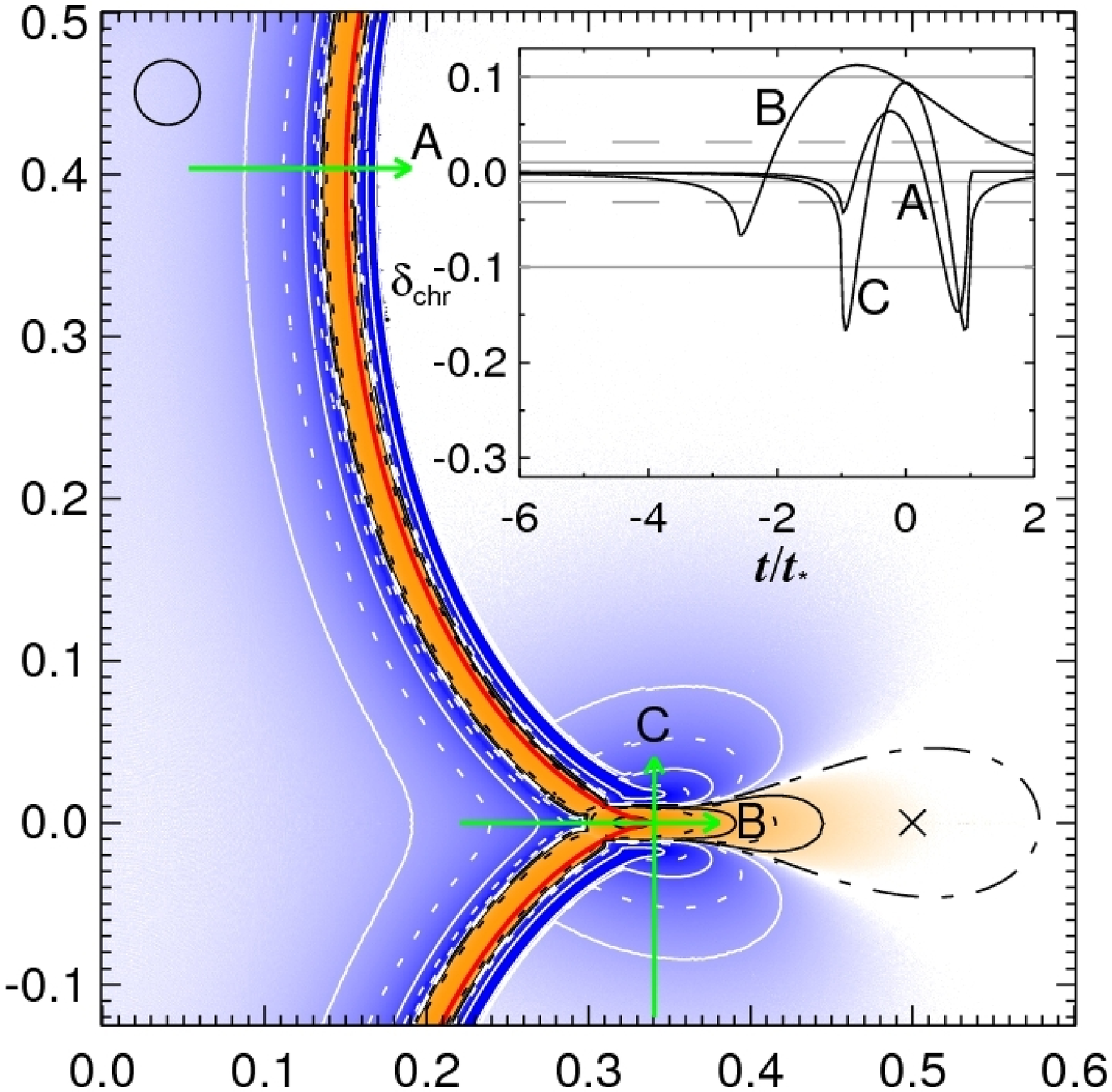}
\includegraphics[scale=0.22]{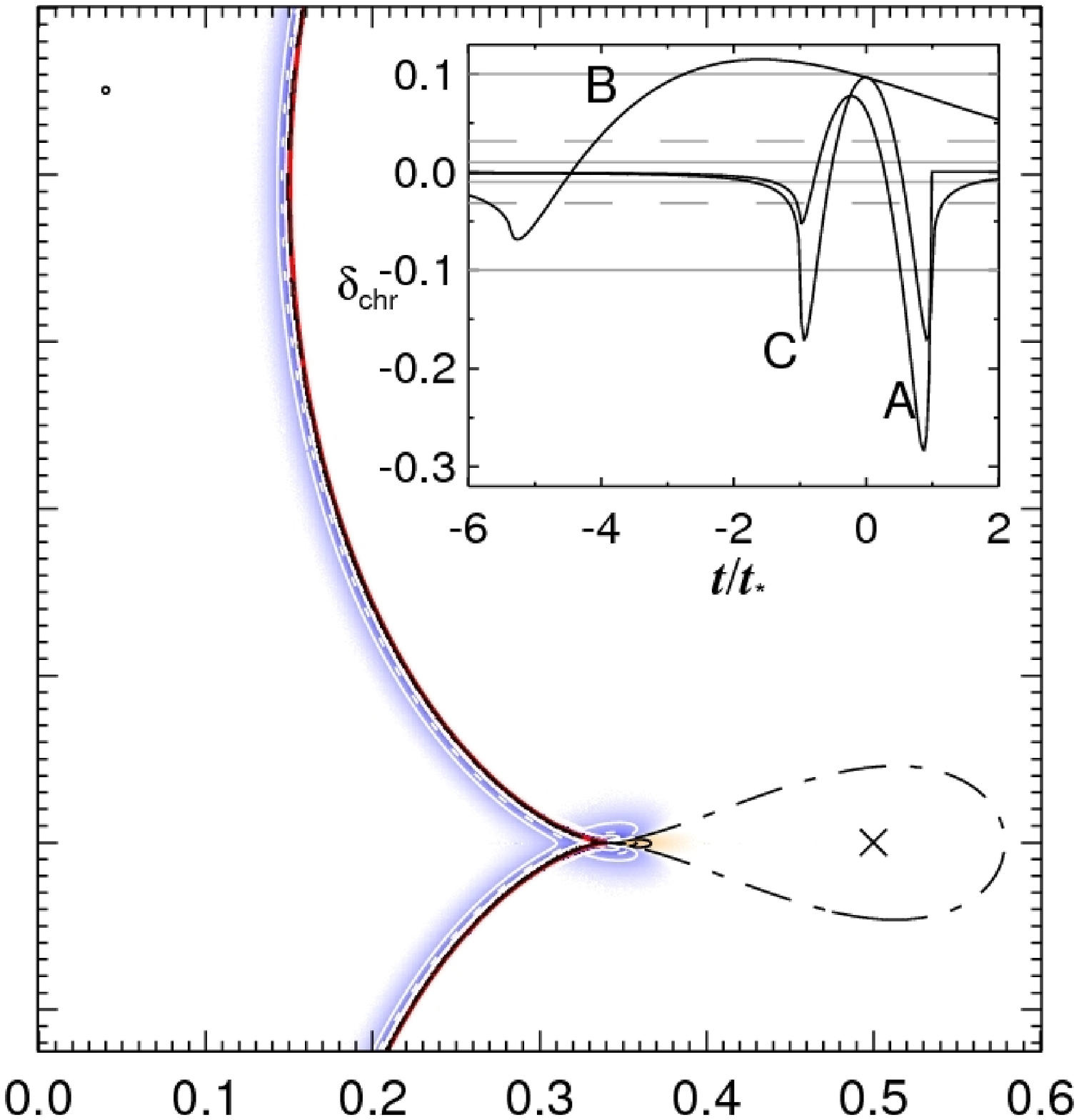}\\
\vspace{0.3cm}
\includegraphics[scale=0.44]{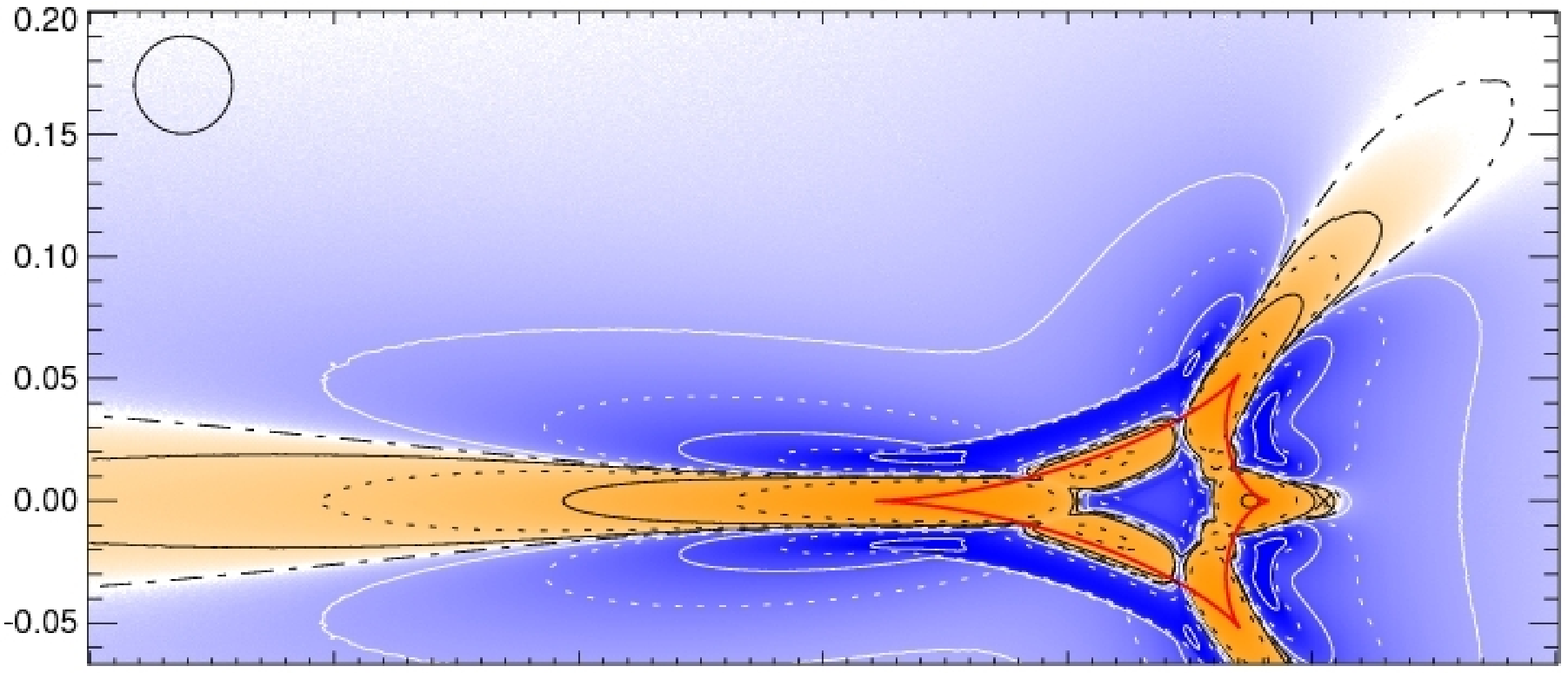}
\includegraphics[scale=0.44]{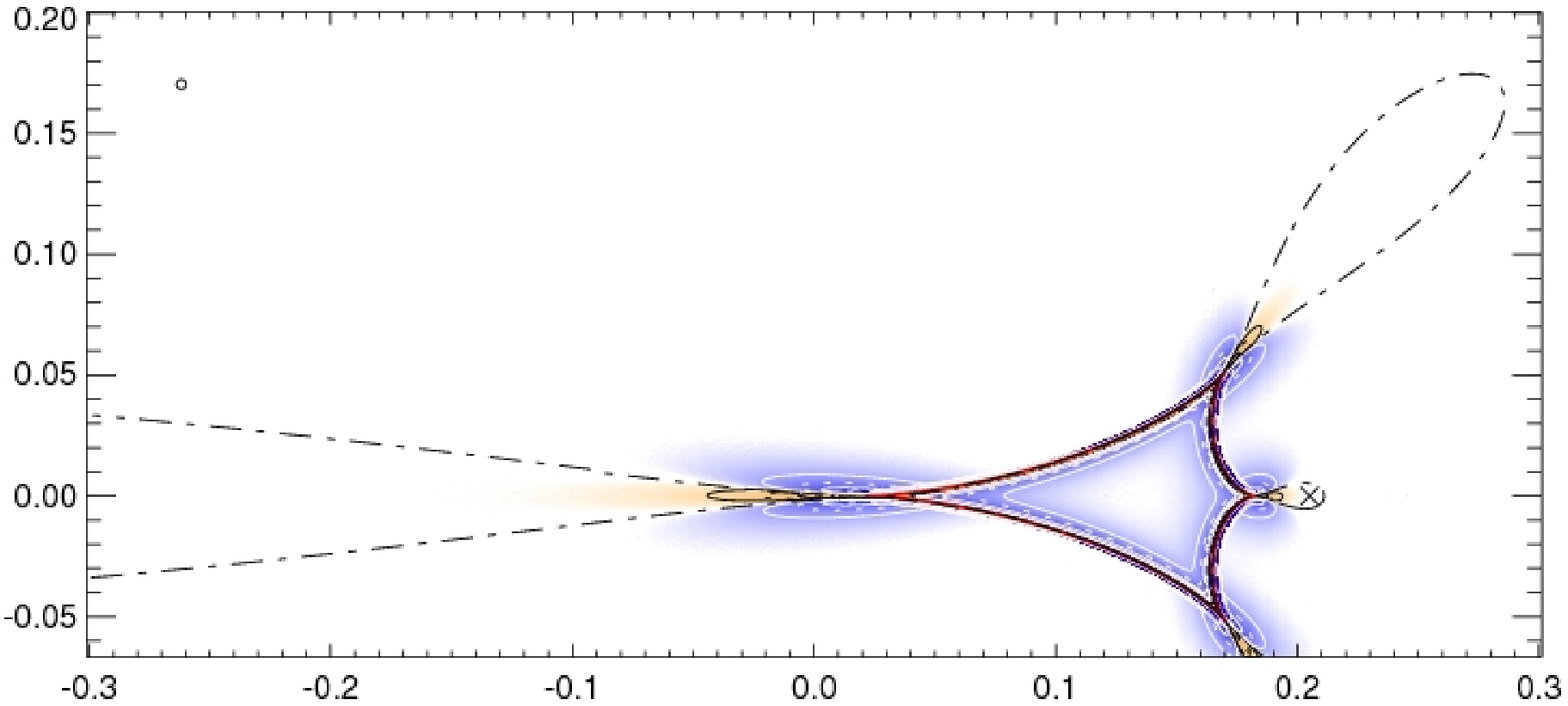}\\
\end{center}
\caption{Details of chromaticity $\delchr({\bf y}_{\rm c})$ for a $d=1,$ $q=1$ lens (top panels), and a $d=2.05$, $q=1/9$ lens (middle and bottom panels) for source radii $\rho_*=0.02$ (top left and middle panels) and $\rho_*=0.002$ (top right and bottom panels). Meaning of symbols and lines as in Figure~\ref{fig:delta_chr}, with four additional dotted contours at $\pm 10^{-2.5}$ and $\pm 10^{-1.5}$. Top left inset: cuts for source tracks marked by arrows; top right inset: cuts for the same number of source radii along the same tracks. Caustic crossing occurs at time $t=0$; $t_*$ is the source-radius crossing time. Gray horizontal lines in the insets mark contour values $\delchr=\pm10^{-2},\,\pm10^{-1.5},$ and $\pm10^{-1}$.
\label{fig:delta_chr_det}}
\enf

When computing $\delchr({\bf y}_{\rm c})$ or, more generally, when computing $A_*({\bf y}_{\rm c},\kappa)$ for different values of $\kappa$, it is sufficient to separately evaluate fluxes corresponding to $f_1$ and $f_2$, and combine them in the numerator and denominator of equation~(\ref{eq:amp_integral}) to get the amplification. In this way we obtained $A_*$ for $\kappa_{\rm pk}$, $\kappa_{\rm fl}$, and $\kappa=0$ and plotted $\delchr({\bf y}_{\rm c})$ in Figure~\ref{fig:delta_chr} for the nine lens geometries of Figure~\ref{fig:point_mag} and a source with $\rho_* = 0.02$.

\begin{figure*}
\begin{center}
\vspace{0.3cm}
\includegraphics[scale=0.3]{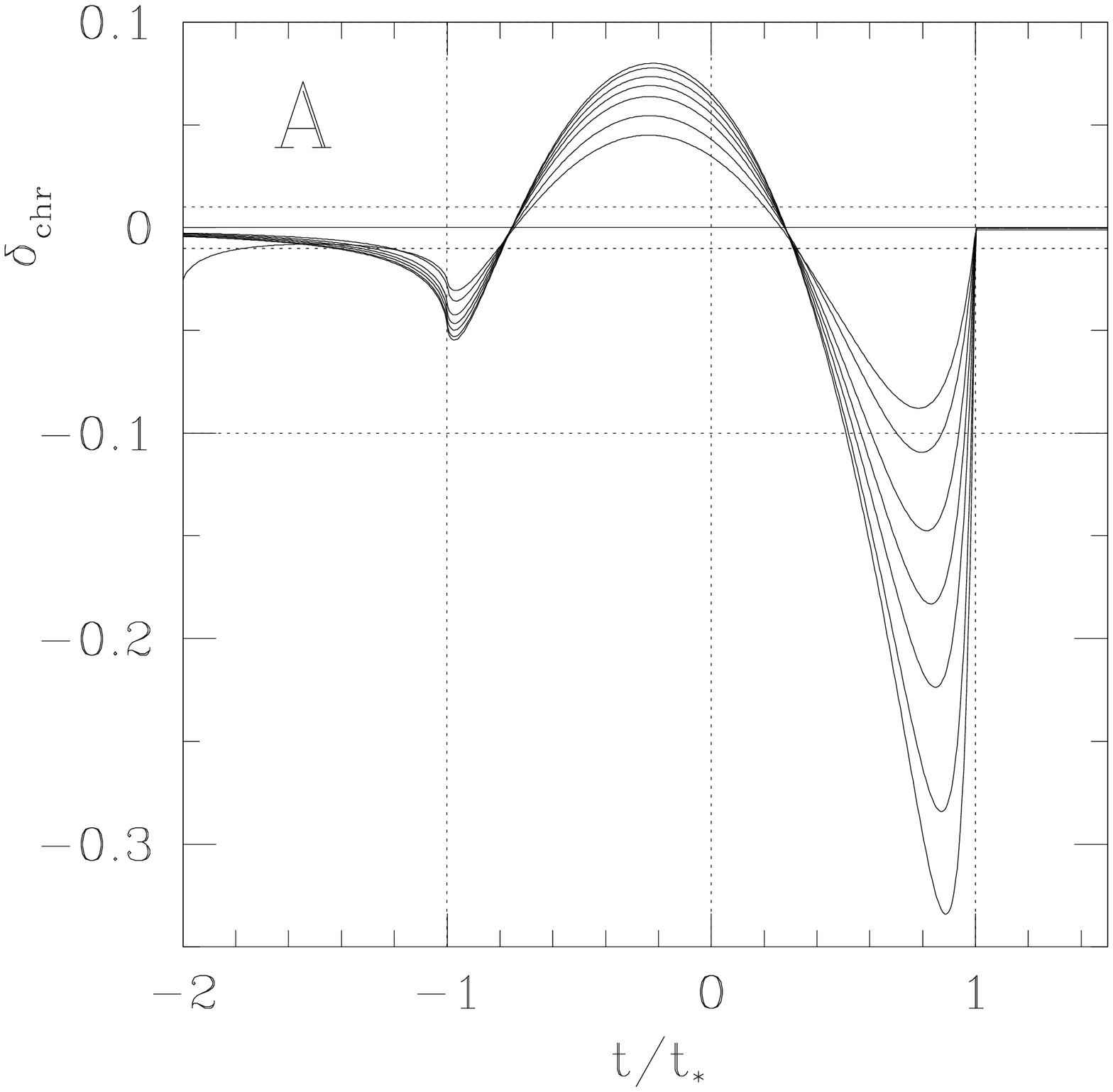}
\includegraphics[scale=0.3]{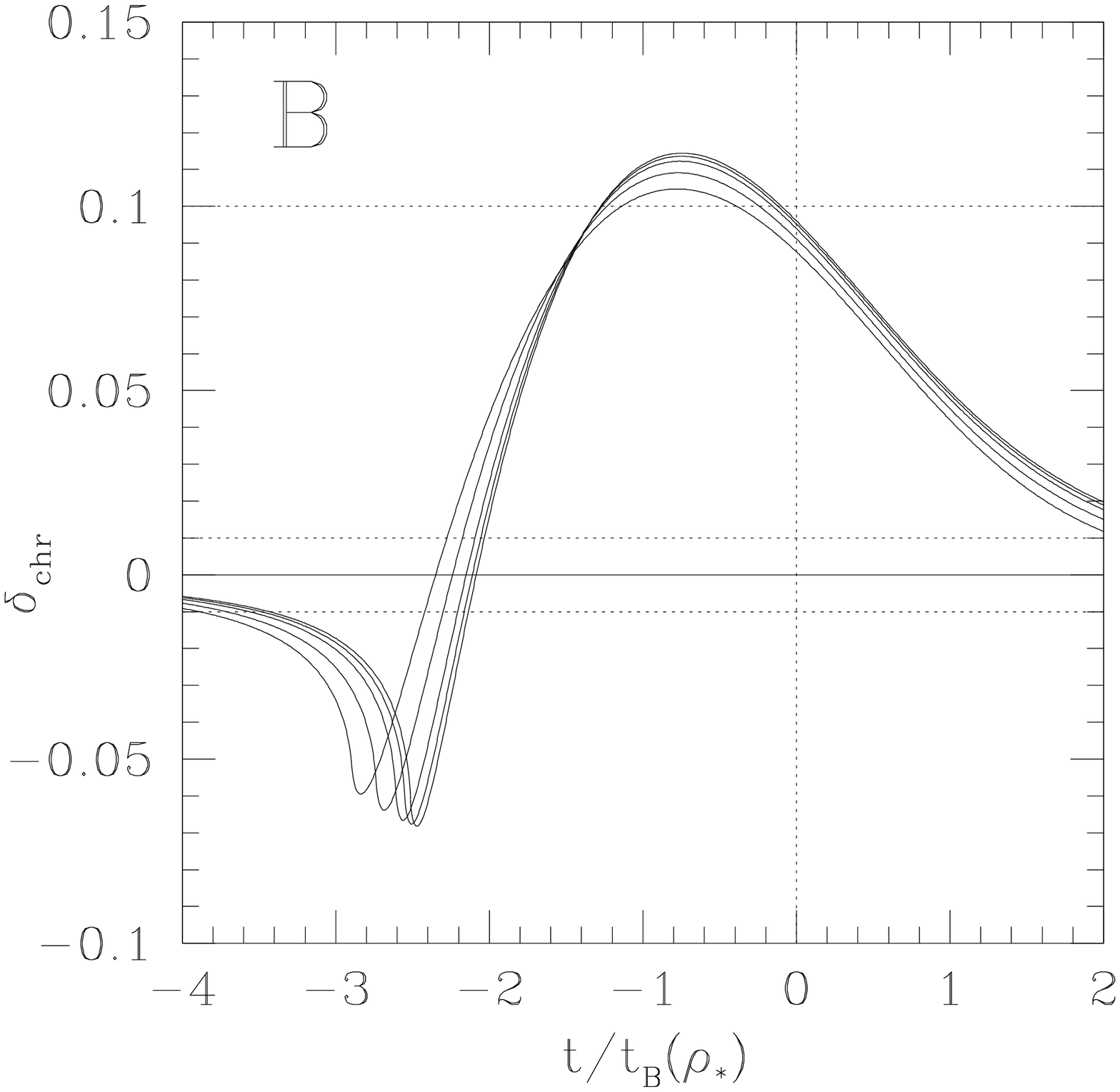}\\
\vspace{0.3cm}
\includegraphics[scale=0.3]{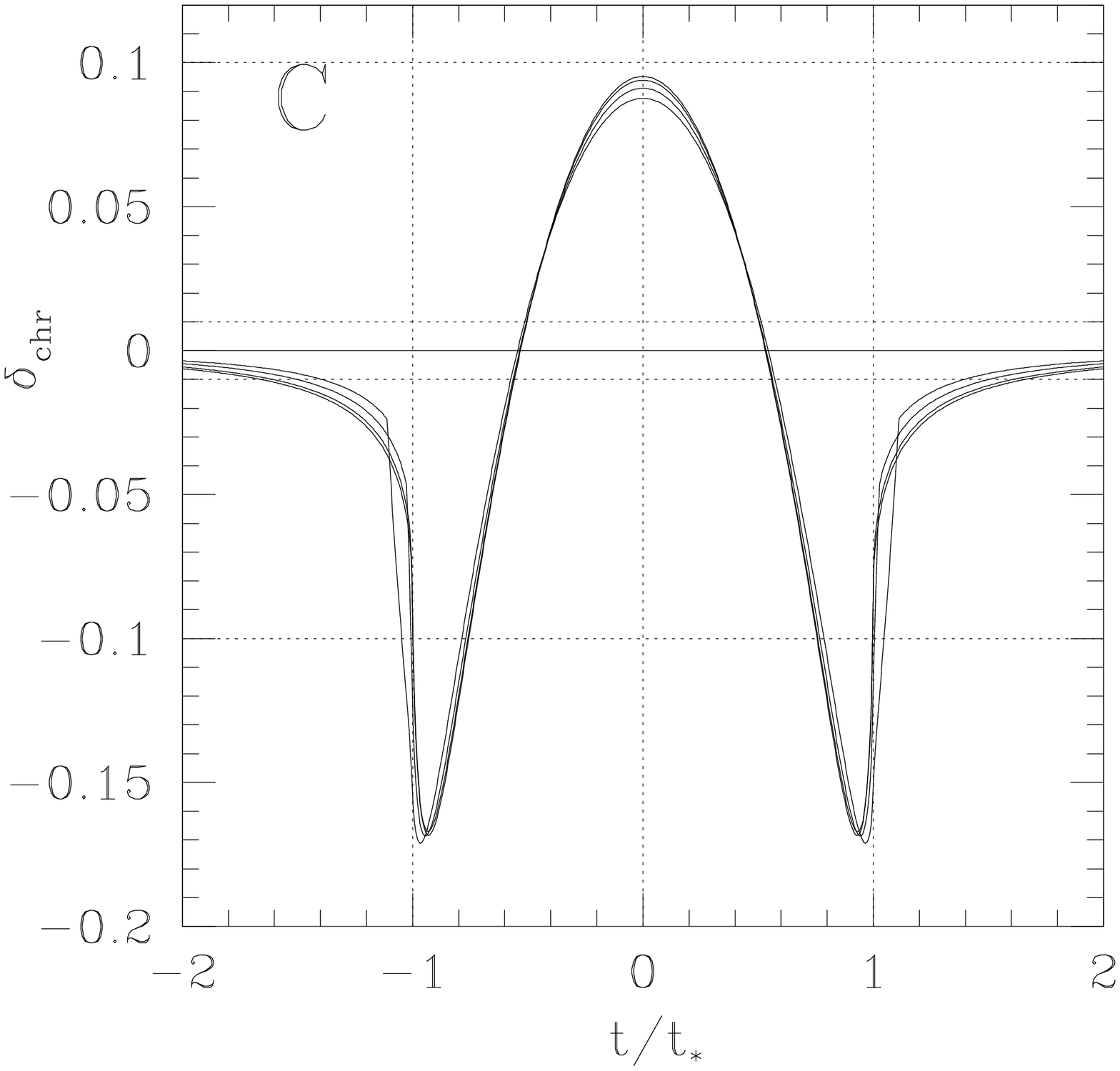}
\includegraphics[scale=0.3]{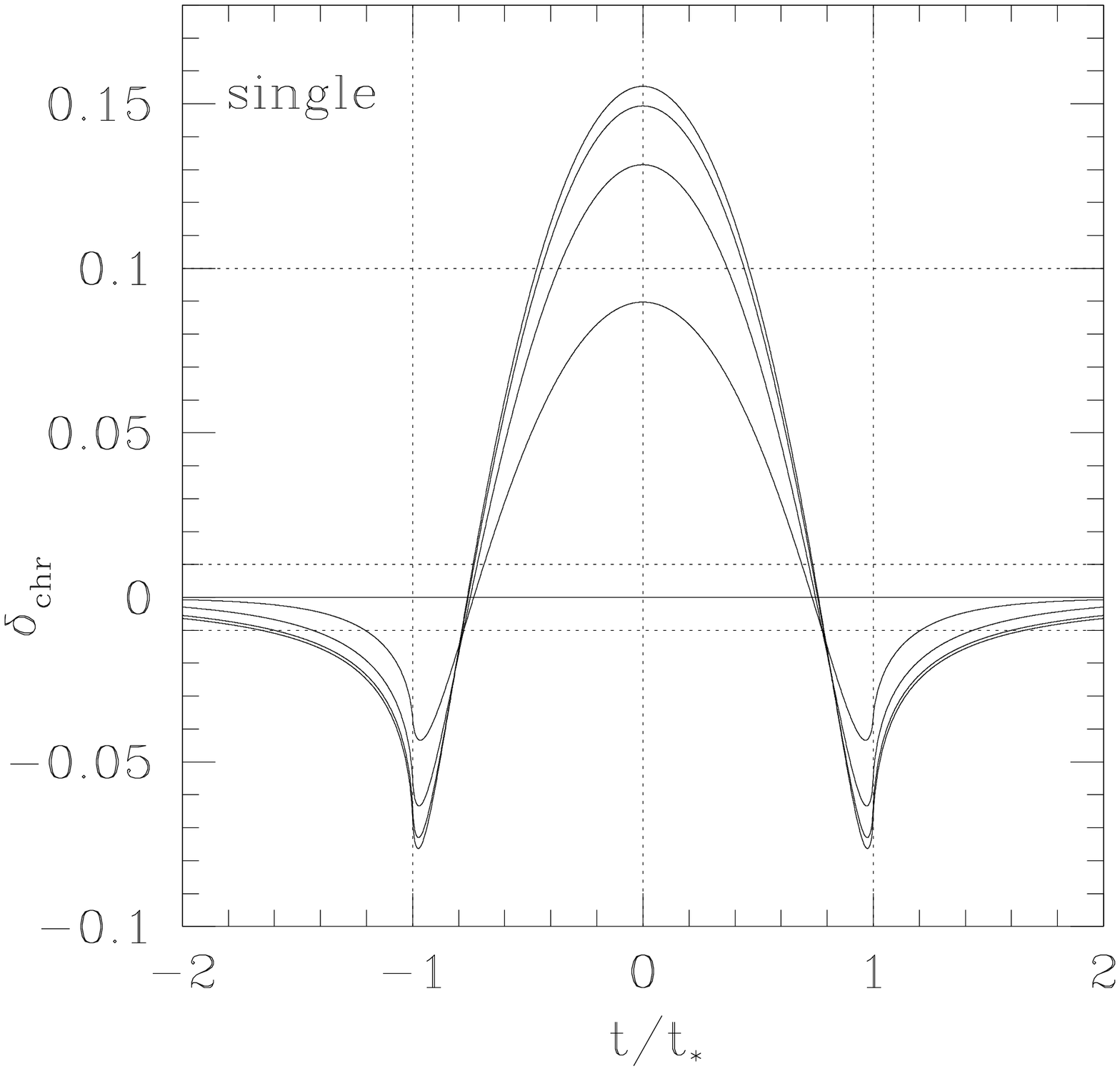}\\
\end{center}
\caption{Source-size dependence of chromaticity $\delchr$. Panels correspond to the source tracks marked in Figure~\ref{fig:delta_chr_det} as noted in the top left corner, and to a source-center-crossing single-point-mass lens (bottom right) for comparison. Curves with increasing peak height correspond to source radii $\rho_*=\{0.1,0.05,0.02,0.01,0.005,0.002,0.001\}$ for panel A, $\rho_*=\{0.1,0.05,0.02,0.01,0.005\}$ for panel B, $\rho_*=\{0.1,0.05,0.02,0.01\}$ for panel C, and $\rho_*=\{2,1,0.5,0.2\}$ for the single lens. The time intervals, time units, and vertically marked times are the same as in Figure~\ref{fig:delex_size}. Horizontal dotted lines mark $\delchr=\pm0.01,\pm0.1$.
\label{fig:delchr_size}}
\end{figure*}

As discussed above, the geometry of the $\delchr$ contours is similar to the geometry of $\delex\approx -5\,\delchr$ contours and the dot-dashed achromatic curve $\delchr=0$ coincides with the zero-effect curve of Figure~\ref{fig:delta_ex}, except in the immediate vicinity of the caustic. Staying away from the caustic, regions with positive $\delchr$, where the amplification of a source with a peaked limb-darkening profile exceeds that of a flat-profiled source, thus coincide with $\delex<0$ regions of Figure~\ref{fig:delta_ex}, where the point-source amplification is higher than the extended-source amplification. Similarly, regions with negative $\delchr$ coincide with $\delex>0$ regions of Figure~\ref{fig:delta_ex}. The explanation of the positive / negative region pattern is analogous to the explanation of the $\delex$ pattern in \S~\ref{sec:region}, taking into account that the peaked profile gives relatively more weight to the source center and less to the limb than the flat profile. Just as in Figure~\ref{fig:delta_ex}, extended areas sensitive in this case to limb-darkening differences can be found connecting the components of compound caustics, even though with lower amplitude $|\delchr|$. Not surprisingly, the dependence of these areas on $d$ and $q$ is the same as for the $\delex$ areas discussed in \S~\ref{sec:region}.

The detailed views presented in Figure~\ref{fig:delta_chr_det} correspond to the same two lens geometries as in Figure~\ref{fig:delta_ex_det} with source sizes $\rho_*=0.02$ (top left and middle panels) and $\rho_*=0.002$ (top right and bottom panels). The contour geometry close to the caustic is clearly different from the extended-source effect $\delex$ contours, as seen best in the middle panel. Unlike the $\delex=0$ curve, the outer branch of the achromatic curve does not coincide precisely with the caustic and the $\delchr>0$ band along the caustic is substantially broader than its $\delex<0$ counterpart in Figure~\ref{fig:delta_ex_det}, especially close to cusps. The most striking feature in Figure~\ref{fig:delta_chr_det} is the negative break at ${\rm y}_{\rm c1}=0.146$ across the positive chromaticity bands in the middle panel. A source on a vertical trajectory extended through the break would cross the caustic without changing the negative sign of its chromaticity. At the break the center of the source is highly amplified due to its position on the fold caustic. However, the limb of the source is amplified even more because a part of it lies along the right vertical fold caustic and another part crosses the caustic close to its cusp. The less striking dents in the positive band along the caustic on both sides of other cusps have a similar explanation.

Regions with maximum positive chromaticity are located just inside of cusps, where the source center achieves maximum relative amplification. Regions with maximum negative chromaticity are located about one source radius outside folds and outside cusps to either side of the cusp axis. At these positions the source limb achieves maximum relative amplification, as it passes through the cusp or its vicinity and has a large part aligned with the caustic.

The insets in the top panels include cuts through the plots along the same three source trajectories as in Figure~\ref{fig:delta_ex_det}, selected to demonstrate the behavior of chromaticity when crossing the caustic at a fold and a cusp, as well as its scaling with source size. The length of the source paths in the top right panel with $\rho_*=0.002$ is adapted to extend the same number of source radii on either side of the caustic. For a source approaching the caustic along the fold-crossing trajectory A, the negative chromaticity peaks as the limb touches the caustic, then reverts to a positive maximum before the center exits the caustic, and drops back to a major negative peak before fully exiting the caustic, after which the chromaticity is practically zero. The first two peaks increase slowly with decreasing source size, but the third peak doubles from $\delchr=-0.14$ for $\rho_*=0.02$ to $\delchr=-0.28$ for $\rho_*=0.002$. These high values are caused by the steep increase of the extended-source amplification when the limb just enters the caustic (in the direction opposite to A), with the slope given by the relative brightness of the limb. The region with $|\delchr|\gtrsim 0.01$ is limited to a band from $1\,\rho_*$ outside the caustic to $1.3\,\rho_*$ and $1.4\,\rho_*$ inside for the larger and smaller source, respectively.

For sources moving along the cusp-exiting trajectories B, the first two extremes are higher than in the corresponding A trajectories, reaching $\delchr=-0.065$ and $\delchr=0.11$ for both source sizes. The first negative peak occurs $2.6\,\rho_*$ from the cusp for the larger source, after the limb of the source crosses the caustic above and below the axis. As in the case of the finite-source effect, due to the curvature of the caustic for the smaller source the peak occurs already $5.3\,\rho_*$ from the cusp. For similar reasons, the positive peak occurs $0.75\,\rho_*$ inside the cusp for the larger source, and $1.6\,\rho_*$ from the cusp for the smaller source. Outside the caustic the curves differ from the fold-caustic case A by slowly dropping towards zero, with the larger source passing $\delchr=0.01$ at a distance of $2.6\,\rho_*$ from the cusp. The smaller source passes $\delchr=0.01$ at $6\,\rho_*$ from the cusp. The cusp-exiting curves subsequently drop into negative values and approach asymptotic 0 from below, as indicated by the achromatic curve. Inside the caustic the larger source crosses $\delchr=-0.01$ at $3.5\,\rho_*$, and the smaller source at $7.3\,\rho_*$ from the cusp.

The symmetric curves of the cusp-grazing trajectories C have a positive peak value $\delchr=0.09$ with the source center directly at the cusp, skirted by negative peaks with $\delchr=-0.16$ at positions $\pm0.95\,\rho_*$ with the limb at the cusp, decreasing in amplitude to $-0.01$ at $\pm1.7\,\rho_*$ from the cusp. As in the case of the extended-source effect curves, the decay rate is intermediate between cases A and B. The chromaticity curves for the two source sizes are practically identical.

Figure~\ref{fig:delchr_size} illustrates the source-size dependence of chromaticity in more detail for the three cuts from Figure~\ref{fig:delta_chr_det} and a single-point-mass lens for comparison. Following the layout of Figure~\ref{fig:delex_size}, the graphs for cuts A, C, and for the single lens are plotted as a function of time in units of the source-radius crossing time $t_*$, while for cut B time is shown in units of $t_{\rm B}(\rho_*)=(\rho_*/0.02)^{-1/3}\,t_*$. In each panel the curves with increasing peak height correspond to decreasing source size. For cut A, the values of $\rho_*=\{0.1,0.05,0.02,0.01,0.005,0.002,0.001\}$; for cut B, $\rho_*=\{0.1,0.05,0.02,0.01,0.005\}$; for cut C, $\rho_*=\{0.1,0.05,0.02,0.01\}$; for the single lens, $\rho_*=\{2,1,0.5,0.2\}$. All the $\delchr$ curves converge for $\rho_*\rightarrow 0$ to limit curves, so that curves for smaller source radii would be indistinguishable in the plots. Only the fold-exiting $t/t_*\rightarrow 1$ part of our cut A curves would still differ --- here the convergence is the slowest. The limiting value of this negative peak can be obtained from equation~(\ref{eq:delta_chr_separated}) if we replace the ratio of the amplified PCA terms by the ratio of limb values $f_2(1)/f_1(1)$. The limiting peak value for our PCA basis is $-1.46$, for linear limb darkening $-1$. Just as in the case of the extended-source effect, the fastest convergence to a generic chromaticity curve is found in the case of the single-point-mass lens \citep[see][]{heysalo00,hey08}, and the limit curves in binary-lens cuts B, C, and most of A are valid for sources two orders of magnitude smaller than in the single-lens case. Reaching the limit during the exit of the trailing source limb in cut A would require reducing the source size by several more orders of magnitude. As discussed previously for the extended-source effect, due to the universality of the local lensing behavior the limit curves A, B, C are valid for arbitrary fold, parallel cusp, and perpendicular cusp crossings, respectively. Only the rate of convergence depends on the geometry of the specific crossing.

Returning to Figure~\ref{fig:delta_chr_det}, the extent of the region with significant chromaticity along the lens axis in the middle and bottom panels is even larger than in the top panels. For the larger source in the middle panel the $\delchr=0.01$ contour extends $6.4\,\rho_*$ out from the leftmost cusp, but only $0.95\,\rho_*$ out from the rightmost cusp. For the smaller source the contour extends $13.8\,\rho_*$ to the left and $2.6\,\rho_*$ to the right. The asymmetry has the same cause as the $\delex$ contour asymmetry in Figure~\ref{fig:delta_ex_det} --- the large difference in the narrowness parameters $K$ of the cusps, as shown in \S~\ref{sec:analytical_cusps}.

For an interesting comparison we refer to \cite{han_park01}, who present similar plots in their Figure 3 of the $B-I$ color-change map and its cuts for the same caustic detail and a larger source with $\rho_*=0.1$. Their blue peaks correspond to our $\delchr>0$, their red peaks to $\delchr<0$. As expected from the results above, the peaks for their cusp-exiting cut are shifted closer to the cusp. A color-change map for the whole caustic of this $q=1$, $d=1$ lens and the same large source in their Figure 2 can be compared with our chromaticity map for a five times smaller source in the top central panel of Figure~\ref{fig:delta_chr}.

To summarize the results of this section, observationally significant chromaticity occurs within regions sensitive to the extended nature of the source, but as a generally weaker effect it is concentrated closer to the caustic. As in the case of the extended-source effect, for smaller sources chromaticity is relevant in regions smaller in terms of Einstein radii, which nevertheless extend further out from cusps in terms of source radii. Positive regions reach highest $\delchr$ values just inside cusps, from where they extend along the caustic and along the outer cusp axes. Negative regions reach a local extreme inwards of the positive band along the caustic, but achieve the overall highest $|\delchr|$ in a band skirting the caustic from outside. These outer high-negative-chromaticity regions are also particularly sensitive to the source size, with a smaller source permitting higher differential amplification when its limb enters the caustic. The significance of these areas is somewhat reduced by the low amplification at the caustic-crossing onset or end of exit. However, in current microlensing monitoring projects caustic exits are generally observed with better photometry and sampling than caustic entries, thus giving ample opportunity for measuring the limb darkening of the source.

The chromatic sensitivity of microlensing is even more pronounced if we examine the wavelength dependence in more detail, that is, if we proceed from photometry to spectroscopy. Although microlensing spectroscopic effects are closely related to broad-band chromaticity \citep{heysalo00}, their exploration in full two-point-mass lensing is beyond the scope of this paper. Instead of a simple two-term limb-darkening law such a study requires the usage of a full model atmosphere with its wavelength-dependent center-to-limb specific-intensity variation at the necessary spectral resolution.

\section{VALIDITY OF LINEAR-FOLD APPROXIMATION}
\label{sec:fold}

The analysis of static binary fold-caustic-crossing microlensing events \citep[e.g.,][]{afonsoetal00,albrowetal01a} with extended-source effects often follows a procedure similar to the one outlined by \citet{albrowetal99b}. An analogous approach has been used for more complicated events such as OGLE-2002-BLG-069 \citep{kubasetal05}, which involves the parallax effect, and in two early papers on EROS-BLG-2000-5 \citep{albrowetal01b,afonsoetal01}, a cusp-approaching event that manifests both binary rotation and parallax effects.

\begin{figure*}
\begin{center}
\vspace{0.3cm}
\includegraphics[scale=0.43]{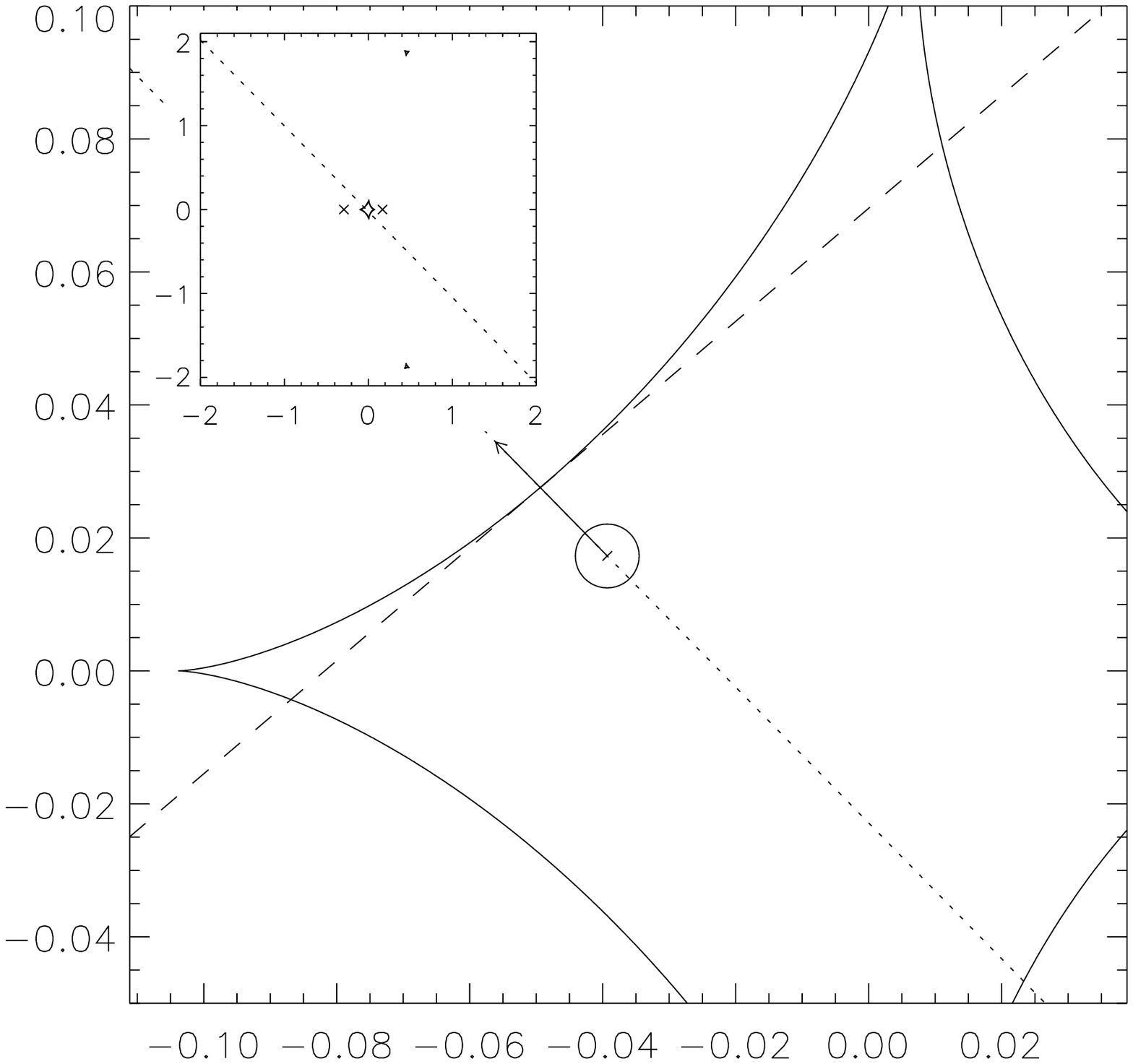}
\includegraphics[scale=0.42]{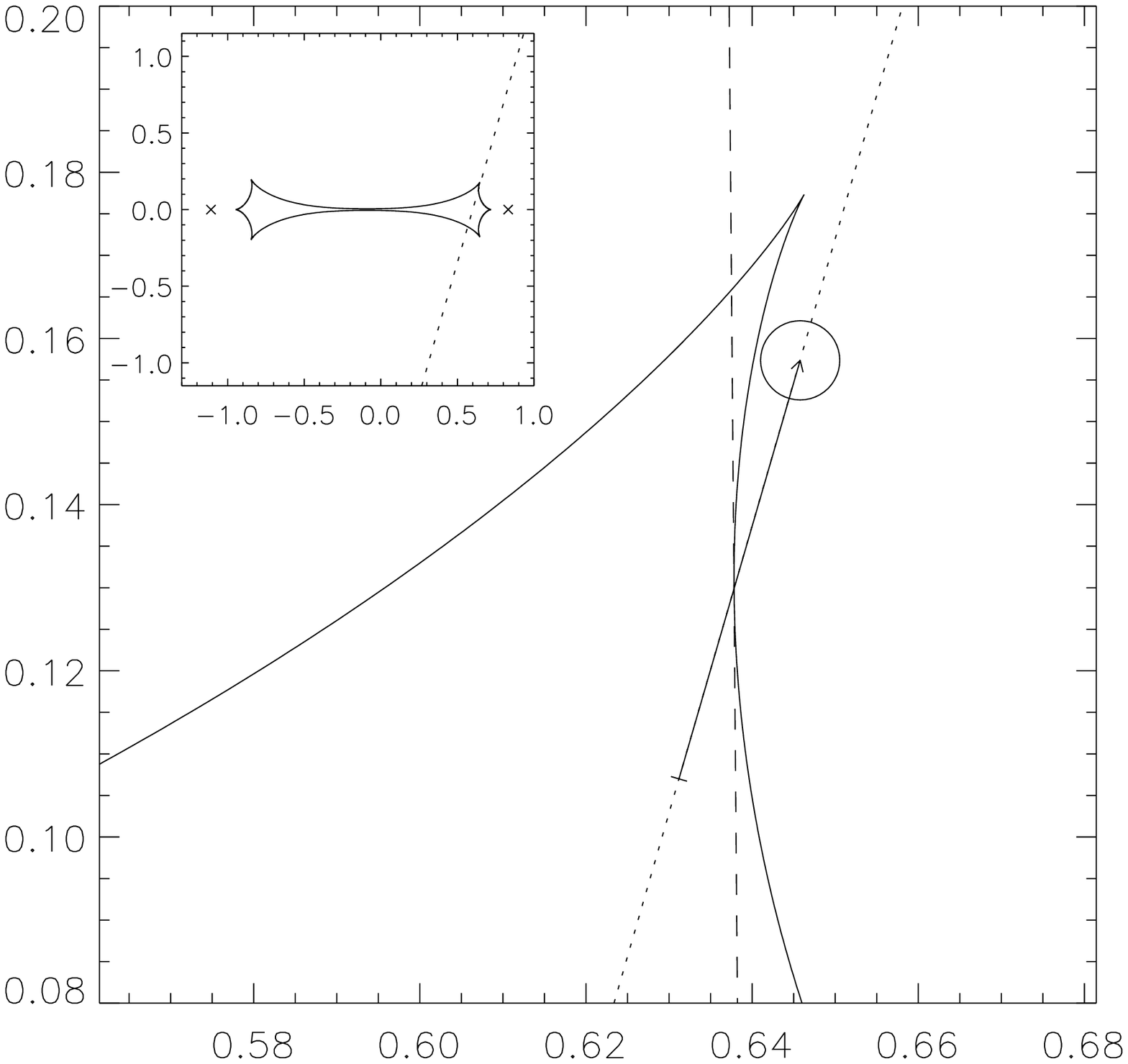}\\
\vspace{0.3cm}
\includegraphics[scale=0.45]{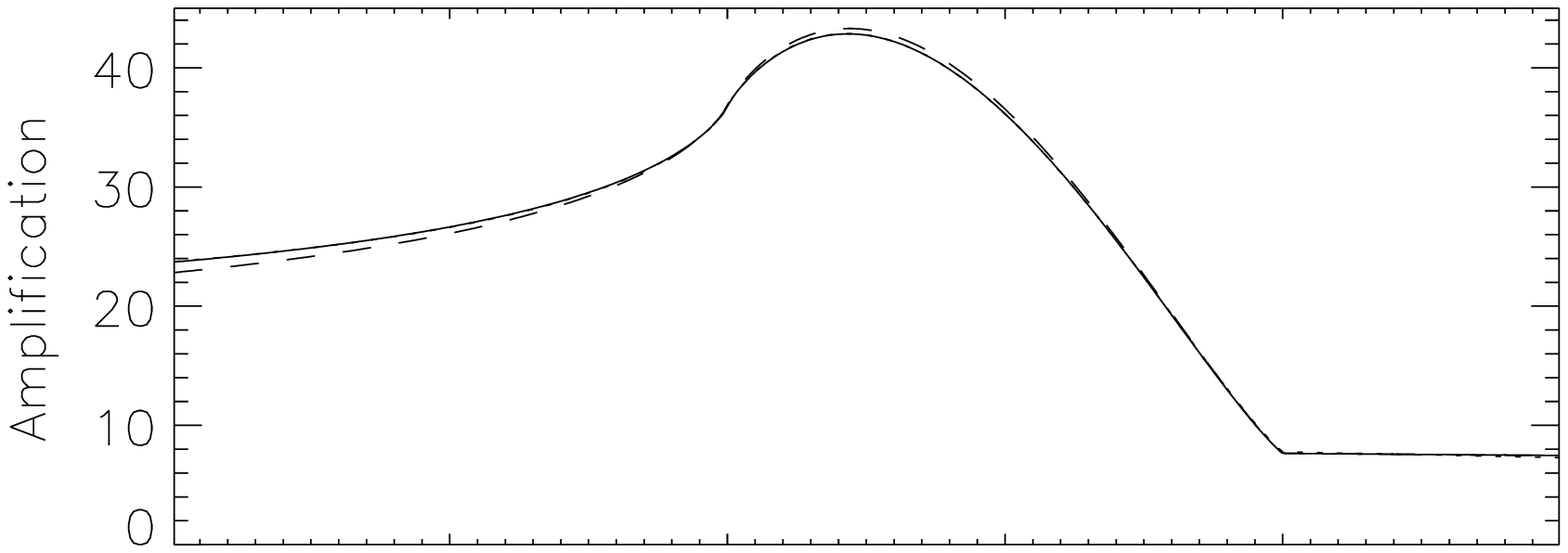}
\includegraphics[scale=0.42]{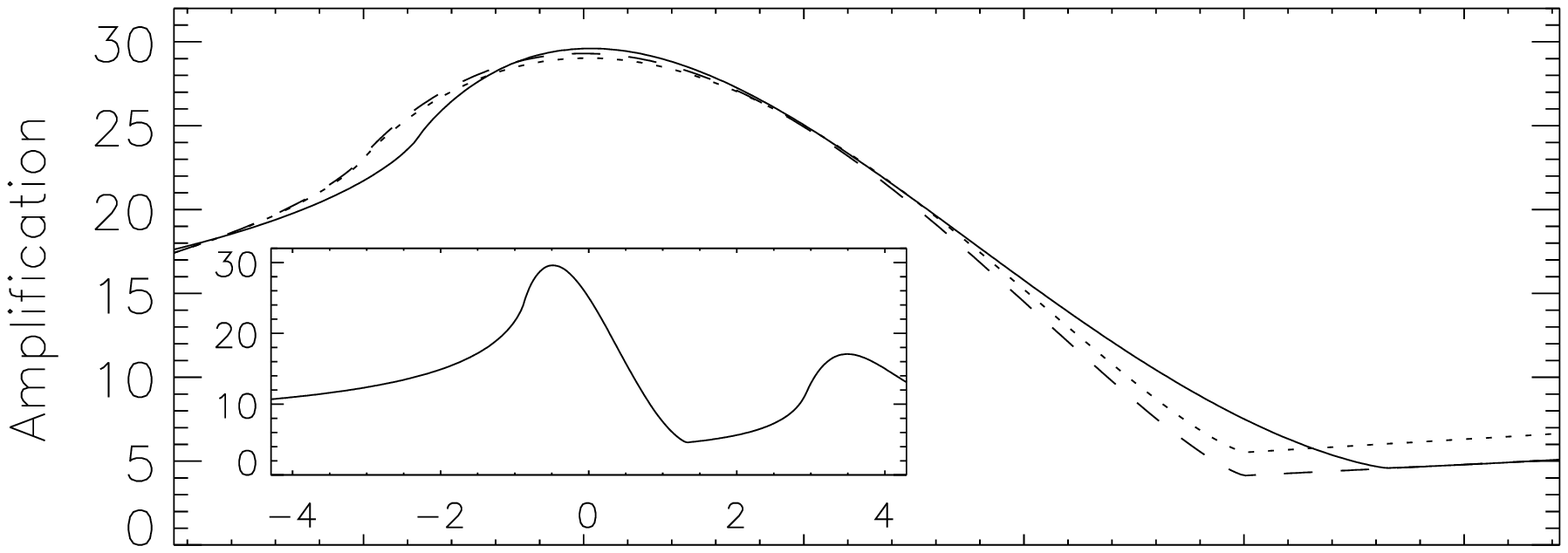}\\
\vspace{0.1cm}
\includegraphics[scale=0.45]{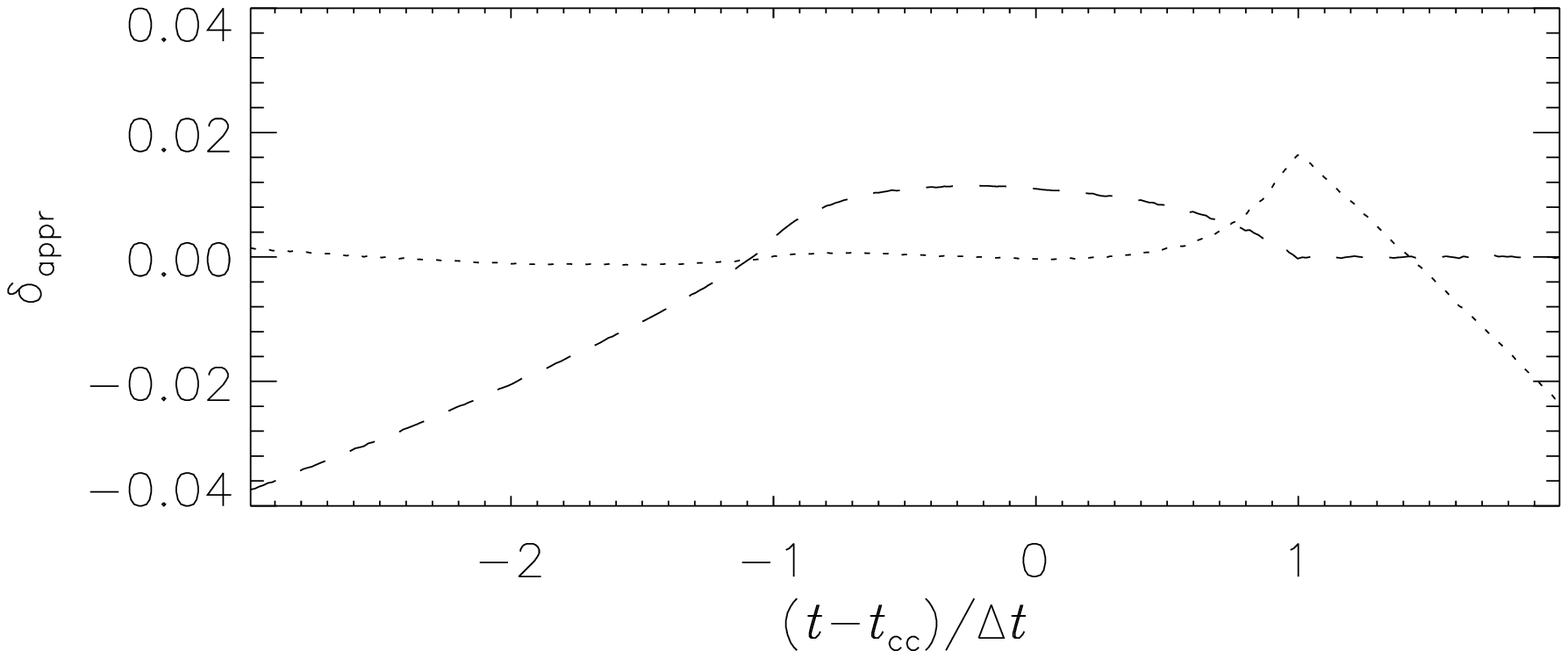}
\includegraphics[scale=0.42]{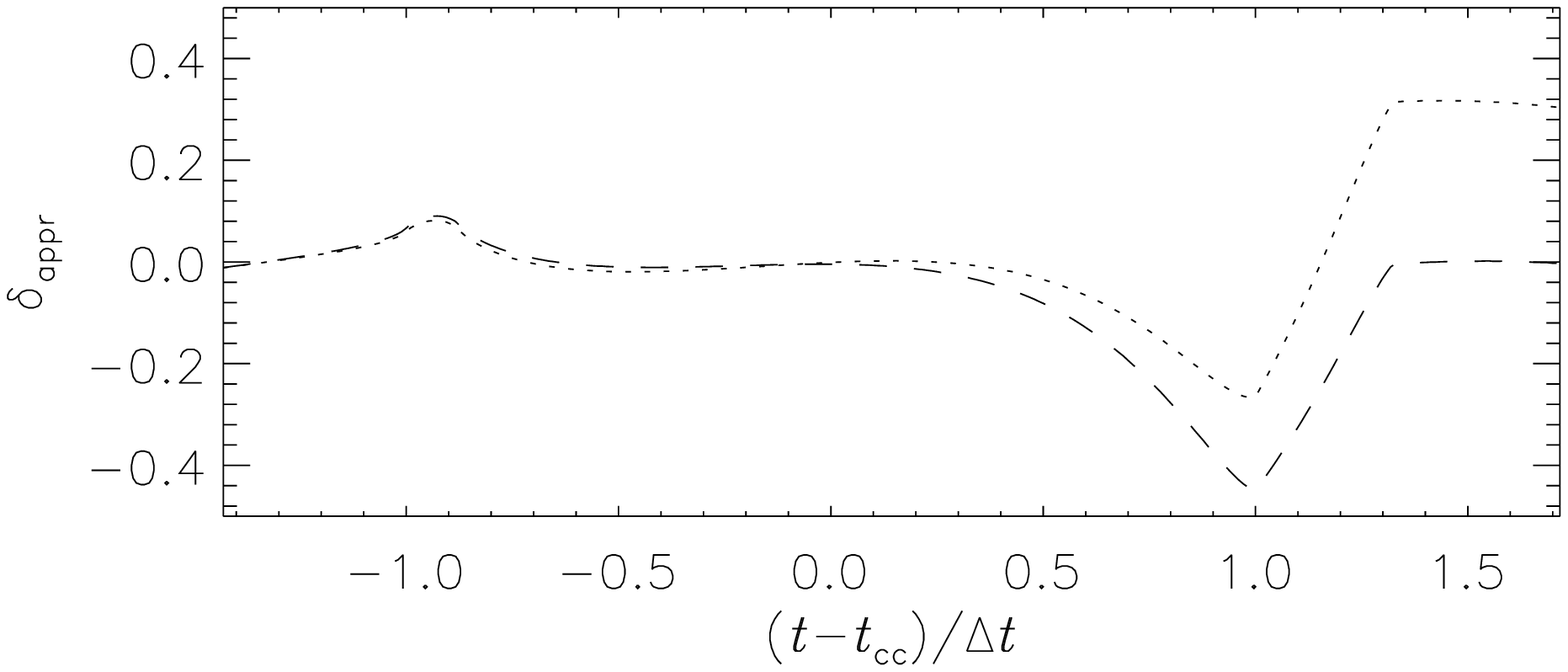}\\
\end{center}
\caption{Caustic-crossing details of events with main parameters taken from OGLE-2002-BLG-069 (left column) and EROS-BLG-2000-5 (right column). Top panels: caustic-crossing geometry with overall event geometry in insets. Solid line: caustic; dotted line: source trajectory with arrow marking the section used in the bottom panels; dashed line: linear fold tangent to the caustic at crossing point. Bottom panels: light-curve detail with fits and their relative residuals $\delappr$. Solid line: exact light curve $A_*$; dotted and dashed lines: best-fit fold-approximation light curves $A_*^{\rm appr}$ for fitting methods I and II, respectively. Bottom right inset: wider part of the light curve including the following cusp-axis approach.
\label{fig:crossing_details}}
\end{figure*}

When a source exits the inner region of the binary caustic, two of the five images disappear. As a point source approaches a fold section of the caustic from inside, the amplification of the two vanishing images diverges in a square-root singularity \citep{bland_nara86,schn_weiss86}. Assuming that the amplification of the three remaining images changes only linearly near the caustic and approximating the caustic by a tangent at the point of crossing, the approximate total amplification $A_*^{\rm appr}(t)$ of an extended source as a function of time $t$ is given by
\beq
A_*^{\rm appr}(t) = U_* G_I\left( \frac{t-t\cacr}{\Delta t}\right) + A\cacr + \omega(t-t\cacr),
\label{eq:caus_approx}
\eeq
where $U_*$ is a scaling factor depending on the source size and the characteristic rise length of the caustic, $A\cacr$ and $\omega$ describe the linear variation of the amplification of the three non-vanishing images, and $t\cacr$ is the time when the source center crosses the caustic\footnote{For a fold-caustic entrance it is sufficient to change the sign of the argument of $G_I$ in equation~(\ref{eq:caus_approx}).}. The timescale of the caustic crossing is measured by the time from first contact to $t\cacr$,
\beq
\Delta t = t_* \csc \phi,
\label{eq:time_scale}
\eeq
where $t_*=\rho_* t_{\rm E}$ is the source-radius crossing time, $t_{\rm E}$ is the Einstein-radius crossing time, and $\phi$ is the angle between the trajectory and the caustic at the point of crossing. $G_I(\eta)$ is a characteristic profile function which depends only on the limb darkening of the source but not on its size,
\beq
G_I(\eta) = 2\,H(1-\eta) \!\!\!\!\!\!\!\! \int\limits_{\max (\eta, -1)}^1 \int\limits_{0}^{\sqrt{1-r_1^2}} \frac{I(\sqrt{r_1^2+r_2^2})}{\sqrt{r_1-\eta}}\, \rmd r_2 \rmd r_1\,,
\label{eq:g_function}
\eeq
where $H(1-\eta)$ is the Heaviside step function, and the limb-darkening function $I(r)$ is usually normalized to unit total flux. For several classes of analytic limb-darkening functions equation~(\ref{eq:g_function}) can be expressed in terms of elliptical integrals; in more general cases $G_I(\eta)$ can be computed numerically. \citet{albrowetal99b} explain the linear-fold approximation and its parameters in more detail.

A fold-caustic crossing with a fixed limb-darkening profile can thus be fitted by equation~(\ref{eq:caus_approx}) with five free parameters: $U_*$, $t\cacr$, $\Delta t$, $A\cacr$, and $\omega$. The procedure of \citet{albrowetal99b} then continues with a fit to the non-caustic-crossing part of the light curve constrained by the pre-fitted parameters to determine the lens geometry. Finally, the full data set is used to refine the solutions. However, equation~(\ref{eq:caus_approx}) is still used for the caustic-crossing part of the light curve.

The outlined methodology has several shortcomings that can hinder correct reconstruction of the lens geometry and the retrieval of source parameters. First, as we showed in \S \ref{sec:sensitivity}, extended-source effects are not limited to the immediate vicinity of the caustic curves. Instead, affected areas may be substantially larger (see Figure~\ref{fig:delta_ex} and Table \ref{tab:cross}). Second, not every assumption of equation (\ref{eq:caus_approx}) is justified under all circumstances. For example, the curvature of the caustic becomes important for trajectories with small $\phi$, and so does the change of $A_0$ parallel to the caustic. As shown in \S~\ref{sec:region} and Figure~\ref{fig:delex_size}, even in perpendicular crossings the source has to be small enough for the light-curve shape to be independent of $\rho_*$. Furthermore, the magnification of the three images unaffected by caustic crossing cannot be always regarded as a linear function of time, for example when the caustic crossing occurs in the vicinity of a cusp. Close to cusps even the inverse-square-root approximation of the divergence breaks down. There have been several cases of fold-crossing events, in which equation (\ref{eq:caus_approx}) turned out to be inadequate for the analysis. For example, in the case of OGLE-1999-BUL-23 \citet{albrowetal01a} found that the difference between the approximation given by equation (\ref{eq:caus_approx}) and amplification computed by the method of \citet{gould_gaucherel97} amounted to as much as $4\%$. To cope with this problem they used pre-calculated tabulated corrections to improve the approximate expression.

In the following we compute $n-$point light curves of fold-caustic crossings spanning several $\Delta t$ around the time of caustic crossing. For simplicity, we assume that $\Delta t$ and $t\cacr$ are known from the event geometry. Thus, we are left with a linear fit of the three parameters $U_*$, $A\cacr$, and $\omega$. There are two natural ways how to perform the fit. We may use standard linear least-squares fitting for all $n$ points (hereafter method I). Alternatively, realizing that $G_I(\eta) = 0$ for $\eta > 1$, we let such points determine $A\cacr$ and $\omega$, and points with $\eta < 1$ are then used to fit $U_*$ (hereafter method II).

We illustrate the drawbacks of the linear-fold approximation on two caustic-crossing events chosen to resemble actually observed events, OGLE-2002-BLG-069 and EROS-BLG-2000-5. The former event was selected as a case favorable for the method of \citet{albrowetal99b}, based on the analysis by \citet{kubasetal05}. We used their close-binary parameters ($q=0.58$, $d=0.46$) but we omitted annual parallax because it has a negligible effect during the short interval of caustic crossing, and our objective is to study the method rather than the event itself. The source star with radius $\rho_* = 0.0048$ followed a linear trajectory parameterized by its closest distance to the center of mass $u_0 = Min({\rm y}_{\rm c}) = 0.016$ and the angle $\alpha = 134.4\degree$ between the ${\rm y}_1$ axis and the trajectory (see the top left panel of Figure~\ref{fig:crossing_details} for a visualization of the geometry). We use here linear limb darkening (eq.~[\ref{eq:lin_limb}]) and set $\upsilon = 0.62$ to reproduce the $I$-band limb-darkening coefficient of \citet{kubasetal05}. The second intersection of the trajectory and the caustic occurs under an angle $\phi \simeq 86\degree$, thus making it a nearly perpendicular crossing. We plot the relative residuals $\delappr$ of the approximation given by equation~(\ref{eq:caus_approx}) from the exact amplification $A_*$ computed from equation~(\ref{eq:amp_integral}) as described in \S~\ref{sec:method},
\beq
\delappr = \frac{A_*^{\rm appr} - A_*}{A_*}\, ,
\label{eq:delta_appr}
\eeq
in the bottom left panel of Figure~\ref{fig:crossing_details}. Even in this favorable case the residuals may exceed $2\%$ with either fitting method, a deviation from the linear-fold approximation that is observationally significant. Such deviations would be deleterious especially for limb-darkening measurement, taking into account that the chromaticity mostly remains below the level of $10\%$. With method I, $\delappr$ peaks when the edge of the source exits the caustic (i.e., when the two highly magnified images disappear) and the amplitude of the residual keeps growing over $2\%$ as the source moves more than $2\Delta t$ away from the caustic. When using method II, $\delappr$ keeps growing over $2\%$ when the source lies more than $2\Delta t$ inside the caustic.

In the case of the latter event, EROS-BLG-2000-5, initial works \citep{castroetal01,albrowetal01b,afonsoetal01} concentrated on the second caustic crossing, during which spectra had been measured. The event was subsequently thoroughly analyzed by \citet{anetal02}. \citet{fieldsetal03} refined the event parameters further and compared the measured limb darkening to stellar atmosphere models. In our analysis we adopt the parameters obtained by \citet{fieldsetal03}, namely $d=1.940$, $q=0.75$, $\rho_* = 0.004767$, $\alpha=73.851151\degree$, but we do not incorporate binary rotation and annual parallax. From the distance of the closest approach to the cusp given by \citet{fieldsetal03} we compute $u_0 = 0.7096$, which yields $\phi \simeq 16.6\degree$. The event geometry is shown in the top right panel of Figure~\ref{fig:crossing_details}. Based on the reported $V$-band square-root limb darkening, we use a linear limb-darkening profile with $\upsilon = 0.792$\footnote{Computed by a least-squares fit to the two-parameter square-root profile, although details of the limb darkening have little impact on the point raised in this section.}. A comparison of exact and best-fit approximated light curves is given in the bottom right panels of Figure~\ref{fig:crossing_details}. Obviously, the match is poor, locally exceeding $30\%$ for both fitting methods, and the reason is clear. Due to the curvature of the caustic and the small value of $\phi$ the crossing takes longer and, most notably, the caustic exit occurs significantly later than would be expected from the linear-fold approximation. Furthermore, just after exiting the caustic the source passes in the immediate vicinity of a cusp, therefore the amplification of the three non-vanishing images cannot be regarded as a linear function of time. More generally, the amplification strongly changes in the direction parallel to the tangent linear fold.

From the bottom right plot in Figure~\ref{fig:crossing_details} we see that $\delappr$ peaks around $|t-t\cacr| = \Delta t$. One might try to improve the fit by slightly adjusting $t\cacr$ to account for the obvious shift which can be seen in the light curve in the right column of Figure~\ref{fig:crossing_details}. We tried to fit $t\cacr$ together with $U_*$, $A\cacr$, and $\omega$ for both events but found only a marginal improvement, with the relative residuals staying in the percent level for OGLE-2002-BLG-069 and tens of percent in the case of EROS-BLG-2000-5. Obviously, one could improve the fits further by varying also $\Delta t$. However, this would amount to modifying the physical parameters of the event in order to maintain internal consistency, and such an analysis is beyond the scope of this paper. While it might slightly improve the fit for OGLE-2002-BLG-069, the complex pattern of the large residuals in the case of EROS-BLG-2000-5 practically eliminates any hopes for a plausible linear-fold approximation to the crossing. More importantly, it is clear that releasing the fixed parameters in the demonstrated examples leads to a bias in their recovered values, by trying to account for the residuals caused by the inadequate approximation.

Despite the preceding discussion, the linear-fold approximation could still be useful for analyzing caustic-crossing events similar to OGLE-2002-BLG-069 in combination with an appropriate limb-darkening model --- but for substantially smaller sources. However, any usage of the approximation should be justified by a test of its applicability in the particular case, at least by comparing the exact light curve for the obtained parameter values with its best fit by the fold model from equation~(\ref{eq:caus_approx}).

\section{COMMENTS ON OBSERVED EVENTS}
\label{sec:events}

According to \citet{mao_paczynski91} $\sim 10\%$ of all microlensing events in the Galactic bulge should display signatures of strong binary lensing, i.e., they should undergo caustic crossing. In many non-caustic-crossing binary events the influence of lens binarity leads merely to perturbations of the point-source-point-lens (PSPL) light curve \citep{di_stefano_perna97,nightetal07}. Nevertheless, as we showed in \S~\ref{sec:sensitivity}, even in non-caustic-crossing events light curves may be affected by extended-source effects. In this section we first comment on several such observed events that do or potentially could illustrate points raised in previous sections.

\citet{udalskietal05} analyzed the extensive observations of the event OGLE-2005-BLG-071, and concluded that the deviations from the PSPL light curve are caused by the source passing close to three cusps of a caustic induced by a Jovian-mass companion to the primary lens. Here we assume the following parameters for this event: $q=0.0071$, $d=1.294$, $u_0=0.0236$, and $\alpha=274.23\degree$. In Figure~\ref{fig:ogle05071} we illustrate the dependence of amplification $A_*(t)$, extended-source effect $\delex(t)$, and chromaticity $\delchr(t)$ on the source radius $\rho_*$. The precision of the photometric measurements of \citet{udalskietal05} was better than $1\%$. We can see from Figure~\ref{fig:ogle05071} that such a precision permits the detection of the extended-source effect for sources with $\rho_* \gtrsim 0.0003$, and $\delchr$ reaches $1\%$ for sources with $\rho_* \gtrsim 0.001$. This does not necessarily mean that meaningful limb-darkening measurement would be possible in this event, but rather that one should select a more-or-less appropriate limb-darkening profile for the modeling. The new analysis of the event by \cite{dongetal08}, which appeared after our manuscript submission, confirms our conclusions. Source radii $\rho_* > 0.0009$ are ruled out at $>3\,\sigma$, the best-fit solution yields $\rho_*=(5.5\pm1.2)\times 10^{-4}$, and varying the limb darkening had no significant effect on the results.

\citet{jaroszynski02}, \citet{jaroszynskietal04,jaroszynskietal06}, and \citet{skowronetal07} published models of 53 binary microlensing events observed by the OGLE project. Although the number of observations of many of the weak binary events is not enough to constrain extended-source effects or chromaticity, with better sampling some of them would be suitable. For example, trajectories of events OGLE-2004-BLG-280 \citep{jaroszynskietal06} and OGLE-2002-BLG-099 \citep{jaroszynskietal04} cross the regions between the caustic components with enhanced sensitivity to the extended nature of the source. The same is true for MACHO 97-BLG-41 \citep{albrowetal00}, but in this case the situation is more complicated due to the rotation of the binary.

\bef
\plotone{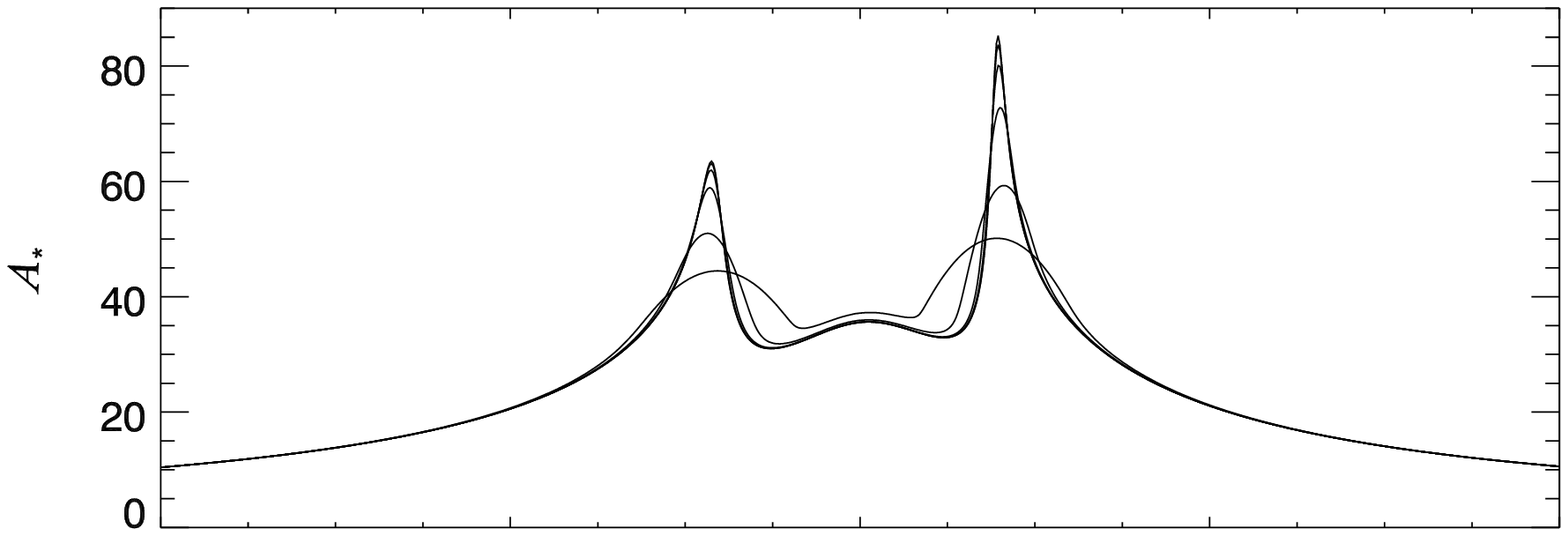}
\plotone{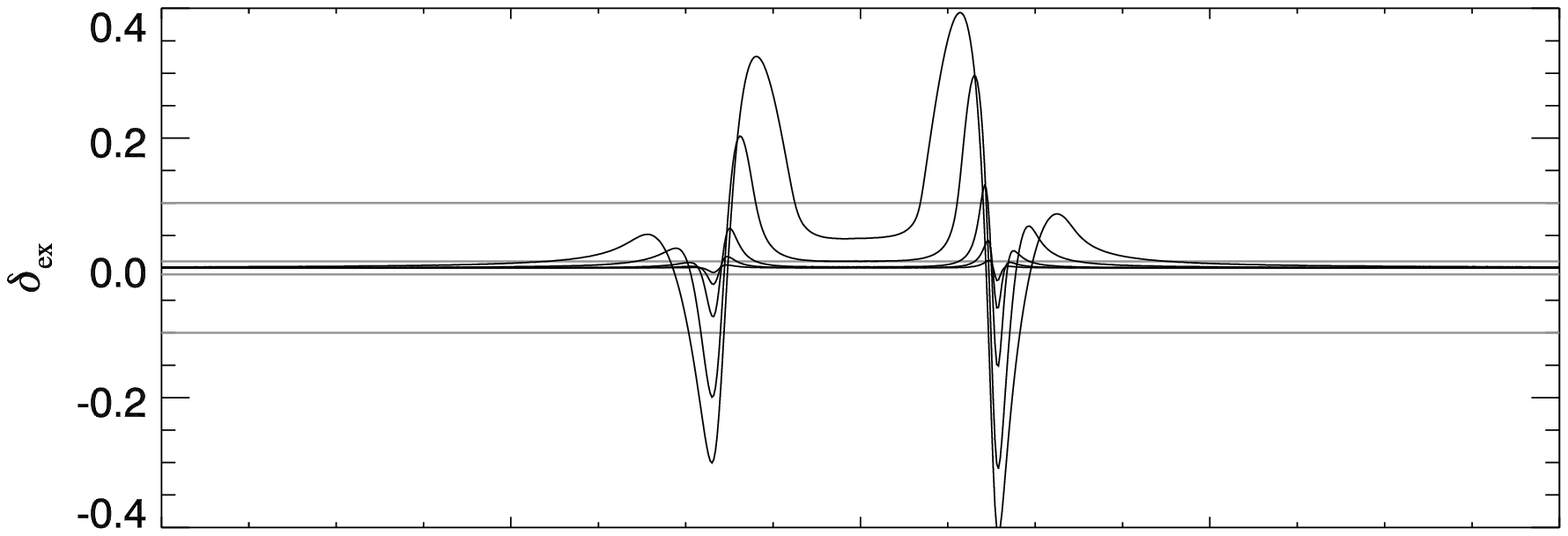}
\plotone{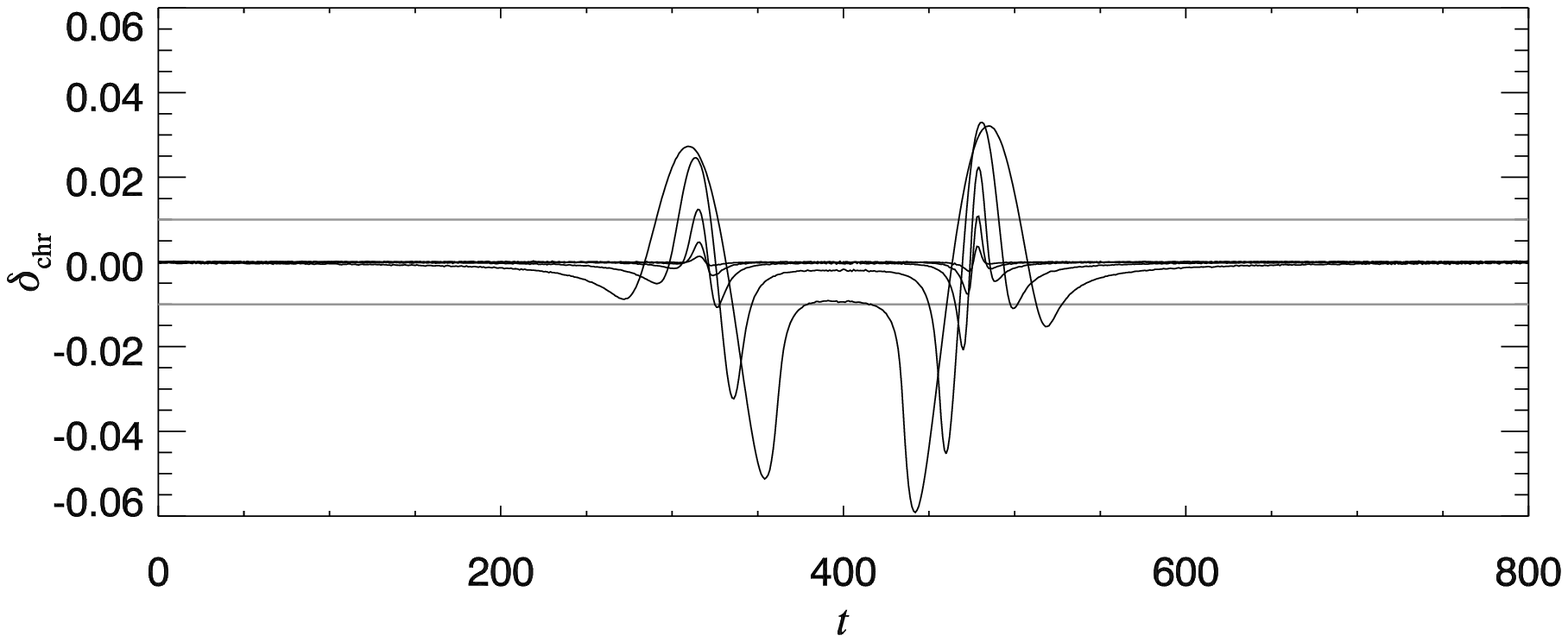}
\caption{Light-curve detail of a planetary-mass non-caustic-crossing event approaching three cusps, with main parameters taken from OGLE-2005-BLG-071.  Top panel: amplification $A_*$; middle panel: extended-source effect $\delex$; bottom panel: chromaticity $\delchr$. Individual curves correspond to different source radii $\rho_*$. In the top panel curves with highest to lowest peaks correspond to a point source (barely distinguishable), $\rho_*=0.0005,$ 0.001, 0.002, 0.005, and 0.01. The order is reversed and the point source omitted in the lower two panels (with the largest source producing largest deviations). The time axis is plotted in arbitrary units; gray horizontal lines in the middle panel mark $\pm10\%$ and $\pm1\%$ effects, in the bottom panel a $\pm1\%$ effect.
\label{fig:ogle05071}}
\enf

Turning to fold-caustic-crossing events, in \S~\ref{sec:fold} we discussed OGLE-2002-BLG-069 and EROS-BLG-2000-5. We showed that even in the case of a nearly perpendicular caustic crossing well separated from cusps the bias introduced by using the linear-fold approximation is observationally significant. We performed a similar analysis for several other binary events with good data coverage. The MACHO-98-SMC-1 analysis by \citet{afonsoetal00} yielded two viable solutions, either a close binary (potentially rotating) or a rotating wide binary, and one non-viable static wide binary solution. Similarly to EROS-BLG-2000-5, there is a cusp approach just after the second caustic crossing. Our inspection of the second caustic crossing for the static close and (non-viable) wide binary models shows that $\delappr$ stays on the percent level with two peaks at $|t-t\cacr|=\Delta t$ reaching as high as $8\%$. For OGLE-1999-BUL-23 we assume the preferred wide binary model of \citet{albrowetal01a}. Again, $\delappr$ stays below $2\%$ during most of the second caustic crossing with peaks at $|t-t\cacr|=\Delta t$ reaching $4\%$ at most. For this event, \citet{albrowetal01a} also found the $4\%$ discrepancy from the true amplification. As mentioned earlier, they retained the analytic approximation of equation~(\ref{eq:caus_approx}) and improved it using pre-calculated corrections.

The crossing of a cusp precludes the applicability of simple approximations such as in the fold case. Hence, the analysis of all such events required the more arduous computation of the exact light curve, usually by variants of the ray-shooting method. This approach has the clear benefit of avoiding potential biases introduced by such approximations.

\section{DISCUSSION}
\label{sec:discussion}

In our analysis we neglected blending, which is a significant effect in observed microlensing events. In general we expect the observed relative amplification excess over the point-source amplification to be less prominent. In high-amplification regions near the caustic the blending effect will be suppressed for all but the most extreme contributions of blended light. The situation is more subtle on the axis connecting the caustic components. The point-source amplification here is often rather low, e.g., $A_0 \sim 5$, so that adding an $\sim80\%$ blend to the baseline may lower $\delex$ by a factor of two. There are several reasons why not to worry about blending too much. Commonly used image-subtraction techniques automatically remove all significant (non-variable) blending. Even without image subtraction, being an additive constant blended flux is easy to fit and eliminate from light curves with decent photometry, leaving the possibility of detecting $\delex$ at least in the pattern of residuals. Based on the recovered event geometry it should be obvious where to look for the effect. In addition, improvements in crowded-field photometry and especially any potential future space-based microlensing surveys will reduce the frequency and impact of strong blending for observations towards the Galactic bulge \citep[see Figure 3 of][]{hanetal06}.

The current approach to observations of gravitational microlensing anomalies is optimized for the detection of lenses with exoplanets. In this setup two survey teams (OGLE and MOA) image the Galactic bulge at most several times per night and provide candidates for followup observations. These are carried out by two other teams (PLANET and $\mu$FUN) equipped with a network of dedicated telescopes. Followup monitoring concentrates on short-lived deviations from the PSPL light curve, especially in high-magnification events, where the chances for discovering a planet are highest \citep{griest_safizadeh98}. Unfortunately, some of the events in which the departure from the PSPL light curve does not match a planetary signal are ignored. However, many of these non-planetary events could provide valuable data on stellar surfaces that are hardly accessible by other means. Based on our results in \S~\ref{sec:sensitivity} we suggest that attention should be paid also to the parts of the light curve that correspond to approaches to any cusp axis, and the regions between facing cusps in particular. An analysis of these parts may constrain the limb darkening of the source even in non-caustic-crossing events and thus enlarge the sample of stars with measured limb darkening.

The present observational setup can be expected to change in the near future. Wide-field survey telescopes will monitor all events in their large field of view with high cadence. Recently, the MOA group started to observe a subset of their fields very frequently during the night \citep{sumi08}. Planned next-generation ground-based survey telescopes \citep{gouldetal07} will continue in the trend of dismantling the alert/followup modus operandi. Proposed space missions such as the {\em Microlensing Planet Finder} \citep{bennettetal07} would provide an almost nonstop flow of high-quality data on microlensing events in the bulge. As we have shown in \S~\ref{sec:sensitivity}, for a giant source the area sensitive to its extended nature can be up to several times wider than the caustic-crossing area. For main-sequence sources this width increase can reach several tens of percent. The developments mentioned above would give excellent prospects for a massive survey of limb darkening of bulge stars with two-point-mass microlensing.

As we demonstrated, modeling of binary microlensing events requires a very accurate computational method. Poor approximations introduce bias not only in the measured limb darkening but also in the recovered binary parameters. All of the events in which modeling adopted the linear-fold approximation given by equation~(\ref{eq:caus_approx}) would benefit from a re-analysis incorporating an accurate light-curve computation method. This applies in particular to events used for source limb-darkening measurement, as the positions with highest chromaticity are largely concentrated at $\pm \rho_*$ around the caustic where the method of \citet{albrowetal99b} tends to produce largest deviations.

\section{SUMMARY}
\label{sec:summary}

We implemented the numerically efficient method of \cite{verm00} for computing the general two-point-mass microlensing amplification of a source with an arbitrary surface-brightness distribution. Using a PCA-based model of stellar limb darkening, we explored the sensitivity of such lenses to the extended nature of the source and to differences in its limb darkening. Traditionally, all extended-source effects were expected to be limited to a narrow band along the caustic. In contrast, we discovered regions of strongly enhanced sensitivity along the outer axes of cusps, which may even bridge the gap between caustic components of close or wide binary lenses. Although for smaller sources the extent of the sensitive region along the cusp axis shrinks in terms of Einstein radii, we show that in terms of source radii it grows, scaling with the inverse cube-root of the source radius.

We derive analytical approximations for the extended-source effect and chromaticity of a small source not positioned directly on the caustic. These expressions can be readily applied not only to the two-point-mass lens, but more generally to any other gravitational lens system. We use exactly computed light curves to check the appropriateness of the linear-fold approximation, which is often used in the analysis of caustic-crossing events. The approximation leaves residuals on the level of a few percent even in near-ideal events such as OGLE-2002-BLG-069, which meet most of the conditions of the approximation. It fails badly in events such as EROS-BLG-2000-5, which violate the conditions excessively. All microlensing events in which the approximation was employed for measuring the limb darkening of the source star therefore deserve a re-analysis, if only as a consistency check in the more favorable cases. Any usage of the linear-fold approximation in event analysis should be accompanied at least by a simple test demonstrating its applicability in the particular situation.

A study of the size of contours of the extended-source effect illustrated that for instance the probability of a microlensed bulge giant causing at least a $1\%$ effect is on average 1.4--1.6 times higher than the probability of its caustic crossing, when averaged over a range of typical lens-component separations. Overall, our results bode optimistically for the prospects of measuring the limb darkening of source stars in a substantially higher number of two-point-mass microlensing events.

\acknowledgements

We thank the referee, Scott Gaudi, for his helpful comments and constructive suggestions. Work on this project was supported by Czech Science Foundation grant GACR 205/07/0824 and by the Czech Ministry of Education project MSM0021620860.

\appendix
\section{AMPLIFICATION OF IMAGES FORMED BY A CUSP}
\label{sec:appendix}
\begin{figure}
\plottwo{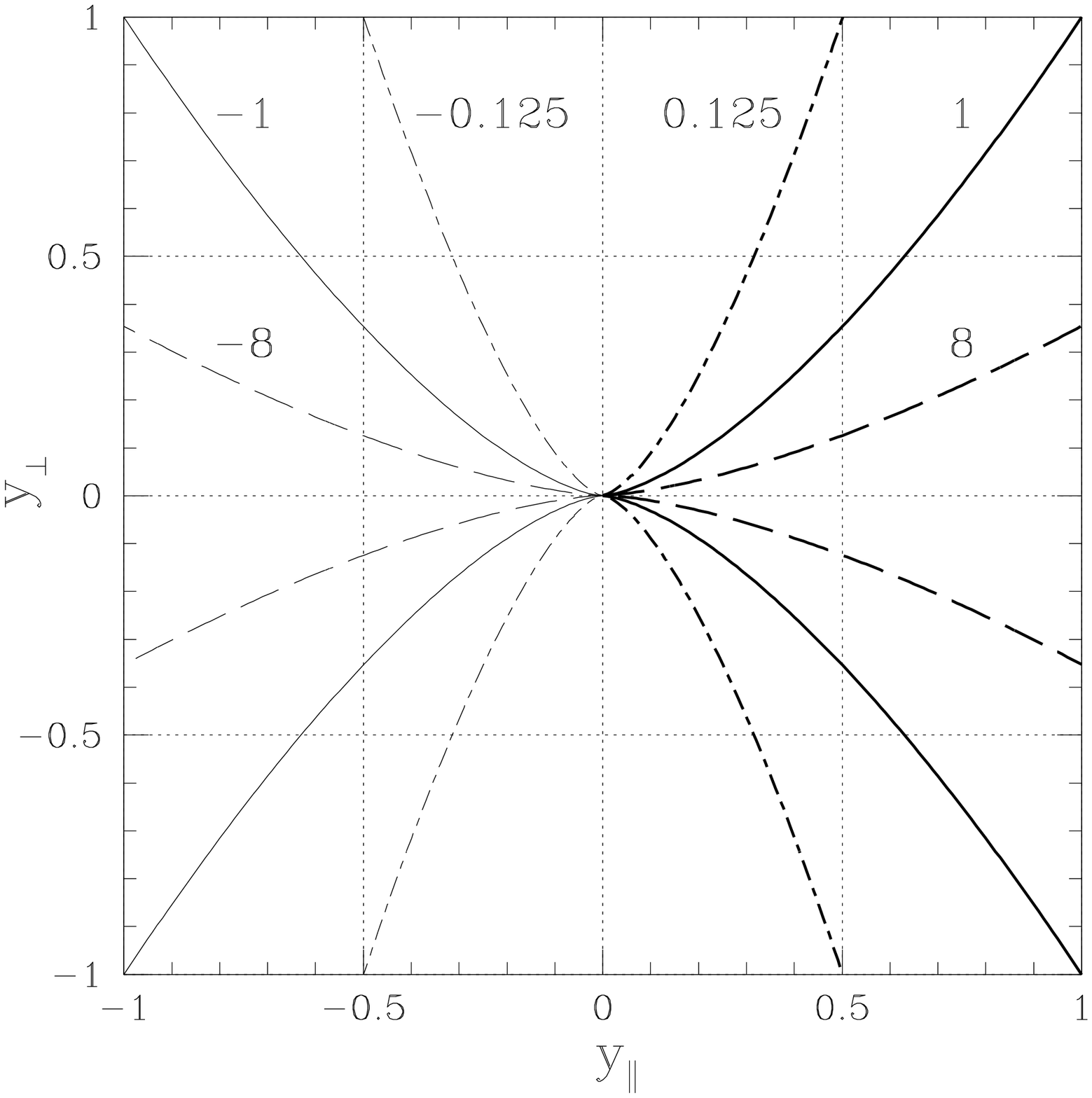}{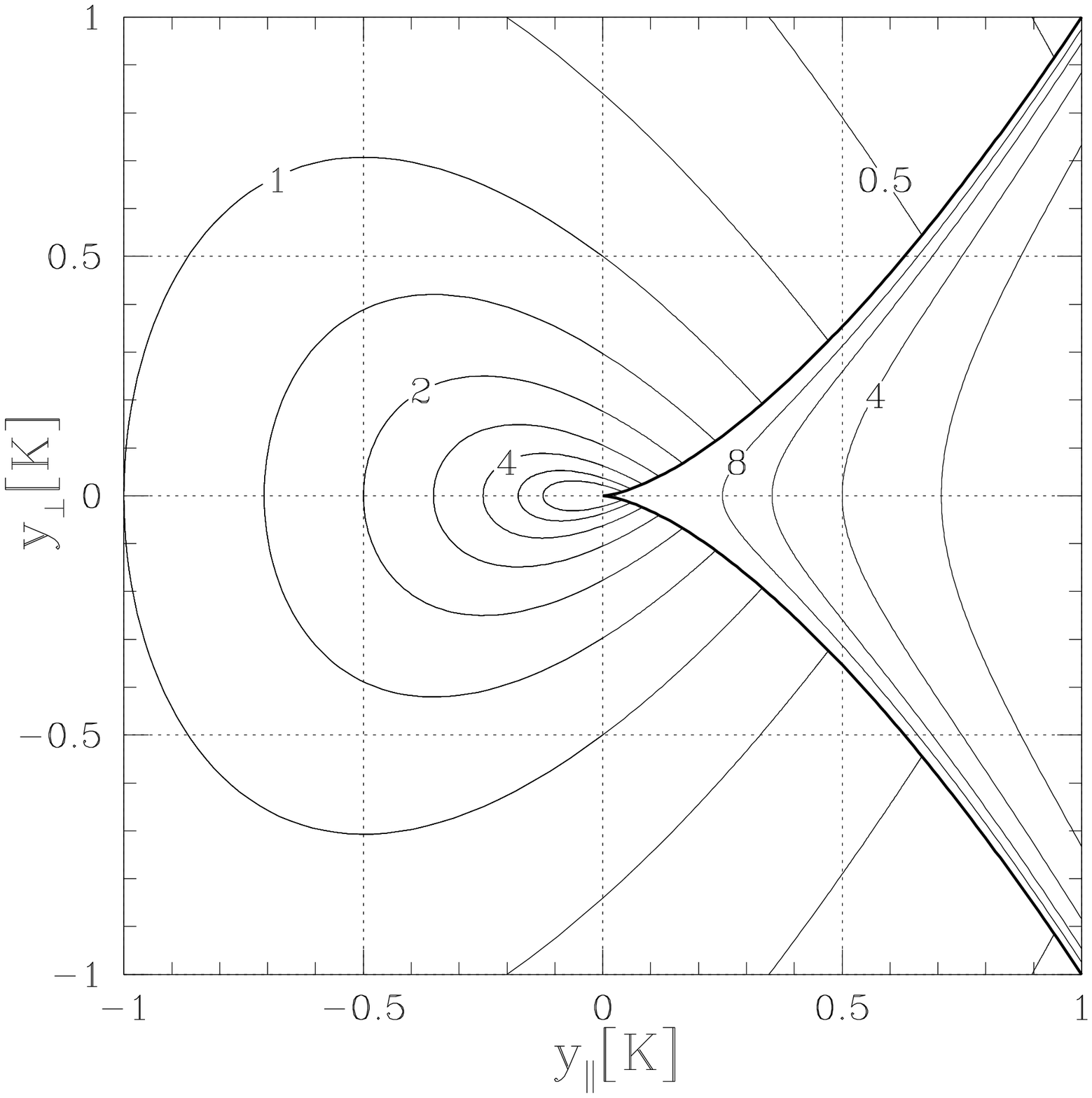}
\caption{Geometry of a generic cusp caustic given by equation~(\ref{eq:cusp_caustic}). Left panel: cusps for six values of $K$, with $K>0$ curves drawn bold and $K<0$ thin. Cusps with $|K|=1$, 8, and 0.125 are marked by solid, dashed, and dot-dashed lines, respectively. Right panel: amplification contours for lensing by a generic cusp (bold curve). Contour values in units of $|K\,b\,|^{-1}$ are spaced by a factor of $2^{1/2}$; outside the cusp levels range from the outermost $2^{-3/2}$ to the innermost $2^{3}$, inside the cusp from the rightmost $2^{3/2}$ to the leftmost $2^3$. With axes marked in units of $K$, the right panel is valid for an arbitrary cusp.
\label{fig:cusp}}
\end{figure}

As shown by \cite{schn_weiss92}, the lens equation near a generic cusp caustic placed at the origin may be written as
\begin{eqnarray}
\nonumber {\rm y_\parallel}&=&c\,{\rm x_\parallel}+\frac{b}{2}\,{\rm x_\perp^2}\\
{\rm y_\perp}&=&b\,{\rm x_\parallel x_\perp}+a\,{\rm x_\perp^3}\,,
\label{eq:cusp_lens}
\end{eqnarray}
where $({\rm y_\parallel},{\rm y_\perp})$ are components of the source position parallel and perpendicular to the axis of symmetry of the cusp, $({\rm x_\parallel},{\rm x_\perp})$ are analogously defined coordinates of the image position, and $a,\,b,\,c$ are constants fulfilling the conditions $b\neq 0$, $c\neq 0$, and $b^2-2ac\neq 0$. The equation of the cusp caustic describes a semi-cubical parabola
\beq
{\rm y_\parallel^3}=K\,{\rm y_\perp^2}\,,
\label{eq:cusp_caustic}
\eeq
where
\beq
K=\frac{27\,c^2}{8\,b^3}(b^2-2ac)
\label{eq:cusp_parameter}
\eeq
is a parameter determining the orientation of the cusp and its narrowness, as shown in the left panel of Figure~\ref{fig:cusp}. A cusp with $K>0$ is pointed to the left (towards negative ${\rm y_\parallel}$), a cusp with $K<0$ is pointed to the right (towards positive ${\rm y_\parallel}$). A cusp with a large value of $|K|$ is narrower than a cusp with a small value of $|K|$. Without a loss of generality we choose for the rest of this section the orientation with $K>0$. If this is not fulfilled automatically by equations~(\ref{eq:cusp_lens}), it can be readily achieved by rotating the coordinate systems ${\bf y}$ and ${\bf x}$ by $180^\circ$, which corresponds to changing the sign of parameter $b$.

Inversion of equations~(\ref{eq:cusp_lens}) leads to a cubic equation for ${\rm x_\perp}$, and the amplification of each image can be expressed as a function of the image position \citep{schn_weiss92}. \cite{zakharov95,zakharov99} presented an elegant alternative method for obtaining image amplifications as a function of source position, without explicitly solving the lens equation. From general properties of the roots of the image equation Zakharov derived a simple cubic equation for the amplification, the roots of which are the amplifications of the images. We reproduce the main results here because of a few typos appearing in the original papers. A point source satisfying ${\rm y^2_\perp}<K^{-1}\,{\rm y^3_\parallel}$ lies inside the cusp and has three images with amplifications
\beq
A^{(j)}({\rm y_\parallel},{\rm y_\perp})= \frac{1}{b}\;\sqrt{\frac{{\rm y_\parallel}}{{\rm y^3_\parallel}-K{\rm y^2_\perp}}}\, \cos\frac{1}{3}\left[\arcsin \sqrt{\frac{K{\rm y^2_\perp}}{{\rm y^3_\parallel}}} + 2\pi(j-1) \right]\,,\;j=1,2,3\;.
\label{eq:cusp_inner_ampl}
\eeq
Note that these amplifications include the appropriate sign corresponding to the parity of the image. The $j=1$ image parity is equal to the sign of $b$, while the $j=2$ and $j=3$ images both have the opposite parity. The total amplification of a source inside the cusp $A^{\rm tot}=\sum_{j=1}^3|A^{(j)}|=2\,|A^{(1)}|$, hence we have
\beq
A^{\rm tot}({\rm y_\parallel},{\rm y_\perp})= \frac{2}{ |\,b\,|}\;\sqrt{\frac{{\rm y_\parallel}}{{\rm y^3_\parallel}-K{\rm y^2_\perp}}}\, \cos\left[\frac{1}{3}\,\arcsin \sqrt{\frac{K{\rm y^2_\perp}}{{\rm y^3_\parallel}}}\,\right]\,.
\label{eq:cusp_inner_total_ampl}
\eeq
A point source lying outside the cusp satisfies ${\rm y^2_\perp}>K^{-1}\,{\rm y^3_\parallel}$ and has a single image amplified by
\beq
A^{(0)}({\rm y_\parallel},{\rm y_\perp})=-\,\frac{K^{-1/3}}{2\,b}\frac{\left[ \sqrt{{\rm y_\perp^2}-K^{-1}\,{\rm y_\parallel^3}}+{\rm y_\perp} \right]^{1/3}+\left[ \sqrt{{\rm y_\perp^2}-K^{-1}\,{\rm y_\parallel^3}}-{\rm y_\perp} \right]^{1/3}} {\sqrt{{\rm y_\perp^2}-K^{-1}\,{\rm y_\parallel^3}}}\,.
\label{eq:cusp_outer_ampl}
\eeq
This image has the same parity as the $j=2$ and $j=3$ images, that is, opposite to the sign of $b$. The total amplification of a source outside the cusp thus is
\beq
A^{\rm tot}({\rm y_\parallel},{\rm y_\perp})=|A^{(0)}({\rm y_\parallel},{\rm y_\perp})|\,.
\label{eq:cusp_outer_total_ampl}
\eeq

In the right panel of Figure~\ref{fig:cusp} we plotted contours of constant total amplification $A^{\rm tot}({\bf y})$ with levels given in units of $|K\,b\,|^{-1}$. By marking both axes in units of $K$ the plot is valid for an arbitrary $K>0$ cusp.


\begin{thebibliography}{dummy}

\bibitem[Abe et al.(2003)]{abeetal03}
Abe, F., et al. 2003, \aap, 411, L493
\bibitem[Afonso et al.(2000)]{afonsoetal00}
Afonso, C., et al. 2000, \apj, 532, 340
\bibitem[Afonso et al.(2001)]{afonsoetal01}
Afonso, C., et al. 2001, \aap, 378, 1014
\bibitem[Albrow et al.(1999a)]{albrowetal99a}
Albrow, M. D., et al. 1999a, \apj, 522, 1011
\bibitem[Albrow et al.(1999b)]{albrowetal99b}
Albrow, M. D., et al. 1999b, \apj, 522, 1022
\bibitem[Albrow et al.(2000)]{albrowetal00}
Albrow, M. D., et al. 2000, \apj, 534, 894
\bibitem[Albrow et al.(2001a)]{albrowetal01a}
Albrow, M. D., et al. 2001a, \apj, 549, 759
\bibitem[Albrow et al.(2001b)]{albrowetal01b}
Albrow, M. D., et al. 2001b, \apj, 550, L173
\bibitem[Alcock et al.(2000)]{alcocketal00}
Alcock, C., et al. 2000, \apj, 541, 270
\bibitem[An et al.(2002)]{anetal02}
An, J. H., et al. 2002, \apj, 572, 521
\bibitem[Beaulieu et al.(2006)]{beaulieuetal06}
Beaulieu, J.-P., et al. 2006, \nat, 439, 437
\bibitem[Bennett \& Rhie(1996)]{bennett_rhie96}
Bennett, D. P., \& Rhie, S. H. 1996, \apj, 472, 660
\bibitem[Bennett et al.(2007)]{bennettetal07}
Bennett, D. P., et al.\ 2007, white paper submitted to the NASA/NSF Exoplanet Task Force, arXiv:0704.0454
\bibitem[Bennett et al.(2008)]{bennettetal08}
Bennett, D. P., et al.\ 2008, \apj, 684, 663
\bibitem[Blandford \& Narayan(1986)]{bland_nara86}
Blandford, R. D., \& Narayan, R. 1986, \apj, 310, 568
\bibitem[Bond et al.(2004)]{bondetal04}
Bond, I. A., et al. 2004, \apj, 606, L155
\bibitem[Cassan et al.(2004)]{cassanetal04}
Cassan, A., et al. 2004, \aap, 419, L1
\bibitem[Castro et al.(2001)]{castroetal01}
Castro, S., Pogge, R. W., Rich, R. M., DePoy, D. L., \& Gould, A. 2001, \apj, 548, L197
\bibitem[Di Stefano \& Perna(1997)]{di_stefano_perna97}
Di Stefano, R., \& Perna, R. 1997, \apj, 488, 55
\bibitem[Dominik(1998)]{dominik98}
Dominik, M. 1998, \aap, 333, L79
\bibitem[Dominik(1999)]{dominik99}
Dominik, M. 1999, \aap, 349, 108
\bibitem[Dominik(2004)]{dominik04}
Dominik, M. 2004, \mnras, 353, 69
\bibitem[Dominik(2007)]{dominik07}
Dominik, M. 2007, \mnras, 377, 1679
\bibitem[Dong et al.(2006)]{dongetal06}
Dong, S., et al. 2006, \apj, 642, 842
\bibitem[Dong et al.(2008)]{dongetal08}
Dong, S., et al. 2008, \apj, submitted (arXiv/0804.1354)
\bibitem[Erdl \& Schneider(1993)]{erdl_schneider93}
Erdl, H., \& Schneider, P. 1993, \aap, 268, 453
\bibitem[Fields et al.(2003)]{fieldsetal03}
Fields, D. L., et al. 2003, \apj, 596, 1305
\bibitem[Gaudi \& Petters(2002a)]{gaudi_petters02a}
Gaudi, B. S., \& Petters, A. O. 2002a, \apj, 574, 970
\bibitem[Gaudi \& Petters(2002b)]{gaudi_petters02b}
Gaudi, B. S., \& Petters, A. O. 2002b, \apj, 580, 468
\bibitem[Gaudi et al.(2008)]{gaudietal08}
Gaudi, B. S., et al. 2008, Science, 319, 927
\bibitem[Gould(1994)]{gould94}
Gould, A. 1994, \apj, 421, L71
\bibitem[Gould(2008)]{gould08}
Gould, A. 2008, \apj, 681, 1593
\bibitem[Gould \& Gaucherel(1997)]{gould_gaucherel97}
Gould, A., \& Gaucherel, C. 1997, \apj, 477, 580
\bibitem[Gould et al.(2007)]{gouldetal07}
Gould, A., Gaudi, B. S., \& Bennett, D. P. 2007, white paper submitted to the NASA/NSF Exoplanet Task Force, arXiv:0704.0767
\bibitem[Gould et al.(2006)]{gouldetal06}
Gould, A., et al. 2006, \apj, 644, L37
\bibitem[Griest \& Safizadeh(1998)]{griest_safizadeh98}
Griest, K., \& Safizadeh, N. 1998, \apj, 500, 37
\bibitem[Han \& Park(2001)]{han_park01}
Han, C., \& Park, S.-H. 2001, \mnras, 320, 41
\bibitem[Han et al.(2006)]{hanetal06}
Han, C., Park, B.-G., Kim, H.-I., \& Chang, K. 2006, \apj, 653, 963
\bibitem[Heyrovsk\'y(2003)]{hey03}
Heyrovsk\'y, D. 2003, \apj, 594, 464
\bibitem[Heyrovsk\'y(2008)]{hey08}
Heyrovsk\'y, D. 2008, in Proceedings of the Manchester Microlensing Conference, ed. E. Kerins, S. Mao, N. Rattenbury, \& L. Wyrzykowski (Manchester: Univ.~of~Manchester), PoS(GMC8)028, \url{http://pos.sissa.it/cgi-bin/reader/conf.cgi?confid=54}
\bibitem[Heyrovsk\'y \& Loeb(1997)]{heylo97}
Heyrovsk\'y, D., \& Loeb, A. 1997, \apj, 490, 38
\bibitem[Heyrovsk\'y et al.(2000)]{heysalo00}
Heyrovsk\'y, D., Sasselov, D., \& Loeb, A. 2000, \apj, 543, 406
\bibitem[Jaroszy\'nski(2002)]{jaroszynski02}
Jaroszy\'nski, M.\ 2002, \actaa, 52, 39
\bibitem[Jaroszy\'nski et al.(2004)]{jaroszynskietal04}
Jaroszy\'nski, M., et al.\ 2004, \actaa, 54, 103
\bibitem[Jaroszy\'nski et al.(2006)]{jaroszynskietal06}
Jaroszy\'nski, M., et al.\ 2006, \actaa, 56, 307
\bibitem[Kubas et al.(2005)]{kubasetal05}
Kubas, D., et al. 2005, \aap, 435, 941
\bibitem[Mao \& Paczy\'nski(1991)]{mao_paczynski91}
Mao, S., \& Paczy\'nski, B.\ 1991, \apjl, 374, L37
\bibitem[Night et al.(2007)]{nightetal07}
Night, C., Di Stefano, R., \& Schwamb, M. 2007, \apj, submitted (arXiv/0705.0169)
\bibitem[Rattenbury et al.(2002)]{rattetal02}
Rattenbury, N. J., Bond I. A., Skuljan, J., \& Yock, P. C. M. 2002, \mnras, 335, 159
\bibitem[Rhie(2002)]{rhie02}
Rhie, S. H. 2002, preprint (astro-ph/0207612)
\bibitem[Rhie \& Bennett(1999)]{rhie_bennett99}
Rhie, S. H., \& Bennett, D. P. 1999, preprint (astro-ph/9912050)
\bibitem[Schneider \& Wei\ss(1986)]{schn_weiss86}
Schneider, P., \& Wei\ss, A. 1986, \aap, 164, 237
\bibitem[Schneider \& Wei\ss(1992)]{schn_weiss92}
Schneider, P., \& Wei\ss, A. 1992, \aap, 260, 1
\bibitem[Skowron et al.(2007)]{skowronetal07}
Skowron, J., et al.\ 2007, \actaa, 57, 281
\bibitem[Sumi(2008)]{sumi08}
Sumi, T. 2008, in Proceedings of the Manchester Microlensing Conference, ed. E. Kerins, S. Mao, N. Rattenbury, \& L. Wyrzykowski (Manchester: Univ.~of~Manchester), PoS(GMC8)025, \url{http://pos.sissa.it/cgi-bin/reader/conf.cgi?confid=54}
\bibitem[Udalski et al.(2005)]{udalskietal05}
Udalski, A., et al.\ 2005, \apjl, 628, L109
\bibitem[Vermaak(2000)]{verm00}
Vermaak, P. 2000, \mnras, 319, 1011
\bibitem[Wambsganss \& Paczy\'nski(1991)]{wambsganss_paczynski91}
Wambsganss, J., \& Paczy\'nski, B.\ 1991, \aj, 102, 864
\bibitem[Witt(1990)]{witt90}
Witt, H. J. 1990, \aap, 236, 311
\bibitem[Witt(1995)]{witt95}
Witt, H. J. 1995, \apj, 449, 42
\bibitem[Witt \& Mao(1995)]{witt_mao95}
Witt, H. J., \& Mao, S. 1995, \apj, 447, L105
\bibitem[Witt \& Petters(1993)]{witt_petters93}
Witt, H. J., \& Petters, A. O. 1993, Journal of Mathematical Physics, 34, 4093
\bibitem[Zakharov(1995)]{zakharov95}
Zakharov, A. F. 1995, \aap, 293, 1
\bibitem[Zakharov(1999)]{zakharov99}
Zakharov, A. F. 1999, Astronomical and Astrophysical Transactions, 18, 17

\end{thebibliography}
\end{document}